\RequirePackage{fix-cm}
\documentclass[natbib,smallextended]{svjour3}       
\smartqed  
\usepackage{graphicx}
\usepackage{bm}
\usepackage{natbib}
\usepackage[colorlinks,allcolors=blue]{hyperref}
\usepackage{amsmath}
\usepackage{amsfonts}
\usepackage{amssymb}
\usepackage{mathrsfs}
\DeclareMathAlphabet{\mathscrbf}{OMS}{mdugm}{b}{n}
\bibpunct{(}{)}{;}{a}{}{,}
\graphicspath{{./fig/}{./png/}}
\usepackage{mathrsfs}
\usepackage{xcolor}

\newcommand{\mcbm}[1]{\bm{\mathcal{#1}}}
\newcommand{\aaa}{{\bm a}}

\newcommand{\bbb}{{\bm b}}

\newcommand{\fff}{{\bm f}}

\newcommand{\jjj}{{\bm j}}

\newcommand{\nnn}{{\bm n}}

\newcommand{\uuu}{{\bm u}}

\newcommand{\xxx}{{\bm x}}

\newcommand{\ooo}{{\bm \omega}}
\newcommand{\AAA}{{\bm A}}
\newcommand{\mAAA}{\overline{\bm A}}

\newcommand{\BBB}{{\bm B}}
\newcommand{\mBBB}{\overline{\bm B}}
\newcommand{\mBBBimp}{\overline{\bm B}^{(\rm imp)}}

\newcommand{\CCC}{{\bm C}}

\newcommand{\EEE}{{\bm E}}

\newcommand{\mcFFF}{\bm{\mathcal{F}}}
\newcommand{\mFFF}{\overline{\bm F}}

\newcommand{\MMM}{{\bm M}}

\newcommand{\JJJ}{{\bm J}}
\newcommand{\mJJJ}{\overline{\bm J}}

\newcommand{\UUU}{{\bm U}}
\newcommand{\mUUU}{\overline{\bm U}}

\newcommand{\mTTT}{\overline{\bm T}}

\newcommand{\mWWW}{\overline{\bm W}}

\newcommand{\gggg}{{\bm g}}
\newcommand{\mUUp}{\overline{U}_\phi}

\newcommand{\mBBi}{\overline{B}_i}
\newcommand{\mBBj}{\overline{B}_j}

\newcommand{\mBBx}{\overline{B}_x}
\newcommand{\mBBy}{\overline{B}_y}

\newcommand{\mBBr}{\overline{B}_r}

\newcommand{\mBBt}{\overline{B}_\theta}
\newcommand{\mBBp}{\overline{B}_\phi}

\newcommand{\mJJj}{\overline{J}_j}

\newcommand{\mEMF}{\overline{\bm{\mathcal{E}}}}

\newcommand{\EMFi}{\overline{\mathcal{E}}_i}

\newcommand{\alppp}{\alpha_{\phi\phi}}
\newcommand{\alptt}{\alpha_{\theta\theta}}
\newcommand{\alpij}{\alpha_{ij}}

\newcommand{\etaij}{\eta_{ij}}
\newcommand{\etaijk}{\eta_{ijk}}

\newcommand{\bcdot}{\bm\cdot}
\newcommand{\Eq}[1]{Eq.~(\ref{#1})}
\newcommand{\Equ}[1]{Equation~(\ref{#1})}

\newcommand{\Eqs}[2]{Equations~(\ref{#1}) and (\ref{#2})} 

\newcommand{\Eqsa}[2]{Eqs.~(\ref{#1}) and (\ref{#2})} 

\newcommand{\EQ}{\begin{equation}}
\newcommand{\EN}{\end{equation}}
\newcommand{\EQA}{\begin{eqnarray}}
\newcommand{\ENA}{\end{eqnarray}}
\newcommand{\brac}[1]{\langle #1 \rangle}
\newcommand{\pd}{\partial}

\newcommand{\DIV}{\bm{\nabla}\bm\cdot }

\newcommand{\mean}[1]{\overline{#1}}

\newcommand{\cP}{c_{\rm P}}
\newcommand{\cV}{c_{\rm V}}

\newcommand{\cst}{c_{\rm s}^2}

\newcommand{\etat}{\eta_{\rm t}}
\newcommand{\etatz}{\eta_{\rm t0}}
\newcommand{\etaT}{\eta_{\rm T}}
\newcommand{\tetat}{\tilde{\eta}_{\rm t}}
\newcommand{\urms}{u_{\rm rms}}

\newcommand{\orms}{\omega_{\rm rms}}

\newcommand{\ocyc}{\omega_{\rm cyc}}
\newcommand{\Pcyc}{P_{\rm cyc}}

\newcommand{\Beq}{B_{\rm eq}}
\newcommand{\Ma}{{\rm Ma}}

\newcommand{\kmax}{k_{\rm max}}

\newcommand{\kf}{k_{\rm f}}

\newcommand{\Peff}{{\mathcal P}_{\rm eff}}
\newcommand{\lammax}{\lambda_{\rm max}}
\newcommand{\chit}{\chi_{\rm t}}

\newcommand{\Co}{{\rm Co}}

\newcommand{\CoF}{{\rm Co}_{\rm F}}

\newcommand{\Ek}{{\rm Ek}}

\newcommand{\Pe}{{\rm Pe}}

\newcommand{\Pra}{{\rm Pr}}

\newcommand{\PrM}{{\rm Pr}_{\rm M}}

\newcommand{\Ra}{{\rm Ra}}

\newcommand{\RaF}{{\rm Ra}_{\rm F}}
\newcommand{\RaFS}{{\rm Ra}_{\rm F}^\star}

\newcommand{\Rey}{{\rm Re}}

\newcommand{\Rm}{{\rm Rm}}
\newcommand{\ReM}{{\rm Re}_{\rm M}}
\newcommand{\Ro}{{\rm Ro}}
\newcommand{\Sh}{{\rm Sh}}
\newcommand{\St}{{\rm St}}
\newcommand{\Ta}{{\rm Ta}}

\newcommand{\Calp}{C_\alpha}
\newcommand{\Calpcrit}{C_\alpha^{\rm crit}}
\newcommand{\tauc}{\tau_{\rm c}}

\newcommand{\kinhel}{\bm\omega\bm\cdot\uuu}
\newcommand{\mkinhel}{\mean{\bm\omega\bm\cdot\uuu}}
\newcommand{\curhel}{\jjj\bm\cdot\bbb}
\newcommand{\mcurhel}{\mean{\jjj\bm\cdot\bbb}}
\newcommand{\maghel}{\aaa\bm\cdot\bbb}

\newcommand{\HelM}{\mathcal{H}_{\rm M}}

\newcommand{\alphaK}{\alpha_{\rm K}}
\newcommand{\alphaM}{\alpha_{\rm M}}
\newcommand{\mOm}{\mean{\Omega}}

\newcommand{\mvOm}{\mean{\bm\Omega}}

\newcommand{\omcyc}{\omega_{\rm cyc}}

\newcommand{\nab}{\bm\nabla}

{}

\newcommand{\Rgas}{{\cal R}}
\newcommand{\Rsun}{R_\odot}

\newcommand{\FFFrad}{{\bm F}^{\rm rad}}

\newcommand{\Ftot}{F_{\rm tot}}

\newcommand{\rhoref}{\rho_{\rm ref}}

\def\onethird{\textstyle\frac{1}{3}}
\def\onehalf{\textstyle\frac{1}{2}}
\def\onesixth{\textstyle\frac{1}{6}}
\newcommand{\Figa}[1]{Fig.~\ref{#1}}

\newcommand{\Figu}[1]{Figure~\ref{#1}}

\newcommand{\Seca}[1]{Sect.~\ref{#1}}
\newcommand{\Sec}[1]{Section~\ref{#1}} 

\newcommand{\Table}[1]{Table~\ref{#1}}

\definecolor{ForestGreen}{RGB}{34,139,34}
\definecolor{AGray}{rgb}{.4,.4,.4}
\definecolor{LightYellow}{rgb}{1.,1.,.8}
\definecolor{LightCyan}{rgb}{0.88,1,1}
\chardef\us=`\_
\begin{document}

\title{Connecting mean-field theory with dynamo simulations}

\author{Petri J.\ K\"apyl\"a}

\institute{Petri J. K\"apyl\"a \at
              Institute for Solar Physics (KIS)\\
              Georges-K\"ohler-Allee 401a\\
              D-79110 Freiburg im Breisgau, Germany\\
              Tel.: +49-761-3198229\\
              \email{pkapyla@leibniz-kis.de}\\
              ORCID: 0000-0001-9619-0053
}

\date{Received: date / Accepted: date}

\maketitle

\begin{abstract}
Mean-field dynamo theory, describing the evolution of large-scale
magnetic fields, has been the mainstay of theoretical interpretation
of magnetism in astrophysical objects such as the Sun for several
decades. More recently, three-dimensional magnetohydrodynamic
simulations have reached a level of fidelity where they capture dynamo
action self-consistently on local and global scales without resorting
to parametrization of unresolved scales. Recent global simulations
also capture many of the observed characteristics of solar and stellar
large-scale magnetic fields and cycles. Successful explanation of the
results of such simulations with corresponding mean-field models is a
crucial validation step for mean-field dynamo theory. Here the
connections between mean-field theory and current dynamo simulations
are reviewed. These connections range from the numerical computation
of turbulent transport coefficients to mean-field models of
simulations, and their relevance to the solar dynamo. Finally, the
most notable successes and current challenges in mean-field
theoretical interpretations of simulations are summarized.

\end{abstract}
\newpage

\tableofcontents
\newpage

\section{Introduction and scope of the review}
\label{intro}

Development of dynamo theory has its motivation in solar observations
starting from Schwabe's realization of cyclicity of sunspots
\citep{1844AN.....21..233S} and Hale's discovery of magnetic fields
\citep{1908ApJ....28..315H} and polarity reversals
\citep{1919ApJ....49..153H} of sunspots. Now we know that the solar
magnetic field shows relatively coherent quasi-cyclic behavior with a
22-year period with superimposed longer term variations and extended
minima such as the Maunder minimum
\citep[e.g.][]{2015LRSP...12....4H}. Direct observations of the
sunspot cycle extend over more than four centuries
\citep[e.g.][]{2020LRSP...17....1A} whereas the activity of the Sun
can be followed over a much longer period using cosmogenic isotopes
gathered from ice cores and tree rings
\citep[e.g.][]{2004Natur.431.1084S,2017LRSP...14....3U}. The apparent
regularity of the solar cycle is remarkable considering the underlying
highly turbulent and chaotic dynamics of the solar plasma
\citep[e.g.][]{MT09,2020RvMP...92d1001S}.

The efforts to explain solar magnetism go back to the seminal work of
\cite{Larmor1919} who proposed that rotation of sunspots maintains the
observed magnetic fields. This road led to an impasse with Cowling's
anti-dynamo theorem \citep{1933MNRAS..94...39C}, stating that a purely
axisymmetric magnetic field cannot be maintained by dynamo action. The
first successful solar dynamo model was presented by \cite{Pa55b} who
considered the combined action of helical convection cells and
differential rotation, ideas which are still central concepts in
current efforts to explain solar and stellar dynamos \citep[e.g.][and
  references therein]{2020LRSP...17....4C,2023SSRv..219...55B}. These
same ideas were incorporated in the mathematically rigorous mean-field
dynamo theory that was developed independently in the former German
Democratic Republic under the leadership of Max Steenbeck
\citep[e.g.][]{1966ZNatA..21..369S,1967ZNatA..22..671K,KR80}. This
theory is based on the idea that turbulent motions do not only diffuse
but also generate large-scale magnetic fields, with differential
rotation and meridional flows contributing to the process. The advent
of mean-field dynamo theory led to a proliferation of dynamo models of
the Sun, planets, and of stars other than the Sun
\citep[e.g.][]{1969AN....291...49S,1969AN....291..271S,1971A&A....13..203S,MB95,2017MNRAS.466.3007P}.

The main difficulty with mean-field dynamo models is that the
turbulent electromotive force $\mEMF = \mean{\uuu \times \bbb}$, where
the overbars denote suitably chosen averages, and where $\uuu$ and
$\bbb$ are the turbulent (small-scale) velocities and magnetic fields,
needs to be known. This information is not available observationally
from the interiors astrophysical objects such as the Sun. Analytic
studies face the turbulence closure problem
\citep[e.g.][]{1991AnRFM..23..107S} and have to resort to
approximations that typically cannot be rigorously justified in
astrophysically relevant parameter regimes
\citep[e.g.][]{2007GApFD.101..117R}. Mean-field models are therefore
susceptible to ad hoc modifications and simplifications, and
very different physical ingredients can be used to construct models
that reproduce, for example, the salient features of the solar cycle
\citep[e.g.][and references therein]{2020LRSP...17....4C}.

Numerical simulations solving the equations of magnetohydrodynamics
(MHD) self-consistently in spherical shells have been around since the
1980s with pioneering works of \cite{1981ApJS...46..211G},
\cite{Gi83}, and \cite{Gl85}. While these simulations produced
differential rotation reminiscent of the Sun with a fast equator and
slower poles, the magnetic field solutions were distinctly non-solar
such that either no cycles were detected \citep{1981ApJS...46..211G}
or the \emph{dynamo waves} propagated toward the poles instead of the
equator as in the Sun \citep{Gi83}. The steady increase of computing
power has enabled more comprehensive parameter studies and changed the
picture to the extent that \emph{ab initio} simulations produce
solutions that are in many respects similar to the Sun within a
limited range of parameters. For example, solar-like large-scale
differential rotation \citep[e.g.][]{BMT04,KKB14,HRY15a} and cyclic
equatorward propagating magnetic fields
\citep[e.g.][]{GCS10,KMB12,ABMT15,SBCBN17,2022ApJ...926...21B} emerge
routinely in such models. Nevertheless, the mechanism producing the
solar-like dynamo solutions in many of these simulations is unlikely
to be the same as in the Sun \citep[][see also
  \Sec{subsec:globalco}]{WKKB14}.

Furthermore, the parameter regimes of even the most recent
state-of-the-art simulations are still far removed from the Sun
\citep[e.g.][]{O03,2023SSRv..219...58K} and it is questionable whether
these simulations have reached an asymptotic regime where diffusion at
small scales no longer affects the results at large scales
\citep[e.g.][]{2017A&A...599A...4K,2022ApJ...933..199H,2022ApJ...940..151G}. Even
though the computing power a modern astrophysicist has access to is
rapidly increasing, the immense disparity of spatial and temporal
scales in the convection zones of the Sun and stars means that direct
simulations of solar and stellar dynamos are still far beyond the
reach of any current or foreseeable supercomputers
\citep[e.g.][]{2017LRCA....3....1K,2023SSRv..219...58K}. Finally, even
in the case that fully realistic simulations of the Sun were
available, it is more than likely that we would not be able to
understand them without resorting to simpler theoretical models that
capture the essential physics. Thus there is still a great demand for
mean-field models that faithfully capture the relevant dynamics of
complex 3D systems.

From the point of view of mean-field dynamo theory, 3D simulations
offer a great advantage over observations of astrophysical objects in
that the full information about the flows, magnetic fields, and the
electromotive force is readily at hand. Thus it is much easier to
construct mean-field models of simulations than of the Sun where the
detailed information of flows and magnetic fields is missing and
therefore subject to speculation. Furthermore, such comparisons can be
viewed as an essential validation step of the mean-field dynamo theory
itself. That is, if mean-field models can accurately capture the
physics of 3D simulations, there is hope to do the same with more
complex systems such as the Sun. However, starting such detailed
comparisons with mean-field theory from simulations of solar and
stellar dynamos is likely to be complicated and the results are not
necessarily easy to interpret. Thus it is often more fruitful for
physical understanding to study systems that are considerably simpler
and which isolate one or a few of the physical effects that are
present in the highly non-linear, strongly stratified, and rotating
convection zone of the Sun. Such simpler setups will also be the
starting point of the comparisons between simulations and mean-field
theory in this review, with systems of increasing complexity following
from there.

The outline of the review is as follows: since the mean-field models
and simulations often seek to capture some physical system such as the
Sun, a brief overview of the relevant observations is therefore
warranted. This is presented in \Seca{sec:obser} with emphasis on the
salient solar observations. \Sec{sec:dynsims} summarizes different
types of 3D dynamo simulations, what they try to accomplish, their
usefulness with respect to comparisons to mean-field theory, and their
relation to the Sun and stars. A brief outline of mean-field dynamo
theory is presented in \Seca{sec:dyntheo}. The necessary prerequisites
for comparing simulations with corresponding mean-field theory are
discussed in \Seca{sec:prereq}. Methods for extracting turbulent
transport coefficients from simulations and results obtained with such
methods are discussed in \Seca{sec:memethods}. Finally, comparisons of
various kinds between simulations and mean-field models are discussed
in \Seca{sec:comp}. Finally, outstanding issues are discussed in
\Seca{sec:issues}, and \Seca{sec:conclusions} sums up the current state
of affairs.

\section{Brief overview of relevant solar observations}
\label{sec:obser}

The ultimate aim of 3D dynamo simulations and mean-field models is to
address parts or the totality of the dynamo in some real
system. Therefore it is in order to briefly summarize the pertinent
observations that the models seek to capture. Here the discussion is
limited to solar global magnetism and large-scale flows that are an
essential ingredient in the dynamo process.

\subsection{Large-scale magnetism of the Sun}

\begin{figure}[t]
\begin{center}
\includegraphics[width=\textwidth]{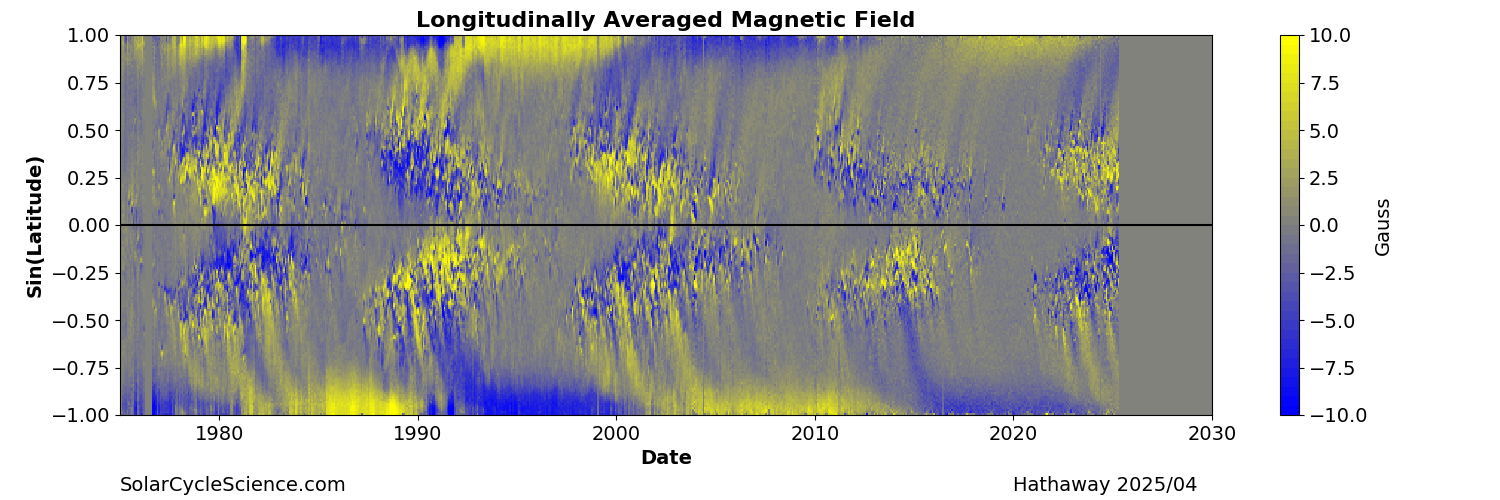}
\end{center}
\caption[]{Time-latitude or butterfly diagram of the longitudinally
  averaged radial magnetic field at the solar surface. The color scale
  is clipped at $\pm10$~G to highlight the weak fields on high
  latitudes. Courtesy of David Hathaway
  (\href{http://solarcyclescience.com}{http://solarcyclescience.com}).}
\label{fig:solarmagbfly}
\end{figure}

The observational knowledge of the Sun both in terms of quality and
quantity far exceeds that of any other star. Therefore it is logical
that efforts at dynamo modeling have to a large extent targeted the
Sun \citep[cf.][]{2023SSRv..219...35C}. More than four centuries of
systematic observations has revealed the approximately 11-year sunspot
cycle and a century of magnetic field observations the 22-year
magnetic cycle. Furthermore, sunspots appear on a latitude strip
$\pm40^\circ$, appearing progressively closer to the equator as the
cycle advances. The surface magnetic fields consists largely of
bipolar regions with opposite polarities at different hemispheres. A
poleward branch, believed to be caused by turbulent diffusion and
advection by the meridional flow \citep[see][and references
  therein]{Jiang_et_al_2014_SSRv_186_491}, is also present at high
latitudes; see \Figa{fig:solarmagbfly}. The polar field changes sign
at sunspot maximum and the radial and longitudinal fields are thought
to be in anticorrelation \citep[][]{St76}. The spatiotemporal behavior
of the large-scale magnetic fields of the Sun captured in this figure
is the primary observable that 3D simulations and mean-field models
seek to reproduce. The reader is referred to other reviews in this
series for the details of the solar cycle and its long-term
variations; see, e.g., \cite{2015LRSP...12....4H},
\cite{2017LRSP...14....3U}, and \cite{Karak_2023_LRSP_20_3}.

\begin{figure}[t]
\begin{center}
\includegraphics[width=0.51\textwidth]{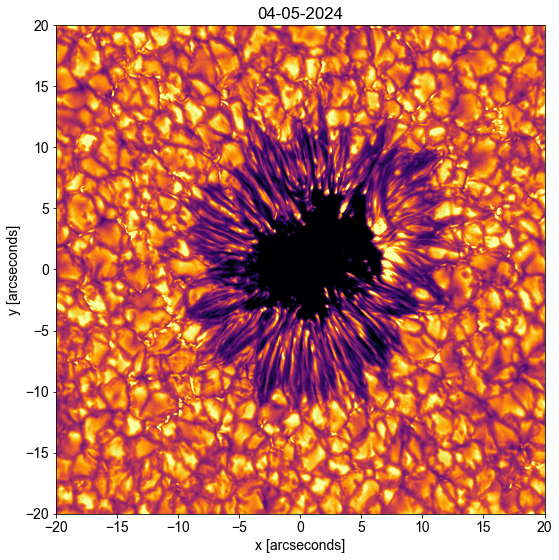}\includegraphics[width=0.49\textwidth]{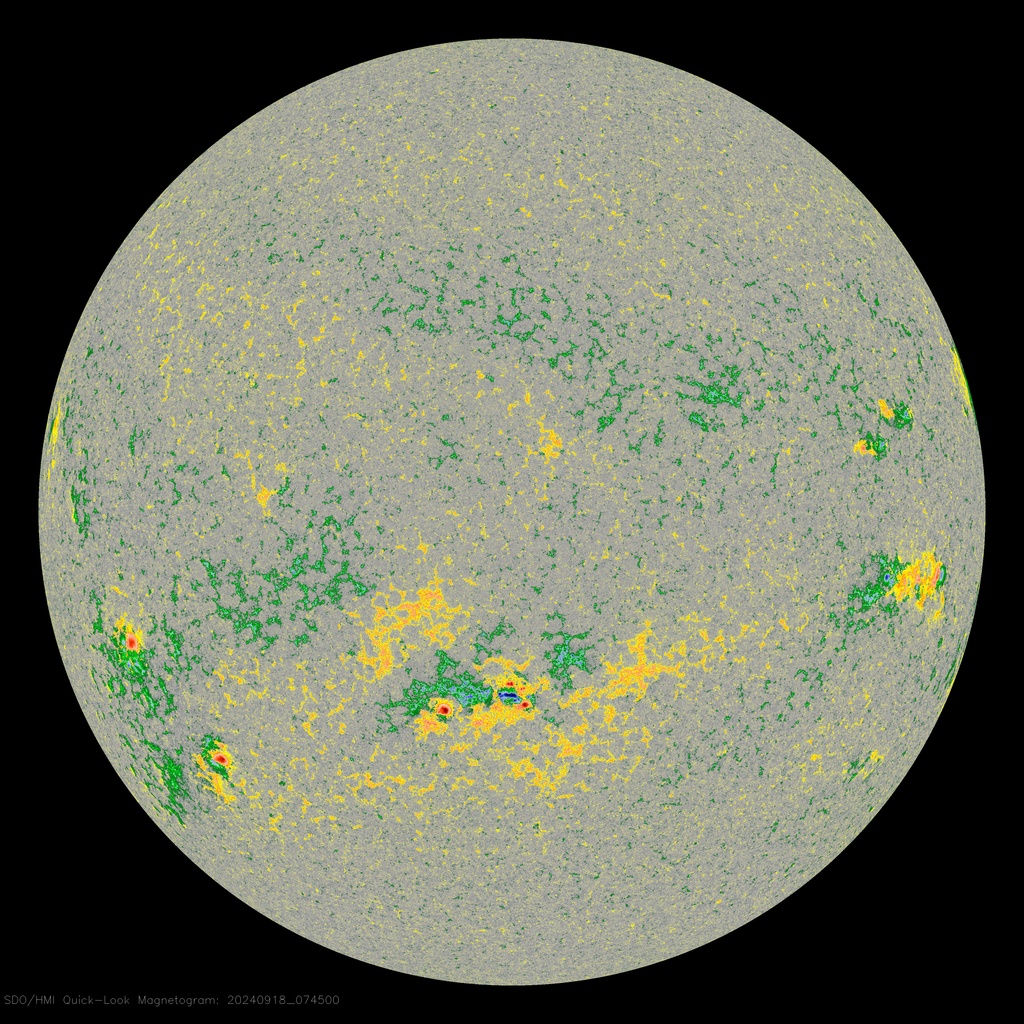}
\end{center}
\caption[]{Left: Sunspot from the GREGOR solar telescope, courtesy of
  Rolf Schlichenmaier and Nazaret Bello Gonz\'alez, Institute for
  Solar Physics (KIS). Right: magnetogram of the solar surface on 18th
  September 2024 from the Solar Dynamics Observatory.}
\label{fig:sunspot}
\end{figure}

\subsection{Sunspots}

The visible manifestation of large-scale magnetism in the Sun are
sunspots which are concentrations of kilogauss strength magnetic
fields at the surface of the Sun
\citep[e.g.][]{2003A&ARv..11..153S,2005LRSP....2....8B}; see also the
left panel of \Figa{fig:sunspot}. Although the visible spots on the
solar surface are highly concentrated, corresponding magnetograms
indicate that the spots are associated with larger scale magnetism
beneath; see the right panel of \Figa{fig:sunspot}. Detecting magnetic
fields in subsurface layers of the Sun is very challenging and
typically relies on indirect methods. One such method is to follow the
rise of active regions in the Sun and in corresponding numerical
simulations of surface convection where magnetic flux tubes are
advected through the bottom boundary
\citep[e.g.][]{2016SciA....2E0557B}. This study suggest that the rise
speeds of active regions are too low for them to originate in the
tachocline at the base of the convection zone. A possible explanation
is that active regions form close to the surface \citep[e.g.][]{Br05}.

This idea arises from the observation that the rotation rate of
sunspots depends systematically on their age: the youngest spots have
a rotation rate that matches that of the base of the near-surface
shear layer (NSSL; see \Seca{sec:LSflows}) at $r=0.95R_\odot$, whereas
the oldest ones are more in line with the surface rotation rate
\citep[e.g.][]{PT98}. This can be interpreted as spots being anchored
at different depths, and that even the youngest spots would be a
shallow phenomenon. Another hint toward this direction is the
strengthening of the surface $f$ mode of the Sun prior to active
region emergence
\citep[e.g.][]{2016ApJ...832..120S,2023SoPh..298...30W}, which can be
seen up to 48 hours before any magnetic fields are detected at the
surface. The effect of subsurface magnetic fields on the $f$ mode has
been studied using idealised numerical simulations
\citep[][]{2014ApJ...795L...8S,2015MNRAS.447.3708S,2020GApFD.114..196S},
where strengthening was found for spatially concentrated fields near
the surface. Nevertheless, the formation mechanism of sunspots is
currently unknown but a successful solar and stellar dynamo theory has
to incorporate this.

\subsection{Differential rotation and meridional circulation in the Sun}
\label{sec:LSflows}

\begin{figure}[t]
\begin{center}
\includegraphics[width=0.53\textwidth]{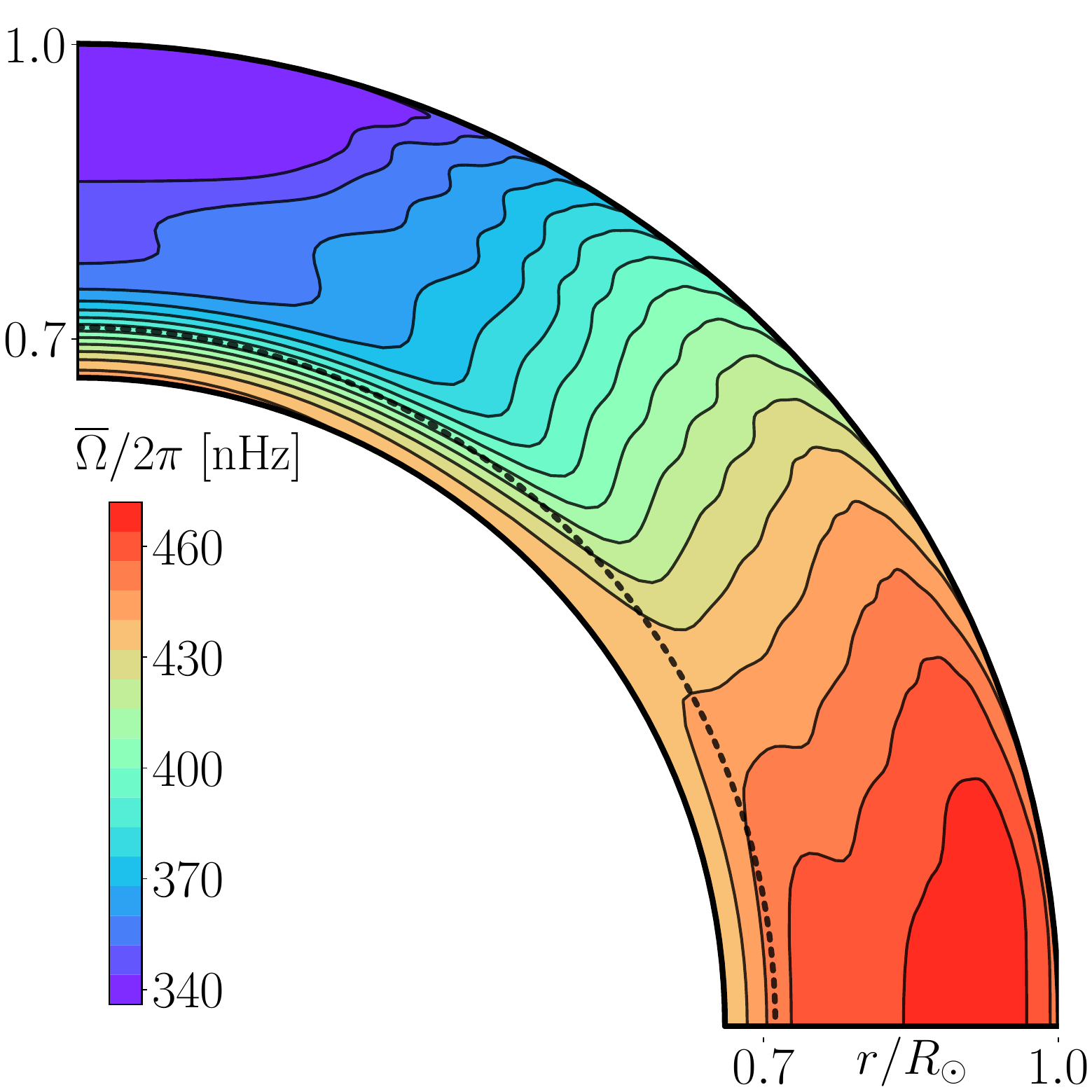}\includegraphics[width=0.47\textwidth]{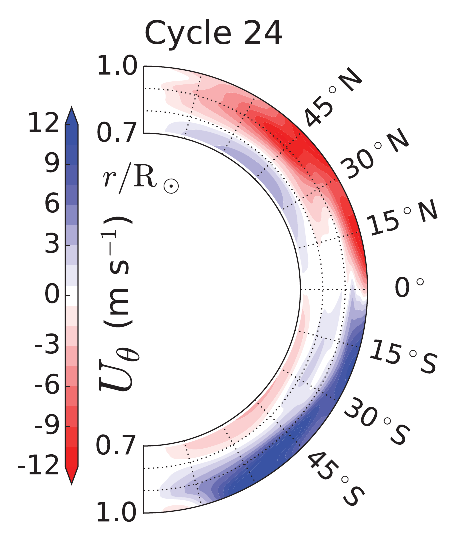}
\end{center}
\caption[]{Left: mean angular velocity in the solar interior from
  global helioseismology using data from
  \cite{Larson_Schou_2018_SoPh_293_29}. The dashed line at
  $r/\Rsun=0.71$ indicates the base of the convection zone. Right:
  latitudinal component of the solar meridional circulation from
  helioseismology from \cite{2020Sci...368.1469G}.}
\label{fig:solarOm}
\end{figure}

Another remarkably well-known characteristic of the Sun is its
interior differential rotation
\citep[e.g.][]{Schouea98,2009LRSP....6....1H}; see the left panel of
\Figa{fig:solarOm}. The interior rotation in the bulk of the
convection zone is roughly constant on radial lines whereas the
boundary layers near the surface (i.e., the NSSL) and at the base of
the convection zone (tachocline) are characterized by strong radial
shear. The radiative interior of the Sun is nearly rigidly
rotating. Solar differential rotation is thought to be driven by the
interaction of convective turbulence and rotation
\citep[e.g.][]{R89,KR95}, although recent simulations suggest that
magnetic fields can also be important
\citep[e.g.][]{2021NatAs...5.1100H,2022ApJ...933..199H,2023A&A...669A..98K,2025ApJ...985..163H}. In
the NSSL the gradients of velocity and density are large and the
convective timescale is much shorter than the solar rotation
period. Therefore the NSSL is challenging to reproduce in global 3D
simulations and can be incorporated self-consistently only at high
resolutions
\citep[][]{HRY15a,2019ApJ...871..217M,2025ApJ...985..163H}. The
transition to rigid rotation in the radiative core is thought to be
mediated by magnetic fields, either of fossil
\citep[e.g.][]{Rudiger_Kitchatinov_1997_AN_318_273,Gough_McIntyre_1998_Nature_394_755}
or dynamo origin
\citep[e.g.][]{2001SoPh..203..195F,2022ApJ...940L..50M}.

A significantly less well constrained part of solar interior
large-scale flows is the meridional circulation
\citep[e.g.][]{2022LRSP...19....3H}. Whereas the poleward flow near
the surface is well-established, the deep subsurface structure and
speed of the meridional flow is still under debate: helioseismic
inversions have yielded both single
\citep[e.g.][]{2020Sci...368.1469G} and multiple cell
\citep[e.g.][]{STR13,ZBKDH13} structures in radius depending on the
methods used. The results from solar cycle 24 from
\cite{2020Sci...368.1469G} is shown in the right panel of
\Figa{fig:solarOm}. Numerical 3D simulations consistently produce
multiple meridional circulation cells per hemisphere in cases with
solar-like differential rotation profile
\citep[e.g.][]{KKB14,PCM15,2017ApJ...836..192B}. In dynamo theory
these large-scale flows contribute to the generation of global
magnetism: differential rotation winds up poloidal fields to produce
the toroidal field ($\Omega$ effect) and meridional flows are
crucially important in a class of mean-field models known as
flux-transport or advection dominated dynamos
\citep[e.g.][]{DC99,CNC04}.

\section{Self-consistent three-dimensional dynamo simulations}
\label{sec:dynsims}

Before delving into the details of the various kinds of dynamo
simulations at large, it is necessary to clarify what is meant by a
self-consistent dynamo simulation. In what follows this is taken to
encompass the simultaneous solution of (at least) the induction and
Navies--Stokes equations in three dimensions without further
approximations, and where dynamo action is due to the resolved
dynamics. These models can be considered as ``direct numerical
simulations'' (DNS) of systems where most dimensionless parameters,
such as Reynolds and Prandtl numbers, are unrealistically small in
comparison to the Sun and stars \citep[see,
  e.g.][]{2017LRCA....3....1K,2023SSRv..219...58K}. Alternatively we
may consider them as ``large-eddy simulations'' (LES) that seek to
capture the large-scale phenomena (possibly of a real system) down to
the grid scale directly, while modeling the subgrid-scale dynamics
typically by highly enhanced diffusion coefficients. This is in
contrast to mean-field models where typically all but the largest
scales are parameterized.

\subsection{Equations of magnetohydrodynamics (MHD)}
\label{sec:mhd}

A non-relativistic charge neutral gas that is sufficiently collisional
that it can be described as a single fluid can be described by the
standard set of MHD equations \citep[see, e.g.][]{davidson_2001}. For
the fully compressible case in a rotating frame these are given by:
\begin{eqnarray}
\frac{\pd \BBB}{\pd t} &=& \bm\nabla \times \left( \UUU \times \BBB - \eta \mu_0 \JJJ \right),\label{equ:indu} \\
\frac{\pd \rho}{\pd t} &=& -\bm\nabla\bm\cdot(\rho\UUU), \label{equ:conti} \\
\rho \frac{\pd \UUU}{\pd t} &=& -\rho\UUU\bm\cdot\bm\nabla\UUU + \rho \gggg - \bm\nabla p - 2\rho\bm\Omega \times \UUU + \JJJ \times \BBB + \DIV (2\rho\nu \bm{\mathsf{S}}) + \fff, \label{equ:NS} \\
\rho T \frac{\pd s}{\pd t}  &=& - \rho T \UUU \bcdot \nab s + \DIV \mcFFF + \eta \mu_0 \JJJ^2 + 2\nu \rho \bm{\mathsf{S}}^2 + {\cal H},\label{equ:entropy}
\end{eqnarray}
where $\BBB$ is the magnetic field, $\UUU$ is the velocity, $\eta$ is
the magnetic diffusivity, $\mu_0$ is the permeability of vacuum, $\JJJ
= \bm\nabla \times \BBB/\mu_0$ is the current density, $\rho$ is the
density, $\nu$ is the kinematic viscosity, $\bm{\mathsf{S}}$ is the
traceless rate-of-strain tensor, $\gggg = -\bm\nabla\phi_g$ is the
acceleration due to gravity, where $\phi_g$ is the gravitational
potential, $p$ is the pressure, $\bm\Omega$ is the rotation vector,
$\fff$ describes additional body forces, $T$ is the temperature, $s$
is the specific entropy, $\mcFFF = \mcFFF_{\rm rad} + \mcFFF_{\rm
  SGS}$ is a flux which often contains radiative $(\mcFFF_{\rm rad} =
-K\bm\nabla T)$ and subgrid-scale (SGS) (e.g., $\mcFFF_{\rm SGS}\ = -
\chit \rho \bm\nabla s$)\footnote{Sometimes the SGS flux is defined
with an additional $T$ factor via $\mcFFF_{\rm SGS} = - \chit \rho T
\bm\nabla s$
\citep[e.g.][]{BMT04,Jones_Kuzanyan_2009_Icarus_204_227,KMCWB13,2018ApJ...856...13O}. \cite{2015JPlPh..81e3904R}
point out that this is formally correct if the internal energy
equation is solved instead of the entropy equation.}  contributions,
where $K$ is the heat conductivity and $\chit$ describes SGS entropy
diffusion
\citep[][]{Braginsky_Roberts_1995_GAFD_79_1,2015JPlPh..81e3904R,2023A&A...669A..98K}. Finally,
${\cal H}$ describes additional heating and cooling from, e.g., from
nuclear reactions. To close the system, an equation of state
$p=p(\rho,T)$ relating pressure to the other thermodynamic quantities
is needed. The mean molecular weight $\mu=\mu(\rho,T)$ enters the
equation of state and depends on the ionization state of the
matter. Very often the equation of state is for simplicity taken to be
that of a monoatomic ideal gas, $p = \Rgas \rho T$, where $\Rgas = \cP
- \cV$ is the universal gas constant, and $\cP$ and $\cV$ are the heat
capacities at constant pressure and volume with $\gamma = \cP/\cV =
5/3$.

However, in many dynamo simulations a reduced set of equations is
considered. For example, the flow can be assumed incompressible or
anelastic in which cases \Eq{equ:conti} is replaced by $\DIV \UUU=0$
or by $\DIV (\rho \UUU)=0$, respectively. In fully compressible
simulations of convection the timestep constraint due to the low Mach
number in the deep convection zone is sometimes alleviated by reducing
the sound speed artificially by multiplying the rhs of \Eq{equ:conti}
by a factor $\xi^{-2} < 1$ \citep[e.g.][]{Re05,HRYIF12}. Often
idealized simulations are done assuming an isothermal equation of
state $p=\rho\cst$, in which case the energy equation drops out. These
models also often neglect density stratification and rotation, and
flows and helicity are injected through an external force $\fff$ at a
pre-defined forcing scale $\kf$.

\subsubsection{Microphysics}
\label{sec:microphys}

In addition to the equation of state, microphysics enter at least in
principle also through the diffusion coefficients $\nu$ and $\eta$,
and via the gas opacity $\kappa$ that goes into the description of
radiation. The dependence of the fluid viscosity and magnetic
diffusivity on the ambient thermodynamic state can be computed using
Spitzer formulas \citep[][]{1962pfig.book.....S} with $\nu \propto
\rho^{-1} T^{5/2}$ and $\eta \propto T^{-3/2}$. These have been used
to compute the microscopic diffusion coefficients for various
astrophysical objects; see \cite{BS05}, \cite{2020RvMP...92d1001S},
and \cite{2022ApJS..262...19J}. However, the Spitzer formulae are
almost never used in numerical dynamo simulations because in
practically all cases of astrophysical interest the resulting
diffusion coefficients either vary too much within the computational
domain and/or are too small to be used in computationally affordable
simulations; see discussions about convection and global dynamo
simulations in, e.g., \cite{2017LRCA....3....1K} and
\cite{2023SSRv..219...58K}. Furthermore, the thermal Prandtl number
$\Pra=\nu/\chi$, where $\chi=K/\cP\rho$, and the corresponding
magnetic Prandtl number $\PrM=\nu/\eta$ are in reality very small in
stars such as the Sun \citep[e.g.][]{2020RvMP...92d1001S}. Therefore
3D dynamo simulations most often use either constant or spatially
varying prescribed diffusivities that are much larger than the
corresponding Spitzer equivalents or seek to minimize the effective
diffusion by the use of implicit or explicit SGS modeling \citep[see,
  e.g.][]{2015SSRv..194...97M}. Values of the thermal and magnetic
Prandtl numbers of the order of unity are used out of numerical
necessity.

A similar situation occurs with the opacity of the gas. Most often the
gas is assumed optically thick is which case $\FFFrad$ can be written
in terms of the diffusion approximation
\begin{eqnarray}
\FFFrad = - K \bm\nabla T,
\end{eqnarray}
where $K$ is the heat conductivity given by
\begin{eqnarray}
K = \frac{16\sigma_{\rm SB}T^3}{3 \kappa \rho},
\end{eqnarray}
where $\sigma_{\rm SB}$ is the Stefan--Boltzmann constant. The opacity
of the gas depends in a complex way on thermodynamics and chemical
composition. In stellar surface convection simulations the opacity is
often taken from tabulated values \citep[e.g.][]{RSK09}, but this
approach is typically not used in most 3D large-scale dynamo
simulations. Instead, global simulations often adopt profiles of $K$
from 1D solar/stellar models \citep[e.g.][]{BMT04,2022ApJ...926...21B}
or coarser approximations such as the Kramers law
\citep[e.g.][]{2019GApFD.113..149K,2020GApFD.114....8K,2021A&A...645A.141V},
or simply assume either constant or fixed spatial profiles for $K$
\citep[e.g.][]{KMB12,MMK15,2015ApJ...810...80S}. The effects of these
different choices for the dynamo solutions can only be indirect, e.g.,
that the vigour and form of the flows and the thermodynamic structure
of the dynamo region are altered. A prominent example is the weakly
stably stratified deep parts of convection zones that have been
detected in simulations where the heat conductivity $K$ connects
smoothly from the convective to radiative zones
\citep[e.g.][]{1993A&A...277...93R,2015ApJ...799..142T,2017ApJ...843...52H,2017ApJ...845L..23K}.

Finally, the additional heating and cooling terms are often used in
models to either mimic radiative cooling near the surface or heating
due to nuclear reactions in the core
\citep[e.g.][]{DSB06,2021A&A...651A..66K,Hidalgo_et_al_2024_AA_691_326}. Sometimes
the effects of radiation are also parameterized in terms of a
heating/cooling term that makes no recourse to heat conductivity or
opacities and is therefore included in ${\cal H}$
\citep[e.g.][]{GCS10,GSdGDPKM15,2020ApJ...892..106M,2022ApJ...933..199H}.

\subsubsection{Dimensionless parameters}
\label{sec:dimpar}

The solutions of MHD equations are characterized by dimensionless
parameters that define the system that arise upon
non-dimensionalization. These include thermal and magnetic Prandtl
numbers, the Taylor number, and in the case of convection the Rayleigh
number:
\begin{eqnarray}
\Pra = \frac{\nu}{\chi},\ \ \PrM = \frac{\nu}{\eta},\ \ \Ta = \frac{4\Omega_0 d^4}{\nu^2}, \ \ \Ra = \frac{g d^4}{\nu \chi} \left(- \frac{1}{\cP}\frac{{\rm d}s}{{\rm d}r} \right),   
\end{eqnarray}
$d$ is a system-scale length scale, e.g., the depth of the convection
zone. The Taylor number is related to the Ekman number
$\Ek=\nu/\Omega_0 d^2$ via $\Ta = 4\Ek^{-2}$. In stars such as the
Sun, where the luminosity $L$ is practically constant on timescales
relevant for the dynamo, it is convenient to use a flux-based Rayleigh
number
\begin{equation}
\RaF = \frac{g d^4 \Ftot}{\cP \rho T \nu \chi^2},
\end{equation}
where $\Ftot = L/4\pi r^2$ is the total flux. The Prandtl, Taylor, and
flux-based Rayleigh number can be combined to a diffusion-free
modified Rayleigh number \citep[e.g.][]{2002JFM...470..115C,CA06}
\begin{equation}
\RaFS = \frac{g\Ftot}{8 \cP \rho T \Omega^3 d^2}.
\end{equation}
This is the only system parameter that 3D simulations of the solar
dynamo can reproduce. Choosing the length scale as $d = \cP T/g \equiv
H$, $\RaFS = \CoF^{-3}$, where $\CoF = 2\Omega H (\Ftot/\rho)^{-1/3}$
is a flux-based Coriolis number
\citep{2023A&A...669A..98K,2024A&A...683A.221K}.

Further dimensionless numbers describing the system are diagnostics
that are available \emph{a posteriori}. These include the fluid and
magnetic Reynolds numbers and the P\'eclet number
\begin{eqnarray}
\Rey = \frac{u\ell}{\nu},\ \ \ReM = \frac{u\ell}{\eta},\ \ \Pe = \frac{u\ell}{\chi},
\end{eqnarray}
where $u$ and $\ell$ are typical velocity amplitude and length
scale. The importance of rotation is given by the Coriolis number
\begin{eqnarray}
\Co = \frac{2\Omega_0 \ell}{u} = 2\Ro^{-1},
\end{eqnarray}
where $\Ro$ is the Rossby number.  In a density-stratified system such
as the solar interior all of the numbers listed above are functions of
radius and can vary several order of magnitude between the base and
the surface of the convection zone. In simulations this needs to be
avoided for numerical reasons \citep[][]{2023SSRv..219...58K}.

\subsubsection{Boundary conditions}
\label{sec:BCs}

An important but often overlooked aspect of numerical modeling are the
boundary conditions that invariably need to be imposed. Most commonly
dynamo models, be it mean-field or 3D, adopt simplified expressions
such as setting the magnetic field parallel or normal to the
boundary. The former corresponds to a perfect conductor and the
magnetic field is confined into the simulation domain, while the
latter allows field lines to cross the boundary. Another condition
similar to the normal field condition is to assume a potential field
outside of the domain. The choice of boundary conditions seems
innocuous but can in fact have far-reaching consequences. First, the
allowed modes and symmetries of magnetic fields depend on the boundary
conditions and can lead to qualitatively different results
\citep[e.g.][]{CBKK16,2017A&A...598A.117B,2018AN....339..631B}. Furthermore,
magnetic boundary conditions have a crucial importance for magnetic
helicity conservation: if magnetic field lines cannot cross the
boundary of the domain, magnetic helicity cannot escape which can lead
to resistively slow growth of magnetic field and saturation of the
dynamo; see \Seca{sec:maghel}.

Numerical simulations of the Sun and stars would in principle also
need to include the surface where the density is decreasing very
rapidly and the gas becomes optically thin. Global dynamo simulations
either do not include the full stratification of the Sun or only model
the star until a manageable outer radius which is smaller than the
actual stellar radius. Yet another approach is to embed the star into
a cube where no explicit boundary condition is imposed at the surface
\citep[e.g.][]{DSB06,2022ApJ...924...75M,2023A&A...678A..82O}. However,
in such cases the surface of the star is typically determined by
spatially fixed profiles of cooling and the transition between the
stellar interior and exterior is much smoother than in
reality. However, there is some evidence to the effect that including
layers outside the star lead to qualitatively different outcomes
\citep[e.g.][]{WKKB16,2022ApJ...931L..17K}.

\subsection{Classification of dynamo simulations}
\label{sec:classification}

The design of dynamo simulations depends to a great extent on the
goal: to study a particular dynamo effect in isolation calls for
models where only the necessary ingredients are retained in a simple
geometry, while study of global dynamos, such as the solar dynamo,
requires spherical geometry and the interplay of convection, rotation,
stratification, and non-linearity due to magnetism. A great variety of
models fits in the spectrum between these extremes. As a general rule
the amount of control of the simulation decreases with increasing
complexity and physical ingredients. At the same time, the complexity
of the corresponding mean-field models increases due to the larger
number and higher dimensionality of turbulent transport coefficients
and mean fields. Dynamo simulations can be categorized in roughly
three classes:
\begin{enumerate}
  \item \emph{Class 1: Forced turbulence simulations} where the
    geometry of the system is simplified and where flows are driven by
    external forcing instead of a natural instability. This class
    corresponds to, for example, fully periodic Cartesian forced
    turbulence simulations of helical
    \citep[e.g.][]{1981PhRvL..47.1060M,B01,CB13,SB14} and non-helical
    \citep[e.g.][]{ISCMP07,2023NatAs...7..662W} dynamos where imposed
    uniform large-scale shear can be further included
    \citep[e.g.][]{YHSKRICM08,KB09,2016MNRAS.458.2885T}. However, also
    systems where physical parameters such as the imposed kinetic
    helicity or large-scale flows have systematic spatial variations
    \citep[e.g.][]{2010AN....331..130M,2013Natur.497..463T,2021PhRvF...6l1701R}
    or more realistic geometry \citep[e.g.][]{MTBM09,MTKB10,JBKMR15}
    belong to this class. Furthermore, simulations where kinetic
    helicity is not necessarily due to forcing but due to the combined
    effects of density stratification and rotation
    \citep[e.g.][]{2012A&A...539A..35B,2013ApJ...762..127B} belong to
    Class 1. Such simulations are typically used to study isolated
    aspects of the dynamo problem such as predictions of mean-field
    theory or the non-linear saturation of dynamos. The simplified
    geometry and idealized physics mean that mean-field theoretic
    interpretation of Class 1 simulations is the most straightforward
    of all models although still not necessarily easy.
  \item \emph{Class 2: Local instability-driven simulations} where the
    geometry is still simple, for example, a Cartesian portion of a
    star, accretion disk, or a Galaxy, where flows are driven by
    physical instabilities such as convection
    \citep[e.g.][]{KKB08,HP09,2014ApJ...794L...6M}, magnetorotational
    instability
    \citep[e.g.][]{1995ApJ...446..741B,2010MNRAS.405...41G}, or
    supernovae
    \citep[e.g.][]{2008A&A...486L..35G,2013MNRAS.430L..40G,2015AN....336..991B}. In
    comparison to simulations in Class 1, Class 2 models are more
    physically consistent in that flows and their statistical
    properties such as turbulence and kinetic helicity are not put in
    by hand via forcing but emerge self-consistently as solutions of
    the MHD equations. A major advantage in this type of models is
    that the mean-field theoretic interpretation of the results
    remains tractable, especially if planar averages are a good
    representation of the mean fields
    \citep[e.g.][]{KKB09a,KKB09b,2014ApJ...794L...6M}. In simulations
    of Class 1 and 2 the large-scale flows are typically either absent
    or externally imposed and therefore a mean-field treatment of the
    Navier-Stokes equations does not need to be considered.
  \item \emph{Class 3: Semi-global and global simulations} are similar
    to the local instability-driven models in terms of physics with
    the distinction that a realistic geometry is assumed. Thus
    comparisons with actual astrophysical objects such as the Sun and
    stars are at least in principle possible. In Class 3 simulations
    the large-scale flows are outcomes of the models and in the
    general case they also would need to be subject to mean-field
    treatment \citep[e.g.][]{R89}. This increases the complexity of
    the mean-field description considerably and typically this task is
    omitted in comparisons of 3D simulations with mean-fields models
    which is also the path taken in the current review. Most relevant
    examples for the current topic of Class 3 models are simulations
    of solar and stellar dynamos
    \citep[e.g.][]{GCS10,KMB12,ABMT15,MMK15,2021ApJ...919L..13W,2022ApJ...926...21B,Warnecke_et_al_2025_AA_696_93}.
    In these cases the mean fields can be considered as averages over
    the longitude, leading to 2D mean-field description with a
    corresponding $(r,\theta)$-dependence of the turbulent transport
    coefficients.
\end{enumerate}
\begin{table}[t!]
\centering
\caption[]{Classification of dynamo simulations. The abbreviations
  denote: for = forcing, sc = self-consistent, loc = local, typically
  Cartesian, and glo = global, typically spherical.}
  \label{tab:classes}
      $$
       \begin{array}{ccccc}
          \hline
          \hline
          \noalign{\smallskip}
          \mbox{Class} & \mbox{Forcing} & \mbox{Helicity} & \mbox{Shear} & \mbox{Geometry} \\
          \hline
          1  &  \mbox{External}  & \mbox{for/sc} & \mbox{for}    & \mbox{loc/glo} \\
          \hline
          2 & \mbox{Instability} & \mbox{sc}     & \mbox{for/sc} & \mbox{loc/glo} \\
          \hline
          3 & \mbox{Instability} & \mbox{sc}     & \mbox{sc}     & \mbox{glo} \\
          \hline
          \end{array}
          $$
\end{table}
The boundaries between the different categories are somewhat fluid but
the big picture is accurate enough. The different classes are
summarized in \Table{tab:classes}. Classes 1 and 2 are useful when
studying individual dynamo effects and have the most straightforward
theoretical interpretation. Simulations of Class 3 are, at least in
theory, the most realistic but also the most challenging to deal with
mean-field theory. The discussion in the current review concentrates
on simulations that are \emph{successful large-scale dynamos}, that
is, they produce an identifiable large-scale magnetic field which is
interpreted to be due to dynamo action. While simulations leading to
small-scale dynamos are mentioned where relevant, especially when
co-existing with a large-scale dynamo, the reader is referred to, for
example, \cite{2023SSRv..219...36R} for an in depth review of this
topic. The following sections summarize examples of each class of
simulations relevant to solar and stellar dynamos.

\subsubsection{Dynamos in simulations with forced flows (Class 1)}

The history of 3D dynamo simulations can be considered to start from
the seminal study of \cite{1981PhRvL..47.1060M} who used forced
turbulence in a fully periodic cube to study dynamos in helical and
non-helical cases, and who were the first to show large-scale magnetic
field growth by helical turbulence by means of a 3D numerical
simulation. This can be considered as the first demonstration of an
$\alpha^2$ dynamo in the mean-field sense using direct solutions of
the MHD equations. However, computational constraints were still
holding back the simulators: \cite{1981PhRvL..47.1060M} had to use
hyperdiffusivity to resolve the helical case and to run the
simulations in the National Center for Atmospheric Research (NCAR)
Cray-1 -- the most powerful supercomputer in the world at the time. A
proliferation of modeling efforts of this kind started in earnest
only much later when the required computational power became more
accessible \citep[cf.][]{1999PhPl....6...89B,B01}. A landmark in this
respect is the study of \cite{B01} where the importance of magnetic
helicity conservation in the non-linear phase of large-scale dynamo
action was demonstrated. Many aspects of the helically forced, or
$\alpha^2$ dynamo, have hence been studied by means of 3D simulations:
for example, \cite{CB13} studied the minimum helicity to drive
large-scale dynamo action, \cite{BD01} and \cite{2002AN....323...99B}
investigated the effects of magnetic helicity losses through
boundaries, \cite{2008ApJ...687L..49B} simulated simultaneous
small-scale and large-scale dynamos to measure the quenching of
turbulent transport coefficients, and \cite{SB14} disentangled the
growth rates of small-scale and large-scale fields.

Another set of models that are conceptually very similar to those
discussed above have the kinetic helicity vary spatially such that it
has a sign change or an ``equator''
\citep[][]{MTKB10,2010AN....331..130M,2021PhRvF...6l1701R}. These
simulations represent another incarnation of an $\alpha^2$ dynamo but
unlike in the case where the kinetic helicity is constant, these
setups lead to oscillatory solutions where the dynamo wave propagates
toward the equator reminiscent of the Sun \citep[e.g.][]{BS87,Ra87}.

\begin{figure}[ht]
\begin{center}
\includegraphics[width=0.45\textwidth]{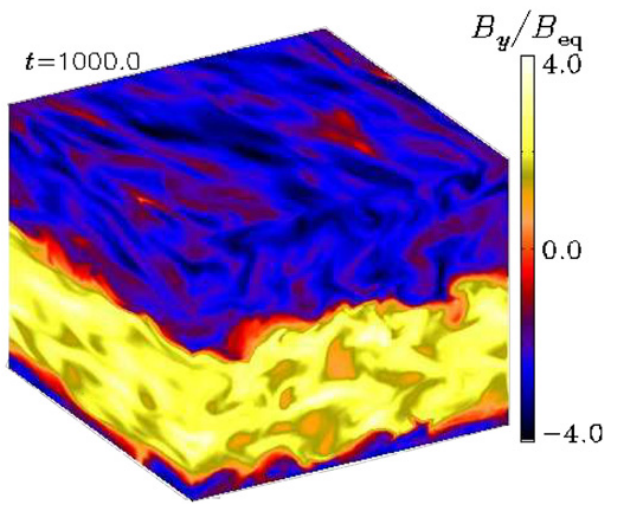}\includegraphics[width=0.55\textwidth]{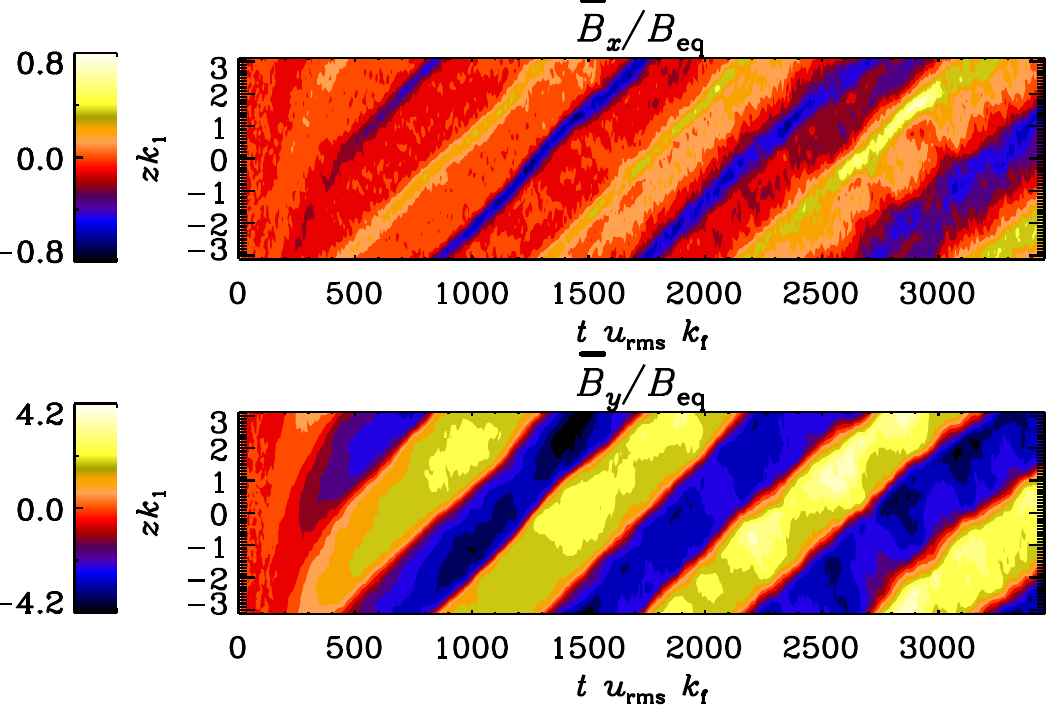}
\end{center}
\caption[]{Left: Instantaneous stream-wise magnetic field $B_y$ in a
  simulation with forced helical turbulence and imposed linear shear
  flow $\mUUU = (0, xS, 0)$ with $\ReM = 209$, $\PrM = 10$, and $\Sh =
  S/(\urms \kf) = -0.18$, where $\urms$ is the turbulent rms-velocity
  and $\kf$ is the wavenumber corresponding to the forcing
  scale. Right: space-time diagrams of the horizontally averaged
  horizontal magnetic fields $\mBBx(z,t)$ and $\mBBy(z,t)$ from the
  same simulation. Adapted from \cite{KB09}.}
\label{fig:forcedsims}
\end{figure}

The addition of imposed large-scale shear flow on top of helical
turbulence arising from the forcing fulfills the minimal requirements
for a classical $\alpha\Omega$ dynamo
\citep[e.g.][]{Pa55b,1969AN....291...49S}. Such systems very often
lead to cyclic solutions
\citep[e.g.][]{KB09,HRB11,2013Natur.497..463T,2016ApJ...825...23P},
see also \Figa{fig:forcedsims}; that can indeed be interpreted in terms
of $\alpha\Omega$ dynamos. A more recent discovery is that also
shearing \emph{non-helical} turbulence can maintain large-scale
magnetic fields
\citep[e.g.][]{YHSKRICM08,YHRSKRCM08,BRRK08,2015ApJ...813...52S,2016MNRAS.458.2885T}.
In such cases, the $\alpha$ effect vanishes on average and
straightforward explanation of the origin of the dynamo in terms of
mean-field theory is not possible. Theoretical interpretation of
shearing non-helical dynamos has turned out to be challenging and it
is still under debate in the community; see more details in
\Seca{sec:nonhelforced}.

The discussion has so far concentrated on simulations where the flow
is driven by an external force in the Navier--Stokes equation leading
to a turbulent solution. For completeness, we mention that there are
also flows that are less complex, but contain the necessary
ingredients for dynamo action such as kinetic helicity. Two prominent
examples include the Roberts flows \citep[e.g.][]{1972RSPTA.271..411R}
and the Galloway--Proctor flow \citep{1992Natur.356..691G}. Often the
Navier--Stokes equations are not solved in such cases. In some cases
these flows allow analytic calculation of mean-field transport
coefficients due to which these flows have been used extensively to
compare with mean-field dynamo theory
\citep[e.g.][]{2009MNRAS.393..113R,CHP10,2014MNRAS.441..116R}.

\subsubsection{Local simulations of dynamos due to convection (Class 2)}

Replacing externally driven flows by ones that are produced by
instabilities occurring in nature while retaining the simplified
geometry is the next step in complexity. Dynamos driven by convection
is naturally the most interesting case for solar and stellar
applications. While small-scale dynamos were reported from local
convection simulations starting in the late 1980s
\citep[][]{1989JFM...205..297M,NBJRRST92,BJNRST96,Cat99}, exciting a
dynamo that produces appreciable large-scale magnetic fields in such
simulations turned out to be significantly more challenging. Rigidly
rotating stratified or inhomogeneous convection produces kinetic
helicity and has therefore all the ingredients of an $\alpha^2$
dynamo. However, simulations often still fail to produce appreciable
large-scale magnetic fields \citep[e.g][]{CH06,HC08,FB12}. Large-scale
dynamos were obtained only when rapid rotation was considered
\citep[e.g.][]{KKB09b,KMB13,2014ApJ...794L...6M,GHJ15,MS16,2017JFM...815..333G,2018A&A...612A..97B}. Often
these dynamos are associated with hydrodynamic states that are
dominated by large-scale vortices
\citep[e.g.][]{Chan07,2011ApJ...742...34K,2013E&PSL.371..212C,GHJ14}
that are possibly due to two-dimensionalization of turbulent flows in
the rapid rotation regime. Simulations including large-scale shear
lead to successful large-scale dynamos much more easily
\citep[e.g.][]{KKB08,KKB10a,KMB13,HP09,HP13}.

\begin{figure}[ht]
\begin{center}
\includegraphics[width=0.38\textwidth]{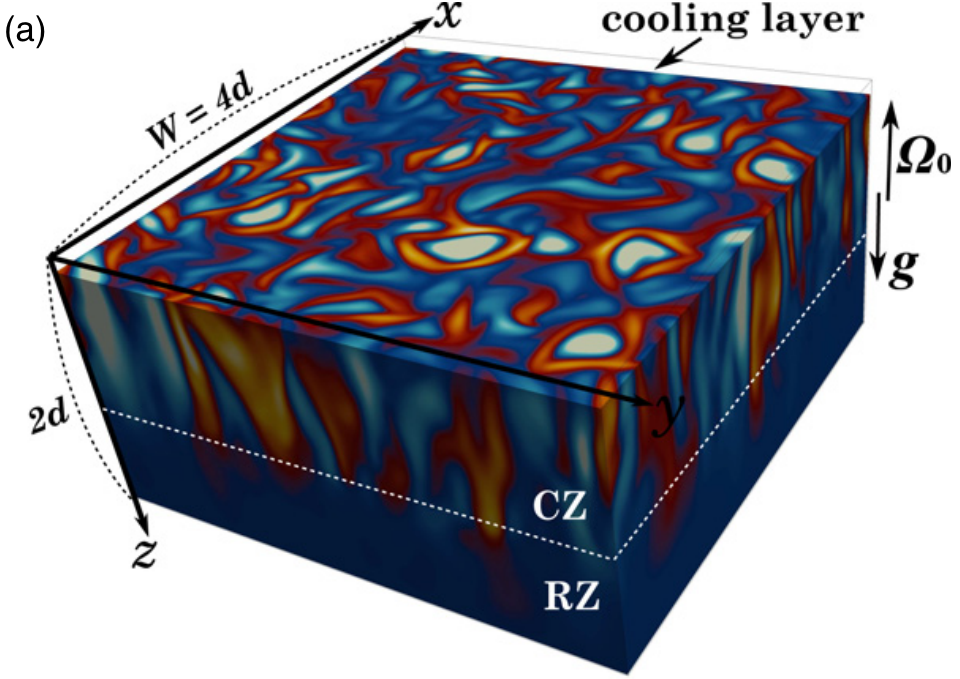}\includegraphics[width=0.62\textwidth]{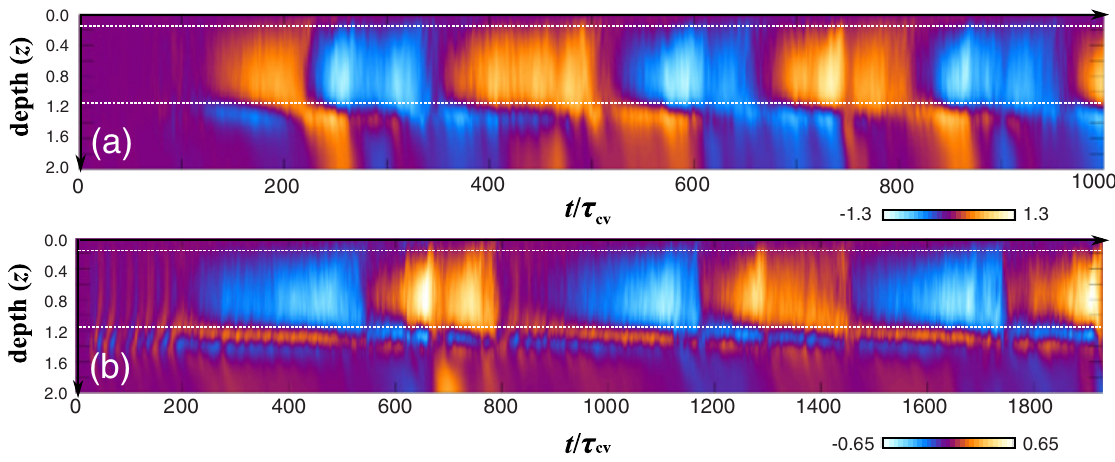}
\end{center}
\caption[]{Left: Instantaneous vertical velocity $U_z$ in a simulation
  of rigidly rotating density-stratified convection in a Cartesian
  slab. Right: time-depth diagrams of the horizontally averaged
  magnetic field $\mBBx(z,t)$ from simulations with $\PrM = 2$ (top)
  and $\PrM=8$ (bottom). Adapted from \cite{2014ApJ...794L...6M}.}
\label{fig:locosims}
\end{figure}

\subsubsection{Representative global solar and stellar dynamo simulations (Class 3)}

Global convective dynamo simulations have been around since the early
1980s when the first large-scale dynamos were obtained from such
models \citep{1981ApJS...46..211G,Gi83}. Initial simulations were made
using the Boussinesq approximation but anelastic models enabling
density stratification were soon also deployed
\citep[][]{1984JCoPh..55..461G,Gl85}. The initial enthusiasm regarding
global simulations waned after the early simulations constantly
yielded solutions where the dynamo wave propagated poleward in
contradiction to the Sun. A renaissance of global modeling started
with the creation of the ASH (anelastic spherical harmonic) code in
the late 90s and early 2000s
\citep[e.g.][]{METCGG00,2000ApJ...533..546E,2002ApJ...570..865B}.
Since then several methods have been developed to model solar and
stellar convection and dynamos either in spherical shells
\citep[e.g.][]{GCS10,FF14,2015ApJ...810...80S,SBCBN17,2019ApJ...880....6G,2020ApJ...892..106M,2021NatAs...5.1100H,2022ApJ...926...21B,2022ApJ...933..199H}
and wedges \citep[e.g.][]{KKBMT10,KMB12,MMK15,2019GApFD.113..149K}, or
in a star-in-a-box setup where a spherical star is embedded into a
Cartesian cube
\citep[e.g.][]{2004A&A...423.1101D,DSB06,2021A&A...651A..66K,2022ApJ...931L..17K,2022ApJ...924...75M,2023A&A...678A..82O,Hidalgo_et_al_2024_AA_691_326}.

To capture global phenomena such as solar and stellar magnetic cycles
on the scale of the convection zone, simulations in global spherical
geometry are required. This inevitably means that the effective
resolution in such models is severely reduced compared to local models
discussed above. Considering the Sun and other late-type stars, the
challenge is exacerbated by the immense density stratification of the
convection zone (more than 20 pressure scale heights), time scales
ranging from 12 days at the base of the convection zone to 5 minutes
near the surface, high fluid and magnetic Reynolds numbers ($10^{12}$
and $10^9$, respectively), low Prandtl numbers ($\Pra=10^{-7}$ and
$\PrM=10^{-3}$), and the strong variation of the Mach number between
the base of the convection zone and the surface ($\Ma\approx 10^{-4}$
in deep convection zone, transonic near the surface)
\citep[e.g.][]{2017LRCA....3....1K,2020RvMP...92d1001S,2022ApJS..262...19J}. None
of these characteristics can be fully reproduced in current global
simulations: the Reynolds numbers are typically of the order of a few
hundred and Prandtl numbers of the order of unity. The only
characteristic that can be accurately modeled is the rotational
influence on the flows, measured by $\RaFS$ or the flux Coriolis
number $\CoF$ \citep[e.g.][]{2024A&A...683A.221K}. Furthermore,
helioseismic and solar surface observations suggest that the velocity
amplitudes at large horizontal length scales are much lower in the Sun
in comparison to simulations
\citep[e.g.][]{HDS12,2020SciA....6.9639H,2021PhDT........26P,Birch_et_al_2024_PhFl_36_117136}. This
is commonly referred to as the convective conundrum
\citep{2016AdSpR..58.1475O} which is most likely the key reason why
there is no global simulation to date that captures the solar dynamo
and interior differential rotation fully self-consistently. Comparison
of observations with global simulations is a vibrant current topic and
the reader is referred to \cite{2023SSRv..219...77H} and
\cite{2023SSRv..219...58K} for more thorough discussions of the
current status of global numerical simulations targeting solar
differential rotation and solar and stellar dynamos, respectively.

\begin{figure}[ht]
\begin{center}
\includegraphics[width=0.27\textwidth]{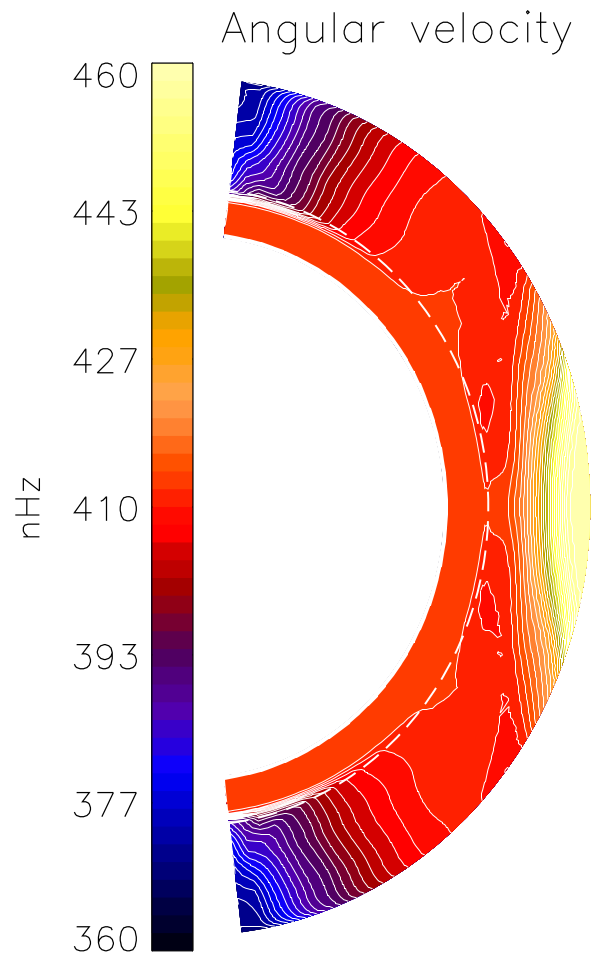}\ \ \ \ \includegraphics[width=0.44\textwidth]{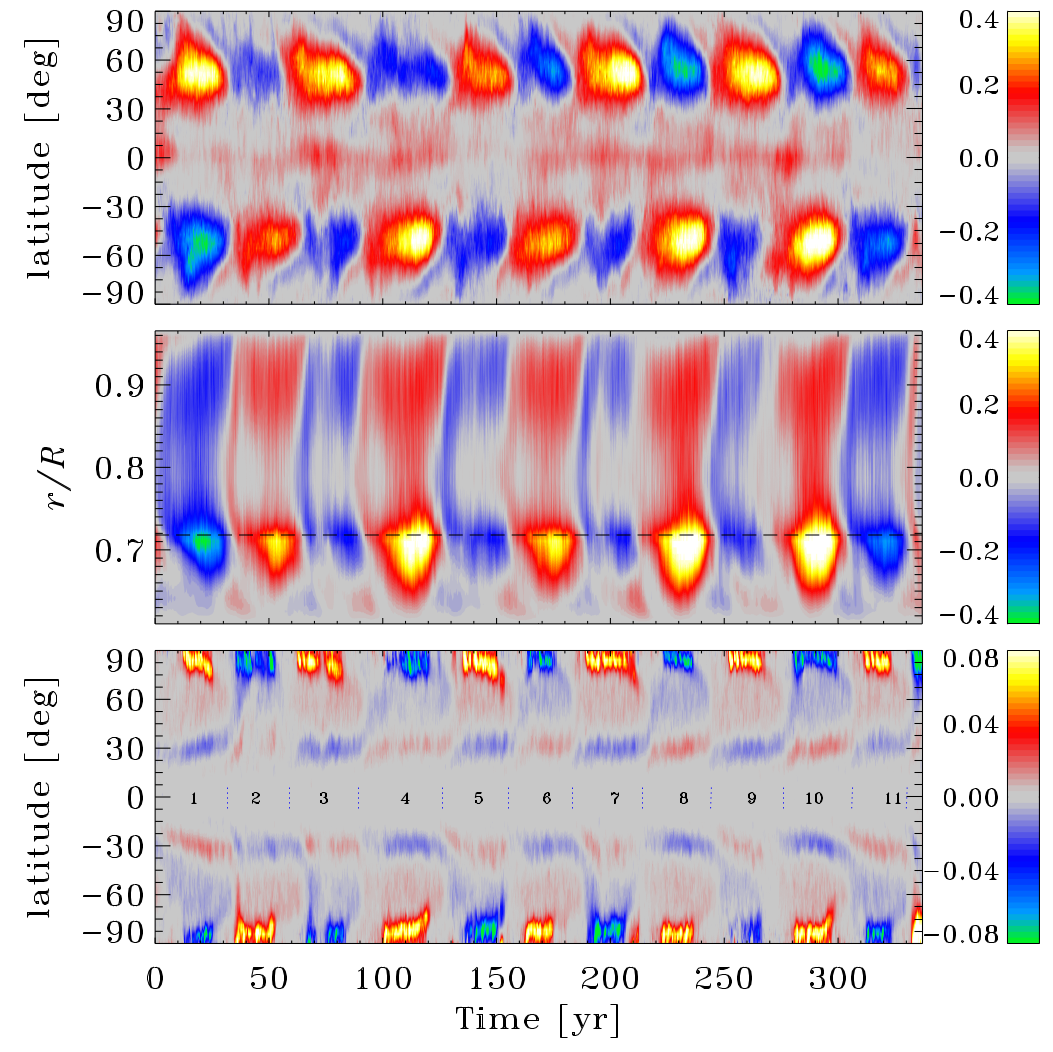}
\hspace*{2cm}\includegraphics[width=0.17\textwidth]{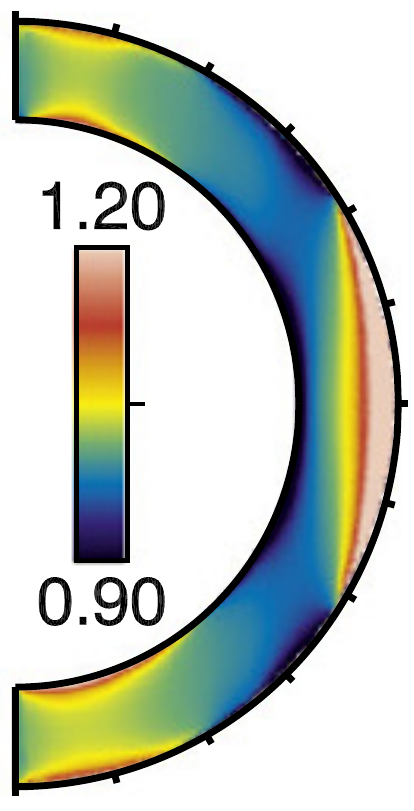}\hspace{.1cm}\includegraphics[width=0.6\textwidth]{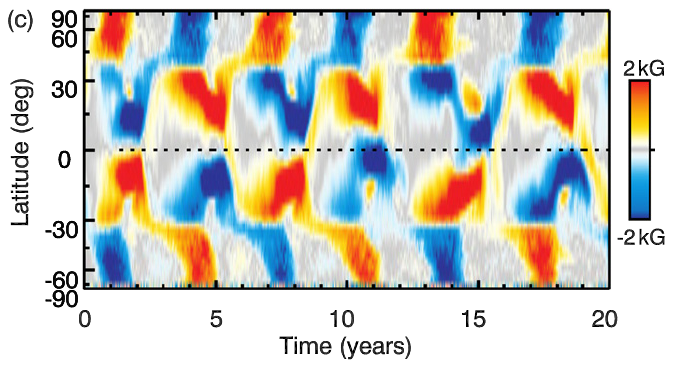}
\hspace*{1cm}\includegraphics[width=0.19\textwidth]{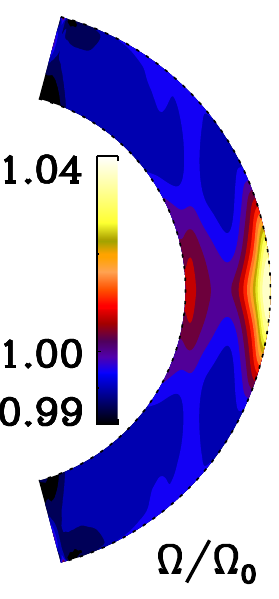}\hspace{1.2cm}\includegraphics[width=0.45\textwidth]{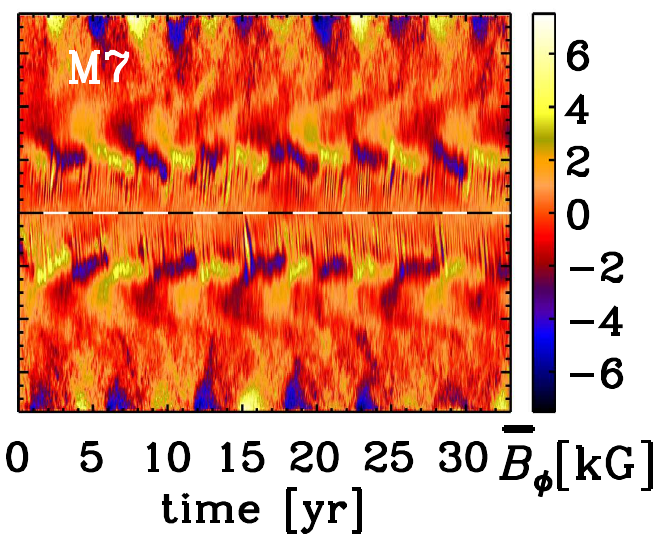}\\
\end{center}
\caption[]{Left: Azimuthally and temporally averaged mean angular
  velocity from global dynamo simulations of \cite{GCS10}, here
  adapted from \cite{RCGBS11} (top), \cite{ABMT15} (middle), and
  \citep{2018A&A...616A..72W} (bottom). Right: corresponding
  time-latitude diagrams of the azimuthally averaged magnetic field
  $\mBBp(\theta,t)$, except in the top panel where $\mBBp(r,t)$ and
  $\mBBr(\theta,t)$ are also shown.}
\label{fig:solarsims}
\end{figure}

Nevertheless, there have been many studies of convective dynamos as a
function of rotation which have revealed that cyclic solutions, akin
to those in the Sun, appear on a relatively narrow range of Coriolis
numbers \citep[e.g.][]{2018A&A...616A.160V,2022ApJ...926...21B}. For
slow rotation -- roughly corresponding to $\Co\lesssim 1$ -- the
large-scale magnetic fields tend to be predominantly axisymmetric and
quasi-steady
\citep[e.g.][]{2017A&A...599A...4K,2018ApJ...863...35S}. This
coincides with \emph{anti-solar} differential rotation: the rotation
rate at the equator is slower than at the poles. A transition to a
solar-like differential rotation occurs around $\Co\approx 1$
\citep[e.g.][]{GYMRW14,KKB14}. The large-scale magnetic fields in such
simulations tend to be still predominantly axisymmetric but the
dominant dynamo mode is oscillatory
\citep[e.g.][]{GCS10,2015ApJ...810...80S,SBCBN17,2018ApJ...863...35S,2018A&A...616A.160V,2023ApJ...947...36B}.
A number of simulations have appeared where a solar-like dynamo with
equatorward migration is found
\citep[e.g.][]{KMB12,ABMT15,2018ApJ...863...35S,2018A&A...616A..72W,2020ApJ...892..106M,2022ApJ...931L..17K,2022ApJ...926...21B}.
Representative examples of such solar-like solution are shown in
\Figa{fig:solarsims}. Comparisons of global convection-driven dynamos
with mean-field theory has concentrated on this particular type of
simulations which will be discussed in detail in
\Sec{subsec:globalco}.

\section{Mean-field dynamo theory}
\label{sec:dyntheo}

The large-scale spatiotemporal coherence of the solar magnetic field
is suggestive that perhaps its evolution can be explained by a model
that has far fewer degrees of freedom than the full 3D
magnetohydrodynamic (MHD) problem where all scales down to the
dissipation scales are solved. The latter is, and will likely remain,
numerically infeasible for the foreseeable future
\citep[e.g.][]{2017LRCA....3....1K,2023SSRv..219...58K}. The premise
behind mean-field theory is to solve only for the large scales while
the small-scale (turbulent) effects are parameterized by tensorial
transport coefficients. There are several textbooks
\citep[e.g.][]{M78,Pa79,KR80,1983flma....3.....Z,RH04,RKH13,moffatt_dormy_2019}
and review articles
\citep[e.g][]{BS05,2019JPlPh..85d2001R,2021JFM...912P...1T,2023SSRv..219...55B}
where mean-field dynamo theory is covered in great detail. Therefore
only the most relevant parts of this theory will be briefly repeated
here.

The starting point of mean-field dynamo theory is the induction
equation which describes the evolution of the magnetic field:
\begin{equation}
\frac{\pd \BBB}{\pd t} = \bm\nabla \times \left( \UUU \times \BBB - \eta \mu_0 \JJJ \right).
\label{equ:induction}
\end{equation}
Inspection of \Eq{equ:induction} shows that a necessary requirement
for the maintenance of magnetic fields is that $\UUU \neq 0$ and that
the induction term $\UUU\times\BBB$ needs to overcome magnetic
diffusion. Therefore the magnetic Reynolds number
\begin{equation}
 \ReM = \frac{u\ell}{\eta}, 
\end{equation}
where $u$ and $\ell$ are typical velocity and length scales, has to
exceed a critical value which depends on the the properties of the
flow and the geometry of the system. The velocity and magnetic fields
$\UUU$ and $\BBB$ must also be sufficiently complex: for example,
Cowling's anti-dynamo theorem states that an axisymmetric field cannot
be sustained by a dynamo \citep{1933MNRAS..94...39C}\footnote{However,
\cite{2020PhRvR...2a3321P} demonstrated recently that axisymmetric
flow can sustain a dynamo provided that the electrical conductivity is
non-axisymmetric.}. Furthermore, the Zeldovich theorem states that a
planar flow cannot sustain a dynamo \citep{zeldovich1957}.

In mean-field theory the idea is to follow the evolution of the
large-scale (mean) fields in detail whereas the small-scale
(turbulent) contributions are described via turbulent transport
coefficients. To obtain closed equations for the large scale
quantities the mean and fluctuations need to be separated. This is
done using the Reynolds decomposition:
\begin{equation}
\BBB = \mBBB + \bbb,\ \ \UUU = \mUUU + \uuu,
\end{equation}
where $\mBBB$ and $\mUUU$ are suitably averaged mean fields and flows,
and $\bbb$ and $\uuu$ are the fluctuations. Azimuthal averaging is
most often used in the case of solar and stellar dynamos whereas
planar averages are often used in simulations of Classes 1 and
2. These averages obey the Reynolds rules:
\begin{eqnarray}
& \mean{\mean{\BBB}} = \mean{\BBB}, \ \ \ \mean{\mean{\BBB}_1 + \mean{\BBB}_2} = \mean{\BBB}_1 + \mean{\BBB}_2, \ \ \ \mean{\mBBB \bbb} = 0, & \\
& \mean{\BBB_1 \BBB_2} = \mean{\BBB}_1 \mean{\BBB}_2 + \mean{\bbb_1 \bbb_2}, \ \ \ \mean{\frac{\pd B_i}{\pd x_j}} = \frac{\pd \mean{B}_i}{\pd x_j}, \ \ \ \mean{\frac{\pd B_i}{\pd t}} = \frac{\pd \mean{B}_i}{\pd t}.&
\end{eqnarray}
The last relation containing the time derivative is exact for ensemble
averaging while the accuracy of spatial averaging improves the longer
the averaging time is. Using the Reynolds decomposition in
\Eq{equ:induction} yields the mean-field induction equation
\begin{equation}
\frac{\pd \mBBB}{\pd t} = \bm\nabla \times \left( \mUUU \times \mBBB +
\mEMF - \eta \mu_0 \mJJJ \right).
\label{equ:meaninduction}
\end{equation}
The additional term that appears in the mean-field induction equation
is the electromotive force (EMF):
\begin{equation}
\mEMF = \mean{\uuu \times \bbb},
\end{equation}
which describes the correlations of small-scale velocities and
magnetic fields. Solving \Eq{equ:meaninduction} thus requires that the
EMF is known. It is customary to represent $\mEMF$ in terms of the
mean magnetic field and its gradients
\begin{equation}
\EMFi = \EMFi^{(0)} + a_{ij} \mean{B}_j + b_{ijk} \frac{\pd \mBBj}{\pd x_k} + \ldots,
\label{equ:EMFi}
\end{equation}
where $\EMFi^{(0)}$ is a contribution that can appear in the absence
of a mean field, $a_{ij}$ and $b_{ijk}$ are tensorial coefficients,
and where the dots indicate possible higher order derivatives
\citep[e.g.][]{KR80}. \Equ{equ:EMFi} is an approximation that is valid
if the mean fields vary slowly in space and time in comparison to the
integral scale and turbulent time, respectively. In other words,
\Eq{equ:EMFi} assumes that the connection between $\mEMF$ and $\mBBB$
is local and instantaneous.

However, in the general case the effects of non-locality cannot be
neglected. In this case the EMF can be formally written as
\citep[][]{2014AN....335..459R}
\begin{eqnarray}
  \EMFi(\xxx,t) &=& \EMFi^{(0)} + \int_0^\infty \int_\infty \big( {\cal A}_{ij} (\xxx,t;\bm\xi,\tau)\mean{B}_j(\xxx+\bm\xi,t-\tau) \nonumber \\ & & \hspace{2cm} + {\cal B}_{ij}(\xxx,t;\bm\xi,\tau) \frac{\pd\mean{B}_j(\xxx+\bm\xi,t-\tau)}{\pd x_k} \big) {\rm d}^3\xi {\rm d}\tau,\label{equ:EMFgeneral}
\end{eqnarray}
where ${\cal A}_{ij}$ and ${\cal B}_{ij}$ are tensorial kernels which,
in the kinematic case, depend on $\UUU$ and $\eta$, and where the
integrals run over all space and past time. In practice, the
contributions from sufficiently large $\bm\xi$ and $\tau$ become
negligible and real flows fall somewhere between the two extremes
\Eqsa{equ:EMFi}{equ:EMFgeneral}. Issues related to non-locality will
be revisited in later sections.

Equation~(\ref{equ:EMFi}) is not particularly informative about the
physical effects that contribute to the EMF via the coefficients
$a_{ij}$ and $b_{ijk}$. An equivalent way of writing the $\mEMF$ in
the absence of $\mEMF^{(0)}$ is \citep[e.g.][]{1980AN....301..101R}
\begin{eqnarray}
\mEMF = \bm\alpha \bm\cdot \mBBB + \bm\gamma \times \mBBB - \bm\beta\bm\cdot(\bm\nabla\times\mBBB) -\bm\delta \times (\bm\nabla\times\mBBB) - \bm\kappa \bm\cdot (\bm\nabla\mBBB)^{\rm (s)},\label{equ:R80EMF}
\end{eqnarray}
where $\alpha$ is the symmetric part of $a_{ij}$ and describes the
generation of magnetic fields through the $\alpha$ effect which is
related to kinetic helicity of the flow
\citep{1966ZNatA..21..369S}. The anti-symmetric part of $a_{ij}$ gives
rise to magnetic pumping, $\gamma_i = -\epsilon_{ijk} a_{jk}/2$, which
describes an advection-like effect. $\bm\beta$ is a second rank tensor
describing turbulent diffusion, and $\bm\delta$ is a vector describing
the R\"adler or $\mean{\bm\Omega} \times \mJJJ$ effect
\citep{1969VeGG...13..131R} that can also lead to dynamo
action. Finally, $\bm\kappa$ is a third-rank tensor acting on the
symmetric part of the mean magnetic field gradient tensor. The
physical interpretation of $\bm\kappa$ is currently unclear. Another
contribution to $\bm\delta$ includes the shear-current or
$\mWWW\times\mJJJ$ effect, where $\mWWW=\bm\nabla\times\mUUU$
\citep{2003PhRvE..68c6301R,2004PhRvE..70d6310R}. This effect is
postulated to occur in shearing nonhelical turbulence and which enters
through an off-diagonal component of the magnetic diffusivity
tensor. Also a magnetic variant of the shear-current effect has been
proposed
\citep{2004PhRvE..70d6310R,2015PhRvE..92e3101S,2016JPlPh..82b5301S}
which is conjectured to occur if the small-scale magnetic fields due
to a small-scale dynamo are sufficiently strong. Furthermore,
fluctuations of kinetic helicity and $\alpha$ effect have also been
shown to support dynamos if large-scale shear is also present in cases
where the mean helicity and $\alpha$ effect vanish
\citep[e.g.][]{1997ApJ...475..263V,1997ARep...41...68S,2007MNRAS.382L..39P,Kleeorin_Rogachevskii_2008_PhRvE_77_036307,2010JFM...664..265S,2021MNRAS.508.5163J}.

Lastly, there is the $\mEMF^{(0)}$ term which occurs in the absence of
$\mBBB$. This term can be thought to encompass battery terms such as
the Biermann battery \citep{1950ZNatA...5...65B} which are needed to
seed the magnetic field in the absence of magnetic monopoles when
$\BBB=0$ initially. However, there are other possible contributions to
$\mEMF^{(0)}$ that can occur if there is a pre-existing magnetic
turbulence \citep[e.g.][]{1976IAUS...71..323R}, i.e., a small-scale
dynamo where $\bbb\neq0$ with $\mBBB=0$, or when a mean flow and
cross-helicity $\mean{\uuu\bm\cdot\bbb}$ are present
\citep[e.g.][]{1990PhFlB...2.1589Y,1993ApJ...407..540Y,2013GApFD.107..207B}.

\subsection{Analytic methods to compute turbulent transport coefficients}

The main difficulty in mean-field dynamo theory lies in computation of
the turbulent transport coefficients such as those in
\Eq{equ:R80EMF}. In the kinematic case, where $\UUU$ is assume given,
it suffices to solve for the fluctuating magnetic field $\bbb$ which
is given by
\begin{eqnarray}
\frac{\pd \bbb}{\pd t} = \bm\nabla \times (\mUUU \times \bbb + \uuu
\times \mBBB + \bm{\mathcal G}) + \eta\bm\nabla^2
\bbb,\label{equ:smallbbb}
\end{eqnarray}
where $\bm{\mathcal G} = \uuu \times \bbb - \mean{\uuu \times \bbb}$,
is the nonlinear term. It is possible to derive an exact equation for
$\bm{\mathcal G}$ but this leads to an infinite chain of equations for
increasingly higher order correlations of $\uuu$ and $\bbb$. To avoid
this, analytic methods truncate the series of equations with typically
computationally convenient rather than physically justified
approximations. The kinematic case considered above is practically
never the case in observed astrophysical systems and the magnetic
backreaction on the flow cannot be omitted. In that case also the
momentum equation needs to be taken into account. A few of the
analytic methods used in mean-field theory are described below; the
reader is referred to \cite{Rogachevskii_2021} for a more thorough
discourse.

\subsubsection{First order smoothing approximation (FOSA)}

In FOSA, $\bm{\mathcal G}$ is neglected altogether
\citep[e.g.][]{M78,KR80}. This approximation is valid in cases when
either
\begin{eqnarray}
\ReM \ll 1,\ \ \ \mbox{or}\ \ \ \St = \frac{\tauc u}{\ell} \ll 1,
\end{eqnarray}
where $\St$ is the Strouhal number and where $\tauc$ is the
correlation time of the flow. In astrophysical systems $\ReM \gg 1$
practically always \citep[e.g.][]{O03,BS05}. On the other hand, often
in numerical simulations $\St\approx 1$
\citep[e.g.][]{2005A&A...439..835B}, and similar estimates can be
obtained, e.g., for the solar granulation. Therefore many simulations
as well as the Sun fall outside the validity range of
FOSA. Nevertheless, the rigorous results obtained using FOSA are very
useful as benchmarks for methods that seek to determine turbulent
transport coefficients. The validity of FOSA estimates of turbulent
transport coefficients is discussed in more detail in
\Seca{sec:memethods}.

In the high conductivity limit, where $\ReM\gg1$ and $\St\ll1$, FOSA
yields in the case of isotropic and homogeneous turbulence
\begin{eqnarray}
  \mEMF = \alpha\mBBB - \etat \mu_0 \mJJJ,\label{equ:EMFFOSA}
\end{eqnarray}
where the scalar $\alpha$ effect and turbulent diffusivity are given
by
\begin{eqnarray}
\alpha = - \onethird \tauc \mean{\bm\omega\bm\cdot\uuu} \ \ \ \mbox{and}\ \ \ \etat = \onethird \tauc \mean{\uuu^2}, \label{equ:coefsfinalFOSA}
\end{eqnarray}
where $\bm\omega=\bm\nabla\times\uuu$ is the vorticity and
$\mean{\bm\omega\bm\cdot\uuu}$ is the kinetic helicity. In the general
anisotropic and inhomogeneous cases the transport coefficients are
tensors but they are also in principle tractable under FOSA
\citep[see, e.g.][]{1980AN....301..101R}. FOSA is also often referred
to as the second-order correlation approximation (SOCA).

\subsubsection{Minimal tau approximation (MTA)}

Another approximation that gained popularity in the early 2000s is the
minimal $\tau$ approximation (MTA)
\citep[e.g.][]{KRR90,2002PhRvL..89z5007B,2003PhRvE..68c6301R}. In this
method time evolution equation of $\mEMF$ is derived:
\begin{eqnarray}
\frac{\pd \mEMF}{\pd t} = \mean{\dot{\uuu} \times \bbb} + \mean{\uuu \times \dot{\bbb}},
\end{eqnarray}
where the dots indicate time derivatives. In the case of isotropic and
homogeneous turbulence the evolution equation for $\mEMF$ is given by
\begin{eqnarray}
\frac{\pd \mEMF}{\pd t} = \tilde{\alpha}\mBBB - \tilde{\eta}_{\rm t}\mu_0\mJJJ + \mTTT,
\end{eqnarray}
where $\mTTT$ contains triple correlations and where
\begin{eqnarray}
  \tilde{\alpha}= - \onethird \left( \mean{\bm\omega\bm\cdot\uuu} - \rho^{-1}\mean{\jjj\bm\cdot\bbb} \right)  \ \ \ \mbox{and}\ \ \ \tilde{\eta}_{\rm t} = \onethird \mean{\uuu^2}.
\end{eqnarray}
Relating the triple correlations to second correlations via $\mTTT =
-\mEMF/\tau$, where $\tau$ is a relaxation time, and assuming
stationarity yields the same expression for $\mEMF$ as in
\Eq{equ:EMFFOSA}. There is, however, a significant difference which is
the appearance of the magnetic contribution to $\alpha$ via the
current helicity $\mean{\jjj\bm\cdot\bbb}$ \citep{PFL76}:
\begin{eqnarray}
  \alpha = - \onethird \tau \left( \mean{\bm\omega\bm\cdot\uuu} -
  \rho^{-1}\mean{\jjj\bm\cdot\bbb} \right)\equiv \alpha_{\rm K} +
  \alpha_{\rm M} \ \ \ \mbox{and}\ \ \ \eta_{\rm t} = \onethird \tau
  \mean{\uuu^2}.\label{equ:MTAcoefs}
\end{eqnarray}
The minimal $\tau$ approximation appears superior to FOSA in that it
takes into account the dynamical velocity through the Navier--Stokes
equation due to which there is the magnetic correction to the $\alpha$
effect. The latter can be interpreted to be a consequence of magnetic
helicity conservation (see \Seca{sec:maghel}). Furthermore, higher than
second order correlations are to some degree retained through
$\mTTT$. However, the drawback is that there is no rigorously defined
validity range for MTA; see the discussion in
\cite{2007GApFD.101..117R}.

\subsubsection{Lagrangian methods for vanishing diffusivity}
\label{sec:lagmet}

Another way to circumvent the issues related to FOSA is to consider a
Lagrangian solution of the dissipationless induction equation, i.e.,
\begin{eqnarray}
B_i(\xxx,t) = B_j(\aaa,0)\frac{\pd x_i}{\pd a_j},
\end{eqnarray}
where $\aaa$ corresponds to the initial position of a particle. Such
approach was taken by, e.g., \cite{1971ApJ...163..279P},
\cite{1974JFM....65....1M}, and \cite{1976JFM....77..753K}, which
yields the EMF as:
\begin{eqnarray}
\EMFi(\xxx,t) = \brac{\uuu\times\bbb}_i = \brac{\uuu\times\BBB}_i = \epsilon_{ijk}\brac{u_j^L(\aaa,t) B_l(a,0)\frac{\pd x_k}{\pd a_l}},
\end{eqnarray}
where $\uuu^L$ corresponds to the Lagrangian velocity $\uuu^L(\aaa,t)
= (\pd\xxx/\pd t)_\aaa = \uuu(\xxx,t)$, and where the averages denoted
by angular brackets are taken over an ensemble of trajectories. In the
homogeneous isotropic case discussed above this leads to
\begin{eqnarray}
\alpha(t) = \onethird \alpha_{ii}(t) = -\onethird \int_0^t \brac{\uuu^L(\aaa,t)\bm\cdot \bm\nabla_\aaa \times \uuu^L(\aaa,\tau)}{\rm d}\tau,\label{equ:alplan}
\end{eqnarray}
and
\begin{eqnarray}
\beta(t) &=& \onesixth \epsilon_{ijk}\beta_{ijk} = \onethird \int_0^t \brac{\uuu^L(t)\bm\cdot\uuu^L(\tau)} {\rm d}\tau + \int_0^t \alpha(t)\alpha(\tau){\rm d}\tau \nonumber \\ && \hspace{-1cm} + \onesixth \int_0^t\!\!\int_0^t \brac{ \uuu^L(t)\!\bm\cdot\!\uuu^L(\tau')\bm\nabla_\aaa\!\bm\cdot\!\uuu^L(\tau)\!-\!\uuu^L(t)\!\bm\cdot\!\bm\nabla_\aaa\uuu^L(\tau)\!\bm\cdot\!\uuu^L(\tau')} {\rm d}\tau {\rm d}\tau'.\label{equ:etalan}
\end{eqnarray}
While \Eq{equ:alplan} closely resembles the FOSA expression, the
Lagrangian result for $\beta$ is quite different from $\etat$ in
\Eq{equ:coefsfinalFOSA}. The first term on the rhs of \Eq{equ:etalan}
corresponds to turbulent diffusion of a scalar which was originally
derived by \cite{taylor1921}. The second term describes the effects of
fluctuations of $\alpha$, which can theoretically lead to reduced
turbulent diffusion resulting enhanced growth of the magnetic field
\citep[e.g.][]{1976JFM....77..753K}. The remaining terms in
\Eq{equ:etalan} involve triple correlations whose physical
interpretation is not as straightforward.

\subsection{Nonlinearity due to direct magnetic back-reaction}
\label{sec:nonlinearity}

The expression \Eq{equ:EMFi} is linear in $\mBBB$ and therefore
applicable in the case where the mean magnetic field is weak. When the
mean field is non-negligible the turbulent transport coefficients need
to be reinterpreted as $a_{ij}=a_{ij}(\BBB)$ and
$b_{ijk}=b_{ijk}(\BBB)$. Arguably the simplest way of incorporating
the nonlinearity is to assume that the mean magnetic fields affect the
velocity field $\UUU$. A common way to deal with the nonlinearity is
to assume that the backreaction ensues when the magnetic energy is
comparable to equipartition with kinetic energy via an algebraic
quenching formula:
\begin{eqnarray}
\alpha = \frac{\alpha_0}{1 + \left(\mBBB/\Beq\right)^2},
\end{eqnarray}
where $\Beq = \sqrt{\mu_0 \rho \UUU^2}$ and $\alpha_0$ is the
kinematic value of $\alpha$. This can be understood as the
back-reaction of the large-scale field on the small-scale flow
$\uuu$. Another form of $\alpha$ quenching includes a factor
containing the magnetic Reynolds number
\begin{eqnarray}
\alpha = \frac{\alpha_0}{1 + \ReM \left(\mBBB/\Beq\right)^2},\label{equ:acata}
\end{eqnarray}
and leads to a negligibly small $\alpha$ even for very weak fields if
astrophysically relevant values of $\ReM$ are used. This is known as
\emph{catastrophic quenching} in the original sense
\citep{CV91,VC92,Gruzinov_Diamond_1994_PRL_72_1651,GD95,Bhattacharjee_Yuan_1995_ApJ_449_739,CH96}. This
is now understood in terms of magnetic helicity conservation in fully
periodic or closed systems; see \Seca{sec:maghel}. The Sun and other
astrophysical objects are not fully closed such that magnetic helicity
can be shed by various kinds of magnetic helicity fluxes. The
preceding discussion has concentrated solely on the nonlinearity of
$\alpha$, but in general all of the turbulent transport coefficients
feel the effects of dynamo-generated magnetic fields
\citep[e.g.][]{KPR94,2008GApFD.102...21P,2014ApJ...795...16K,2024MNRAS.530..382R}.

Models with just $\alpha$ quenching can still considered partly
kinematic because the large-scale flow $\mUUU$ remains unaffected.
Another type of nonlinearity deals with the impact of large-scale
magnetic field on the large-scale flows such as the differential
rotation via the Lorentz force \citep{MP75}
\begin{eqnarray}
\dot\mUUU = \ldots + \frac{1}{\rho} \mJJJ \times \mBBB,
\end{eqnarray}
which quenches the large-scale flows taking part in the dynamo
process. An analogous contribution arises from turbulent Maxwell
stress $\mathcal{M}_{ij} = \mean{b_i b_j}/\mu_0 \mean{\rho}$. The
small-scale Maxwell stress is often the dominant magnetic effect
affecting the large-scale flows in 3D simulations
\citep[e.g.][]{2017A&A...599A...4K,2022ApJ...933..199H,Warnecke_et_al_2025_AA_696_93,2025ApJ...985..163H}. However,
a mean-field theory for the turbulent Maxwell stress in rotating
turbulence has yet to be developed and it is not included in
mean-field modeling.

In the preceding discussion the backreaction was considered to be due
only to the large-scale field $\mBBB$. In astrophysical systems the
magnetic Reynolds numbers are so large that a small-scale dynamo is
very likely to be operating
\citep[e.g.][]{2023NatAs...7..662W,2023SSRv..219...36R}. The
backreaction of $\bbb$ on the turbulent transport coefficients is much
harder to quantify rigorously. First, this involves magnetic helicity
conservation which is discussed below. Second, the $\bbb$ due to a
small-scale dynamo can also influencee $\uuu$ directly without a
recourse to magnetic helicity arguments.  Developments in this
direction are discussed in \Seca{sec:nltf}. Concrete evidence of
nonlinear effects due to small-scale dynamos in numerical simulations
are still quite sparse and sometimes conflicting. \cite{HRY16} found
non-monotonic behavior of the energy of the mean magnetic fields as a
function of effective $\ReM$ and argued that a small-scale dynamo aids
the large-scale field generation at sufficiently high effective
$\ReM$, whereas \cite{2017A&A...599A...4K} found a monotonic increase
of $\mBBB^2$ as a function of $\ReM$. On the other hand,
\cite{2014ApJ...789...70C} suggested that large-scale dynamo action is
facilitated by suppression of the small-scale magnetism. This issue is
still open and awaits for further systematic theoretical and numerical
studies.

\subsection{Magnetic helicity conservation}
\label{sec:maghel}

In ideal MHD, an important conserved quantity is the magnetic helicity
\begin{eqnarray}
\HelM \equiv \int \AAA \bm\cdot \BBB\ dV,
\end{eqnarray}
where $\AAA$ is the magnetic vector potential with $\BBB =
\bm\nabla\times\AAA$. This can be seen from the evolution equation of
$\HelM$:
\begin{eqnarray}
  \frac{\pd \HelM}{\pd t} = -2 \int_V \EEE\bm\cdot\BBB\ dV - \int_S (\phi\BBB + \AAA\times\EEE)\bm\cdot \hat\nnn\ dS.\label{equ:maghelint}
\end{eqnarray}
where $\EEE=-\bm\nabla\phi - \pd\AAA/\pd t$ is the electric field and
$\phi$ is a scalar potential
\citep[e.g.][]{BS05}. In a fully periodic or fully closed system the
surface integral vanishes. Substituting Ohm's law $\EEE =
\mu_0\eta\JJJ - \UUU\times\BBB$ yields
\begin{eqnarray}
\frac{\pd \HelM}{\pd t} = -2 \eta\mu_0 \int_V \JJJ\bm\cdot\BBB\ dV \equiv -2 \mu_0 \eta C,\label{equ:maghelclosed}
\end{eqnarray}
where $C = \int_V \JJJ\bm\cdot\BBB\ dV$ is the current helicity. Thus,
magnetic helicity can change only because of \emph{molecular} magnetic
diffusivity, implying that large-scale fields can only saturate on a
diffusive timescale \cite[e.g.][]{B01}. Astrophysical dynamos
practically always have $\ReM \gg 1$ such that $\HelM$ can be
considered to be very nearly conserved \citep[e.g.][]{BS05}.

Magnetic helicity conservation has important consequences for
non-linear dynamos. Let us again consider isotropic and homogeneous
helical turbulence in a fully periodic system where $\mEMF=\alpha\mBBB
- \etat \mu_0 \mJJJ$, where $\alpha$ and $\etat$ are scalars. The
magnetic helicity density can be separated the mean and small-scale
parts as $\mean{\AAA\bm\cdot\BBB} = \mAAA\bm\cdot\mBBB +
\mean{\aaa\bm\cdot\bbb}$. In the isotropic case, small-scale magnetic
and current helicities are related via $\mean{\maghel} \approx
\mu_0\mean{\curhel}/\kf^2$ where $\kf$ is the typical scale of
turbulent eddies. With this information, an evolution equation for the
total $\alpha$ effect, $\alpha = \alphaK + \alphaM$, can be written as
\citep[e.g.][]{Brandenburg_et_al_2002_AN_323_411}
\begin{eqnarray}
  \frac{\pd \alpha}{\pd t} = -2\etat \kf^2 \left(\frac{\alpha\mBBB^2 - \etat\mu_0 \mJJJ\bm\cdot\mBBB}{\Beq^2} + \frac{\alpha-\alphaK}{\Rm} \right),\label{equ:dynalpquench}
\end{eqnarray}
where ${\rm Rm} = \etat/\eta$ is proportional to
$\ReM$. \Equ{equ:dynalpquench} is a dynamical $\alpha$ quenching
formula which takes into account magnetic helicity conservation in the
absence of magnetic helicity fluxes. Allowing for such fluxes, the
dynamical equation for $\alpha$ reads:
\begin{eqnarray}
  \frac{\pd \alpha}{\pd t} = -2\etat \kf^2 \left(\frac{\alpha\mBBB^2-\etat\mu_0 \mJJJ\bm\cdot\mBBB}{\Beq^2} + \frac{\alpha-\alphaK}{{\rm Rm}} \right) - \bm\nabla\bm\cdot\mathscrbf{F}^{\alphaM}.\label{equ:dynalpquenchf}
\end{eqnarray}
An equivalent representation is
\citep[e.g.][]{Brandenburg_2008_AN_329_725,Brandenburg_2018_JPP_84_735840404}
\begin{eqnarray}
\alpha(\mBBB)\!=\!\frac{\alphaK\!+\!\Rm [\etat \mu_0 \mJJJ\!\bm\cdot\!\mBBB/\Beq^2]\!-\!\bm\nabla\!\bm\cdot\!\mathscrbf{F}^{\alphaM}/(2\kf^2\Beq^2)\!-\!(\pd\alpha/\pd t)/(2\etat\kf^2)}{1+\Rm (\mBBB/\Beq)^2}. \label{equ:alphaB}
\end{eqnarray}
This equation reduces to the catastrophic quenching formula in
\Eq{equ:acata} if the terms proportional to $\Rm$ in the numerator
vanish. However, this is valid only in the stationary case under the
assumption of uniform magnetic fields and vanishing magnetic helicity
fluxes. None of these conditions are expected to be valid in real
astrophysical systems. There are several potential sources of magnetic
helicity fluxes
\citep[e.g.][]{Blackman_Field_2000_ApJ_534_984,Kleeorin_et_al_2000_AA_361_5,2001ApJ...550..752V,Vishniac_Shapovalov_2014_ApJ_780_144,Kleeorin_Rogachevskii_2022_MNRAS_515_5437,2023ApJ...943...66G}. The
most commonly invoked fluxes include
\begin{eqnarray}
\mathscrbf{F}^{\alphaM} = \frac{\etat \kf^2}{\Beq^2} ( - \kappa_\alpha \bm\nabla \alphaM - \mUUU \alphaM + \mFFF^{\rm f}).
\end{eqnarray}
where the first two terms of the rhs correspond to turbulent diffusion
\citep[e.g.][]{Covas_et_al_1998_AA_329_350,2010AN....331..130M} and
large-scale advection \citep[e.g.][]{SSS07} of $\alphaM$, whereas the
term $\mFFF^{\rm f}$ encompasses additional contributions that can be
independent of the large-scale magnetic field \citep[see][and
  references therein]{2023ApJ...943...66G}. The latter can in
principle also lead to dynamo action in the absence of kinetic
helicity
\citep[e.g.][]{Brandenburg_Subramanian_2005_AN_326_400,Vishniac_Shapovalov_2014_ApJ_780_144}.

\section{Prerequisites for accurate mean-field modeling}
\label{sec:prereq}

Before diving into the wealth of comparisons between 3D simulations
and mean-field theory, some statements regarding the procedure and
depth of the comparisons are needed. Roughly three levels of
comparisons between simulations mean-field theory can be readily
distinguished:
\begin{enumerate}
\item Estimates and measurements of inductive and diffusive effects
  from simulations without necessarily trying to use the results to
  explain any specific dynamo simulation.
\item The use of measured or inferred mean-field transport
  coefficients to interpret the simulation results
  \emph{qualitatively} in terms of mean-field concepts.
\item Detailed extraction and parametrization of the transport
  coefficients from 3D simulations and their use in corresponding
  \emph{quantitative} mean-field modeling.
\end{enumerate}
The first kind typically yields only limited insight into the actual
dynamo process in the simulations, whereas the second kind is possibly
useful in characterizing the dynamo in mean-field-theoretic
terms. However, only the most ambitious comparisons of the third kind
can give exhaustive knowledge about the dynamo mechanisms at play. To
make such comparisons, the following questions need to be addressed:
\begin{enumerate}
\item How to reliably measure turbulent transport coefficients such
  that they faithfully reproduce the EMF of the 3D simulation?
\item Do mean-field models using these parametrization reproduce the
  dynamo of the original simulations?
\item What is the uncertainty of the results obtained?
\end{enumerate}
The first question can be reformulated such that it is a necessary
requirement that the reconstructed $\mEMF$ is the same (or
sufficiently similar) as that in the original 3D simulation. The
second point addresses the requirement that the derived turbulent
transport coefficients, when used in a mean-field model, must
reproduce the large-scale fields of the 3D simulation. Only if both of
these conditions are met, can it be relatively confidently stated that
the mean-model captures the behavior of the 3D simulation. However,
this is an ambitious goal and typically such accuracy is difficult to
obtain. The third question refers mostly to \emph{systematic}
uncertainties and is crucial for the assessment of the reliability of
the comparisons. This includes the methods of computation of the
transport coefficients and the assumptions that enter the ansatz used
for the EMF such as the treatment of nonlinearity and nonlocality.

\section{Methods to compute turbulent transport coefficients from simulations}
\label{sec:memethods}

A major technical challenge is to compute all of the tensorial
coefficients appearing in the EMF ansatz such as \Eq{equ:R80EMF}. In
3D simulations the three components of $\mEMF$ are readily available,
but typically a much greater number of tensorial components appear on
the rhs of \Eq{equ:R80EMF}. Often this is not even attempted and only
a subset of the coefficients are retrieved. This is achieved via
\emph{ad hoc} simplifications of the EMF ansatz \Eq{equ:R80EMF}, for
instance, by neglecting terms proportional to derivatives of
$\mBBB$. The most common methods to compute the turbulent transport
coefficients from simulations along with representative results and
issues are discussed next.

\subsection{Imposed field method}

\begin{figure}
\begin{center}
\includegraphics[width=.9\textwidth]{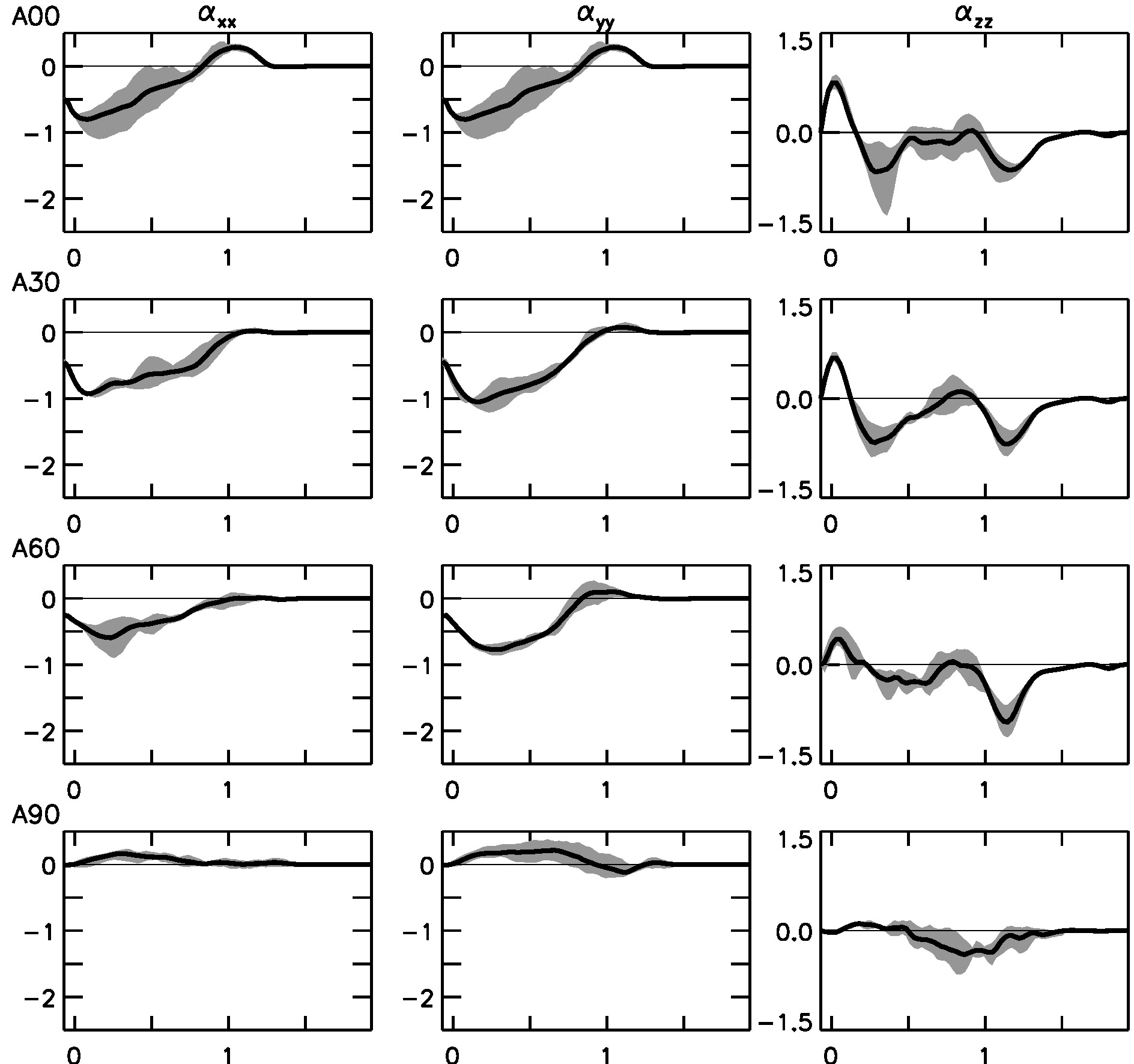}
\end{center}
\vspace{-.35cm}
\caption[]{Diagonal components of $\alpha_{ij}$ in units of $10^{-2}
  \sqrt{dg}$, where $d$ is the depth of the convection zone and $g$ is
  the acceleration due to gravity, as functions of height and latitude
  from Cartesian convection simulations of \cite{OSBG02} with
  $\Co=2.4$, $\Rey=\ReM=260$ that are based on the system scale. The
  different panels correspond to latitude $\theta =0$ (north pole, top
  row), $30^\circ$, $60^\circ$, and $90^\circ$ (equator, bottom
  row). The horizontal axis shows the vertical coordinate $z$ in units
  of $d$. The convection zone spans the range $0 < z/d < 1$.}
\label{fig:Ossendrijver_et_al_2002_alpha}
\end{figure}

Arguably the simplest approach to compute the transport coefficients
is to impose a mean magnetic field $\mBBB=\mBBBimp$ on the solution
and measure the induced $\mEMF$ \citep[see
  e.g.][]{BNPST90,CH96,OSB01,OSBG02,CH06,KKOS06}. In principle it is
possible to choose sufficiently many linearly independent imposed
fields such that all of the turbulent transport coefficients in an
ansatz such as \Eq{equ:R80EMF} can be extracted. However, this is
cumbersome because a separate numerical experiment is needed for each
$\mBBBimp$, and great care has to be taken to ensure that the
additional mean fields $\mBBB(\xxx,t)$ possibly generated in the
simulation remain small compared to $\mBBBimp$. Therefore the imposed
field studies have almost solely been done with uniform fields to
study the $\alpha$ effect \citep[see,
  however][]{1990fas..conf....1B}. Assuming that $\mBBBimp \approx
\mBBB$, \Eq{equ:R80EMF} reduces to
\begin{eqnarray}
\mEMF = \bm\alpha\bm\cdot \mBBBimp + \bm\gamma \times \mBBBimp,\label{equ:mEMFsimple}
\end{eqnarray}
where the latter term is analogous to $\mUUU\times\mBBB$, and where
three experiments suffice to compute all of the components of
$a_{ij}$. The imposed field method was first used to determine the
$\alpha$ effect from simulations of rotating stratified convection
\citep[e.g.][]{BNPST90,OSB01}. These studies showed that the
horizontal and vertical components of $\alpha$ had opposite signs
reflecting the underlying anisotropy of convective flows. Later these
studies have been expanded to map $a_{ij}$ as a function of latitude
and rotation from simulations of mildly turbulent convection by, for
example, \cite{OSBG02} and \cite{KKOS06}. The former found that for
moderate rotation ($\Co\approx 1$) the $\alpha$ effect is roughly
proportional to $\cos\theta$ which is also the lowest order
expectation from mean-field theory; see
\Figa{fig:Ossendrijver_et_al_2002_alpha}. Simulations probing cases
corresponding to the base of the solar convection zone with
$\Co\approx 10$ suggest increasing anisotropy and a latitudinal
maximum of the horizontal $\alpha$ effect around latitude $30^\circ$
\citep{KKOS06}.

The simulations of \cite{OSBG02} and \cite{KKOS06} also extracted the
turbulent pumping effect $\bm\gamma$. Downward pumping of magnetic
fields was already demonstrated by earlier numerical studies
\citep[e.g.][]{NBJRRST92,BJNRST96,TBCT98,TBCT01,2001A&A...365..562D}. \cite{OSBG02}
showed that the vertical pumping of large-scale fields to be dominated
by the diamagnetic effect, and that with sufficiently rapid rotation,
horizontal components of $\bm\gamma$ are non-negligible
\citep[e.g.][]{1991A&A...243..483K}. The latitudinal pumping was found
to be predominantly equatorward and the azimuthal pumping retrograde
in \cite{OSBG02}, both of which can potentially aid in obtaining
equatorward propagation of activity belts. This was confirmed by
mean-field models in \cite{KKT06}, where physically plausible
turbulent diffusivity $\etat$ and meridional circulation of the Sun
along with $\alpha$ and $\gamma$ coefficients from simulations of
\cite{KKOS06} were used.

The main caveat of the imposed field method is that $\mBBBimp$ is a
part of the solution, and that in general $\mBBB^{\rm(imp)} \neq
\mBBB$. This is often the case even if there is no dynamo in the
system, for example, due to inhomogeneity because of boundaries
\citep[e.g.][]{KKB10a}. This is also related to the choice of
averages: even in fully periodic cases with dynamos, for which the
volume averaged $\mJJJ$ vanishes, the relevant mean fields are
Beltrami fields with some scale $k_{\rm m}$ \citep[e.g][]{B01}. These
issues can lead to erroneous estimates of the turbulent transport
coefficients. This can be seen by considering the steady state
solution of \Eq{equ:meaninduction}: $\mEMF - \mu_0 \eta \mJJJ = 0$. If
$\mBBB$ is indeed uniform, then $\mJJJ=0$, further implying that
$\mEMF$ must also vanish if \Eq{equ:mEMFsimple} is assumed. In general
$\mBBB\neq\mBBB^{\rm(imp)}$ in the steady state implying $\mJJJ\neq
0$, and therefore the more general expression \Eq{equ:EMFFOSA} needs
to be used. Furthermore, by order of magnitude the steady state
solution and \Eq{equ:mEMFsimple} yield $\alpha B \approx \eta B/\ell$,
and lead to a normalized amplitude of $\tilde{\alpha} = \alpha/u =
\eta/u\ell = \ReM^{-1}$, where $u$ is a typical velocity
amplitude. Thus in this procedure $\alpha$ appears to be
catastrophically quenched proportional to $\ReM^{-1}$ even in the
kinematic regime \citep{CH06}. However, this is a misconception
arising from the application of the method outside of its range of
validity.

Therefore the imposed field method as described by \Eq{equ:mEMFsimple}
is reliable only if the actual mean field does not deviate greatly
from the imposed field, that is $\mBBB\approx\mBBBimp$. This is
satisfied in 2D where the method is exact and can therefore provide an
important benchmark for other methods. To ensure that
$\mBBB\approx\mBBBimp$ in 3D often involves resetting of the magnetic
field periodically before a steady state is reached
\cite[e.g.][]{OSBG02,KKOS06,2009MNRAS.398.1891H}.

\subsection{Multidimensional regression methods}

\subsubsection{Moments method of \cite{2002GApFD..96..319B}}

\begin{figure}
\begin{center}
  \includegraphics[width=.85\textwidth]{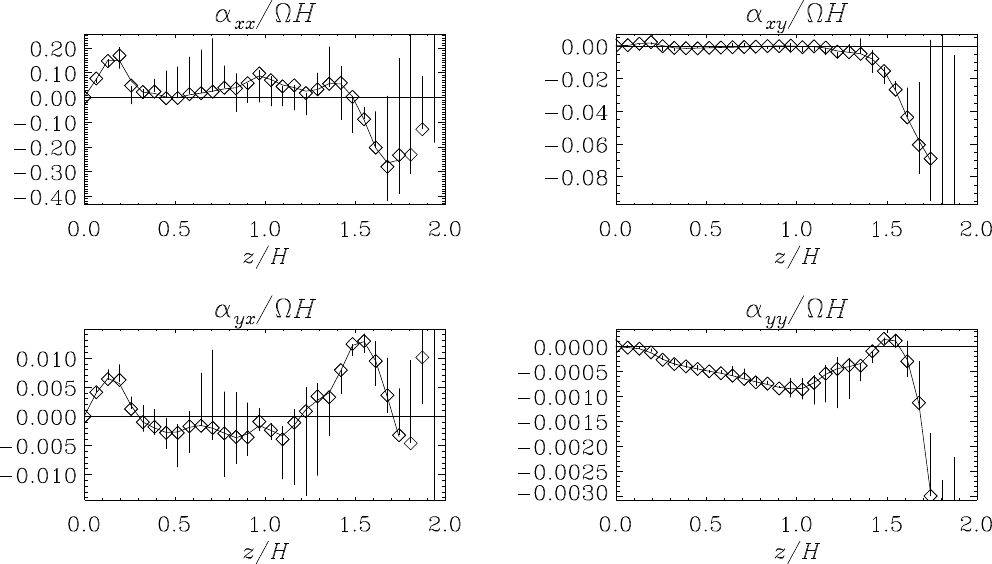}
  \includegraphics[width=.85\textwidth]{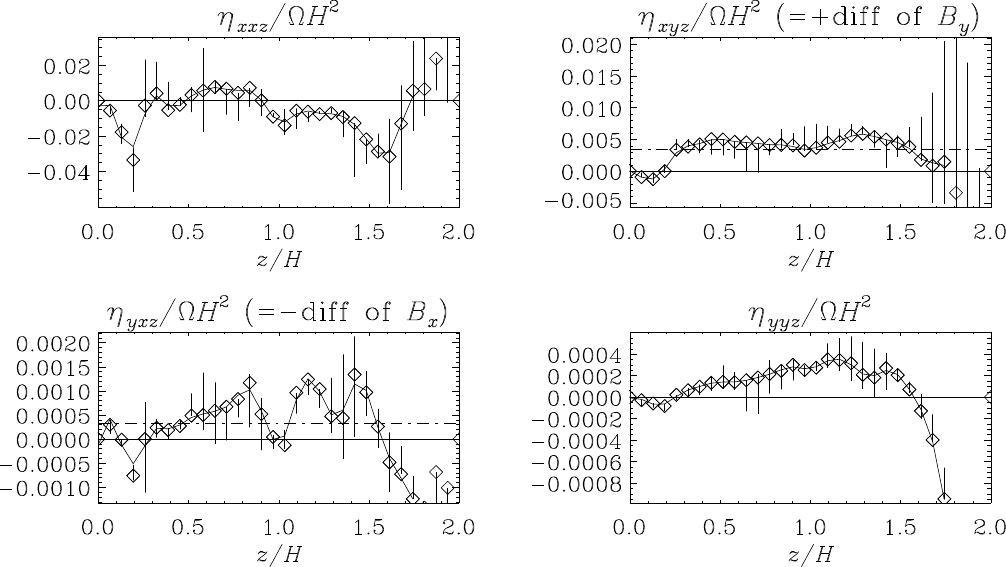}
\end{center}
\caption[]{Upper four panels: $\alpha_{ij}$ from inverting
  \Eq{equ:BS02Matrix} from a shearing box simulation of the
  magnetorotational instability \citep{2002GApFD..96..319B}. Lower
  four panels: $\eta_{ijk}$ from \Eq{equ:BS02Matrix} from the same
  simulation.}
\label{fig:alpeta_BS02}
\end{figure}

Another way to get around the problem of underdetermination is to
consider a temporally varying MHD solution and exploiting the fact
that $\mEMF$ and $\mBBB$ point to different directions at different
times. Using a sufficiently large set of realizations of $\mEMF$ and
$\mBBB$, it is possible to turn an underdetermined problem to an
overdetermined one, where the transport coefficients can be obtained
by fitting \citep{2002GApFD..96..319B}. The potential benefit of this
method is that the transport coefficients can be obtained from the
same calculation without resorting to additional runs with imposed
fields. Assuming local and instantaneous relation between $\mBBB$ and
$\mEMF$ and that the mean fields depend only on one coordinate ($z$),
\Eq{equ:EMFi} reduces to:
\begin{eqnarray}
\EMFi = \alpha_{ij} \mBBj + \eta_{ijz} \pd_z \mBBj, \label{equ:BS1}
\end{eqnarray}
with eight unknowns $\alpij$ and $\eta_{ijz}$. In
\cite{2002GApFD..96..319B} this was circumvented by forming moments of
\Eq{equ:BS1} with $\mBBi$ and $\pd_z\mBBi$. The resulting eight
equations are time averaged and the solution is given by two matrix
equations
\begin{eqnarray}
  \EEE^{(i)}(z) = \CCC^{(i)}(z)\MMM(z),\ \mbox{for}\ \ i=x,y.\label{equ:BS02Matrix}
\end{eqnarray}
The matrices $\EEE^{(i)}$ and $\MMM^{(i)}$ contain the moments of mean
fields and their gradients with $\mEMF$, and with $\mBBB$ and
$\pd_z\mBBB$ themselves, respectively, whereas $\CCC^{(i)}$ contains
the time-averaged transport coefficients.  \Figu{fig:alpeta_BS02}
shows the results from a density-stratified simulation of the
magnetorotational instability (MRI) from
\cite{2002GApFD..96..319B}. It is not a priori known how $\alpha_{ij}$
or $\etaijk$ should look like in this case. However, $-\eta_{yxz}$ is
responsible for diffusing the $\mBBx$ component of the mean field.
This component is predominantly positive, implying anti-diffusion
which is considered unphysical. Further experiments by
\cite{2002GApFD..96..319B} suggest that assuming $\etaijk$ to be
diagonal removes the problem with anti-diffusion, but there is no
physical justification to do this. They also concluded that non-local
effects may play a role in that negative diffusion at large scales can
be compensated by positive one at higher wavenumbers.

\subsubsection{Singular value decomposition (SVD)}
\label{sec:SVDmethod}

The SVD method is similar to the moments method described above and
relies on the availability of data to overcome the problem of
underdetermination of the EMF ansatz. In this method two
time-dependent functions
\begin{eqnarray}
y(t) = \EMFi(t,r,\theta),\ \ \mbox{and}\ \ X_k (t) = [
  \mBBi(t,r,\theta), \pd_r \mBBi(t,r,\theta),
  \pd_\theta\mBBi(t,r,\theta)]
\end{eqnarray}
are constructed from the EMF and magnetic field data. The quantity
$\phi_k = [ a_{ij}(r,\theta), b_{ijr}(r,\theta),
  b_{ij\theta}(r,\theta)]$ contains the turbulent transport
coefficients. The parametrization of the EMF is given by
\begin{eqnarray}
y(t) = \sum_{k=1}^n \phi_k X_k(t),
\end{eqnarray}
where $n$ depends on the ansatz for the EMF. The coefficients $\phi_k$
are required to minimize the least squares fit characterized by a
standard $\chi^2$ approach.

\begin{figure}
\begin{center}
  \includegraphics[width=.7\textwidth]{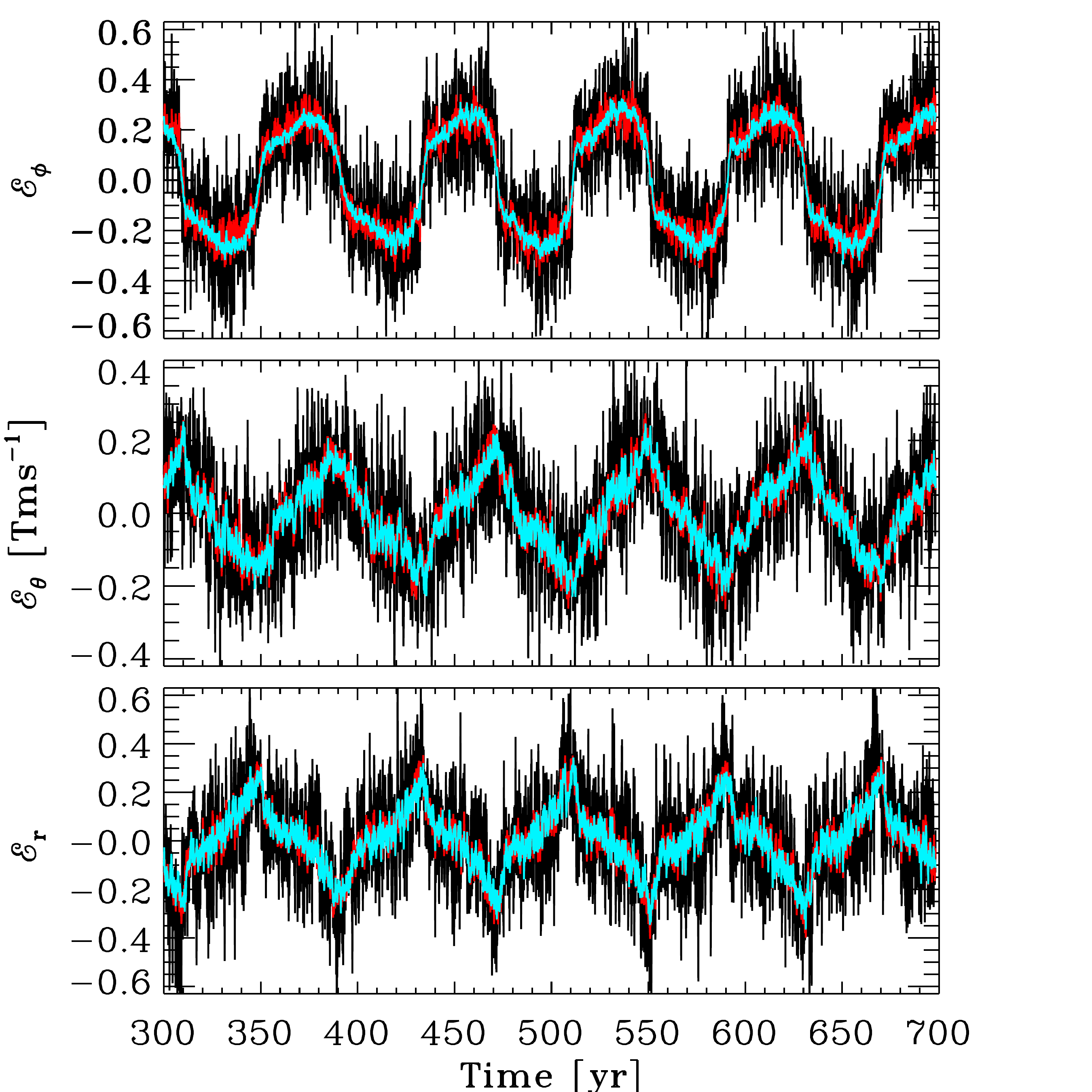}
\end{center}
\caption[]{Components of the EMF (black lines) and fits to it
  using only the $\bm \alpha$ tensor and $\bm\alpha$ (magenta) and
  $\bm\beta$ (red). Adapted from \cite{2016AdSpR..58.1522S}.}
\label{fig:Simard_et_al_2016_EMF}
\end{figure}

\begin{figure}
\begin{center}
  \includegraphics[width=0.9\textwidth]{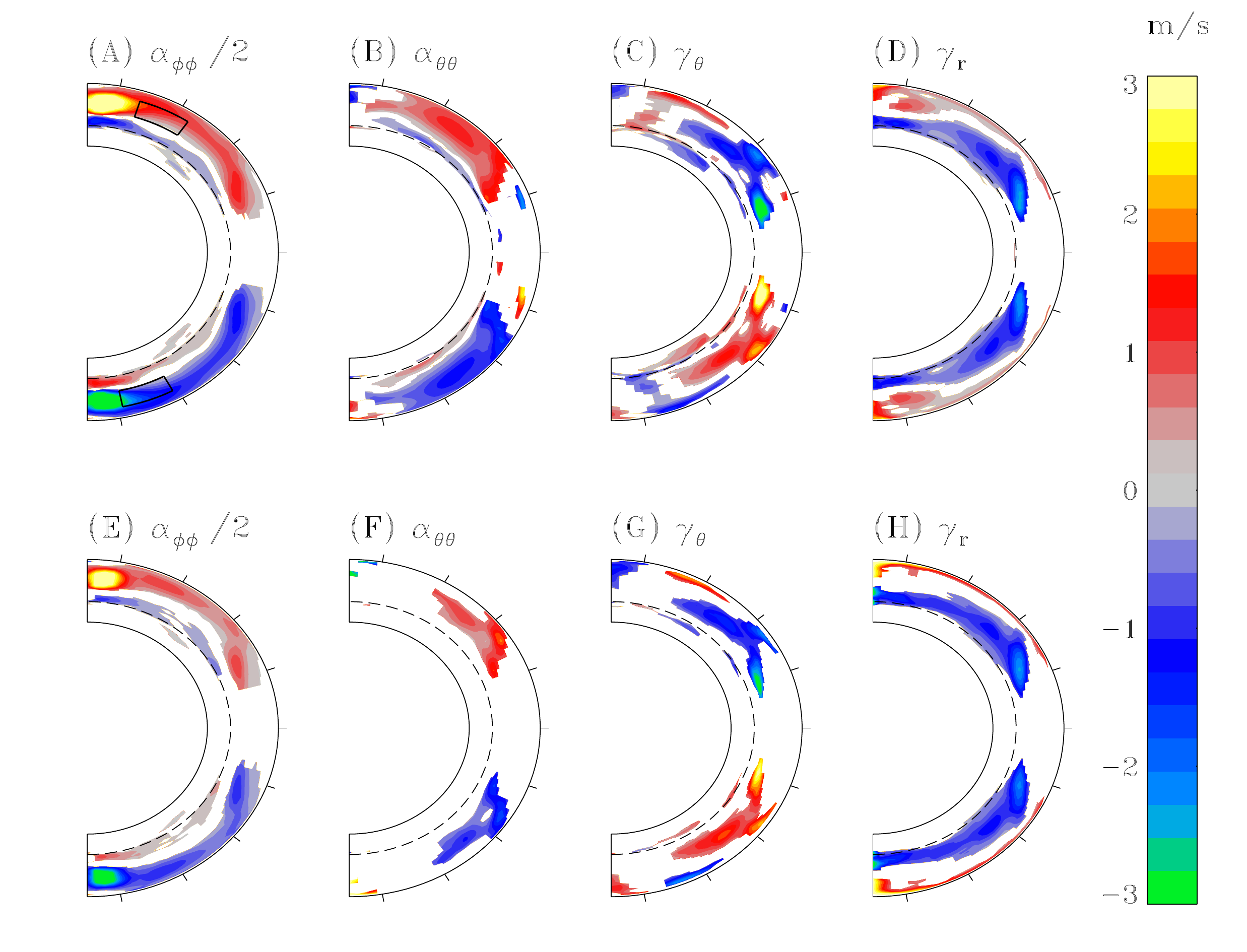}
\end{center}
\vspace{-.25cm}
\caption[]{Top: $\alppp$, $\alptt$, $\gamma_r$, and $\gamma_\theta$
  extracted from an {\sc EULAG} simulation where only the $\bm\alpha$
  tensor is retained in the SVD analysis. Bottom: same as the top
  panel but where the $\bm\beta$ tensor was retained in the
  fitting. Adapted from \cite{2016AdSpR..58.1522S}.}
\label{fig:Simard_et_al_2016_alpgam}
\end{figure}

The SVD method was initially used to extract only the $\bm\alpha$
tensor components from global convection simulations
\citep[e.g.][]{RCGBS11,SCB13,NBBMT13,ABMT15,2022ApJ...926...21B,2022ApJ...935...55S}. In
these studies the EMF was assumed to have the form $\mEMF =
\bm\alpha\bm\cdot\mBBB + \bm\gamma\times\mBBB$. However, the
gradients of $\mBBB$ and the corresponding contributions to the
$\mEMF$ cannot in general be dropped, and in a subsequent study,
\cite{2016AdSpR..58.1522S} generalized the EMF to include all of the
terms in \Eq{equ:R80EMF}. \Figu{fig:Simard_et_al_2016_EMF} illustrates
that with the SVD method the reconstructed $\mEMF$ is in very good
agreement with the actual EMF in both of the aforementioned cases. The
caveat here is that in the SVD method the coefficients are fitting
parameters and the method does not guarantee that they are correct or
physically realizable.

Nevertheless, the results of \cite{2016AdSpR..58.1522S} and the
earlier studies show that the diagonal components of $\alpha$ are in
qualitative agreement with theoretical arguments that $\alpha \propto
- \mean{\kinhel}$, although the magnitude is roughly five times lower
than the FOSA estimate irrespective whether the diffusive contribution
to the EMF were retained. Similarly the diagonal components of $\beta$
are predominantly positive where statistically relevant, but again
about five times lower than the FOSA estimates. The two experiments
discussed above give very similar results for $\bm\alpha$ and
$\bm\gamma$; see \Figa{fig:Simard_et_al_2016_alpgam} and the effects
due to the diffusive contributions were therefore found to be small. A
(although still not fully conclusive) test of the validity of the
coefficients is to use them in a mean-field model of the same system
from which they were extracted. Efforts to this direction are
discussed further in \Seca{sec:comp}.

\subsection{Test field methods}
\label{sec:TFM}

\subsubsection{Quasi-kinematic test field method}
\label{sec:QKTFM}

Another way to avoid the issues with imposed fields is to use the test
field method \citep[][]{SRSRC05,SRSRC07,2005AN....326..787B}, where
the imposed fields are replaced by a sufficient number of linearly
independent test fields.  A separate induction equation for the
fluctuating fields is solved for each of the test fields. That is, for
each test field $\mBBB^{(p)}$,
\begin{eqnarray}
\frac{\pd \bbb}{\pd t}^{(p)} = \bm\nabla \times \left(\mUUU \times \bbb^{(p)} + \uuu \times \mBBB^{(p)} + \mcbm{G}^{(p)}\right) + \eta\bm\nabla^2 \bbb^{(p)},
\end{eqnarray}
is solved, where the velocity fields $\mUUU$ and $\uuu$ come from the
simulation (main run). Neither the test fields $\mBBB^{(p)}$ nor the
small-scale fields $\bbb^{(p)}$ react back on the flow. The non-linear
term $\mcbm{G}^{(p)} = \uuu \times \bbb^{(p)} - \mean{\uuu \times
  \bbb^{(p)}}$ is fully retained such that the method is superior to
FOSA and MTA. This flavor of the test field method is formally
applicable to cases where the small-scale magnetic field $\bbb$
vanishes when $\mBBB\rightarrow 0$, thus excluding cases with $\bbb$
is due to a small-scale dynamo. On the other hand, the velocity field
$\UUU$ can already be affected by a magnetic field $\BBB$ in the main
run. This is why this flavor is referred to as the quasi-kinematic
test field method. Non-linear extensions of the test field method are
discussed in \Seca{sec:nltf}.

In the simplest case the mean fields are assumed to depend only on a
single coordinate, here $z$. The EMF is then
\begin{eqnarray}
  \EMFi = a_{ij} \mBBj - b_{ij} \mu_0 \mJJj,\label{equ:mEMF1D}
\end{eqnarray}
where $b_{i1}=b_{i23}$ and $b_{i2}=-b_{i13}$. The EMF has only $x$ and
$y$ components, and $a_{ij}$ and $b_{ij}$ have four components each. A
sufficient choice of test fields is
\begin{eqnarray}
\mBBB^{\rm 1c} = B_0 (\cos kz,0,0),\ \ \ \mBBB^{\rm 2c} = B_0 (0,\cos kz,0),\label{equ:testfcos}\\
\mBBB^{\rm 1s} = B_0 (\sin kz,0,0),\ \ \ \mBBB^{\rm 2s} = B_0 (0,\sin kz,0),\label{equ:testfsin}
\end{eqnarray}
where $k$ is a wavenumber. This leads to two linear sets of equations
from which $a_{ij}$ and $b_{ij}$ can be solved unambiguously. If
harmonic test fields are used, higher order derivatives of $\mBBB$ can
be considered to have been already included because, for example,
$(\pd^n/\pd^n x)(\cos kx)=k^n \cos \left( kx + \frac{\pi n}{2}
\right)$.
\begin{figure}[ht]
\begin{center}
\includegraphics[width=.6\textwidth]{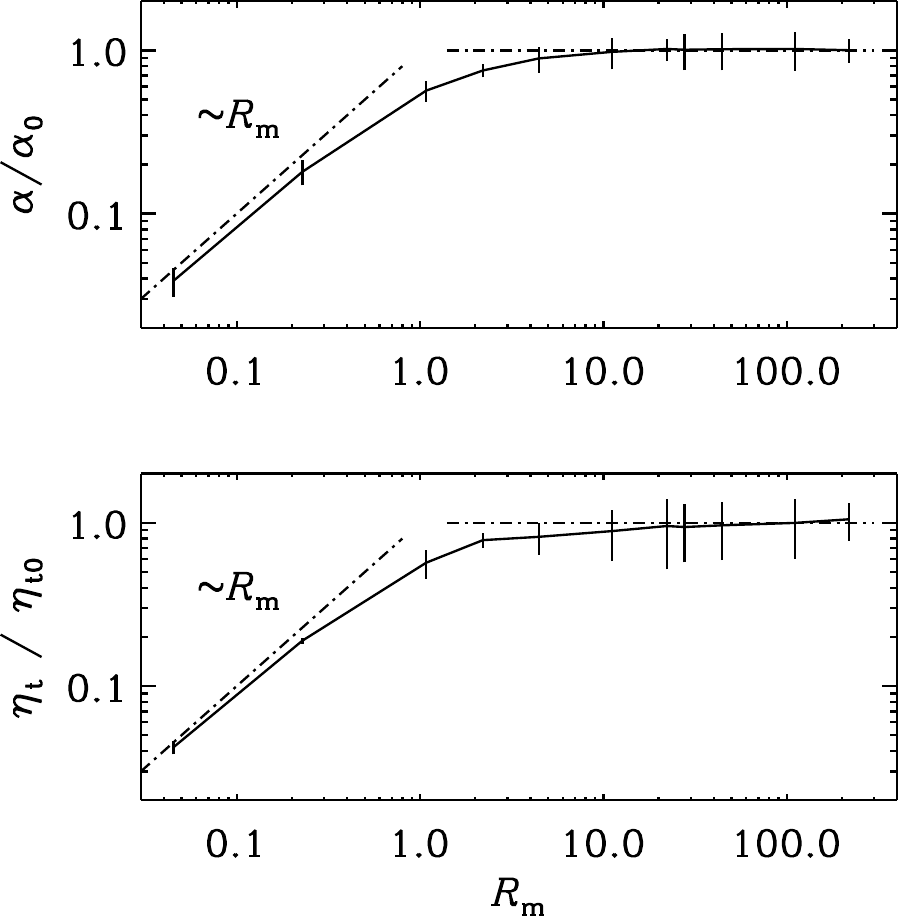}
\end{center}
\caption[]{Top: $\alpha$ normalized by the FOSA estimate $\alpha_0$
  from isotropically forced helical turbulence as a function of the
  magnetic Reynolds number. Bottom: $\etat$ normalized by the FOSA
  estimate $\etatz$ from the same simulations. Adapted from
  \cite{SBS08}.}
\label{fig:alpeta_Sur2008}
\end{figure}
The coefficients $a_{ij}$ and $b_{ij}$ can be recast as
\begin{eqnarray}
& \alpha = \onehalf(a_{11} + a_{22}), \ \ \ \gamma = \onehalf(a_{21}-a_{12}),&\label{equ:zdepco1}\\
  & \etat = \onehalf(b_{11} + b_{22}), \ \ \ \delta = \onehalf(b_{12}-b_{21}),&\label{equ:zdepco2}
\end{eqnarray}
which describe the $\alpha$ effect, turbulent pumping ($\gamma$),
turbulent magnetic diffusivity ($\etat$), and the
R\"adler/shear-current effect ($\delta$), respectively.

Representative results from homogeneous isotropically forced helical
turbulence are shown in \Figa{fig:alpeta_Sur2008}. Here the measured
$\alpha$ effect and $\etat$ turbulent diffusivity are normalized by
the FOSA estimates $\alpha_0 = \onethird \urms$, $\eta_{\rm t0} =
\onethird \urms \kf^{-1}$, where $\kf$ is the energy-carrying scale of
turbulence, and where the former is relevant for fully helical
turbulence. \Figu{fig:alpeta_Sur2008} shows that both $\alpha$ and
$\etat$ are proportional to $\ReM=\urms/\eta \kf$ for $\ReM\lesssim1$
in accordance with the corresponding FOSA result in the low
conductivity limit, see \cite{KR80}. Furthermore, for $\ReM\gtrsim
10$, $\alpha$ and $\etat$ converge to roughly constant values that
agree with within the error estimates with the FOSA estimates. It is
remarkable how good the correspondence is given the rather strict
validity constraints of FOSA. However, the quasi-kinematic test field
method itself is formally valid only when a small-scale dynamo is
absent. This is another stringent condition because the critical
magnetic Reynolds number is around 30 for $\PrM=1$
\citep[e.g.][]{HBD04} and increases to a few hundred for $\PrM
\rightarrow 0$ \citep[][]{KR12,2023NatAs...7..662W}.

This method has been used to study the turbulent transport
coefficients in shearing turbulence without \citep[][]{BRRK08} and
with kinetic helicity \citep[][]{MKTB09}. These studies did not find
conclusive evidence for a dynamo effect through the
R\"adler/shear-current effect in shearing turbulence, which is
mediated via off-diagonal components of $\eta_{ij}$. Furthermore,
\cite{2017AN....338..790B} found a contribution to $\etat$ due to
kinetic helicity from forced turbulence simulations from analysis of
test field data. Such contributions go beyond FOSA and arise only when
fourth order correlations in the fluctuations are considered in the
computation of transport coefficients
\citep[e.g.][]{1988GApFD..43..149N,2025ApJ...984...88B,2025ApJ...985...18R}. A
further new aspect discovered with the help of test field calculations
is the scaling of the $\alpha$ effect in density-stratified systems
\citep{2013ApJ...762..127B}. While earlier theoretical studies yielded
$\alpha \propto \rho^\sigma \urms$ with $\sigma > 1$
\citep[][]{Ruediger_Kichatinov_1993_AA_269_581}, the analytic results
by \cite{2013ApJ...762..127B} suggest that $\sigma = \onehalf$. Such
scaling was also found from forced turbulence and sufficiently
stratified turbulent convection simulations for slow rotation
($\Co\lesssim 0.2$) using the test field method in
\cite{2013ApJ...762..127B}. Results at more rapid rotation were less
coherent but consistently yielded $\sigma < 1$ in contrast to the
earlier theoretical predictions.

\begin{figure}[ht]
\begin{center}
\includegraphics[width=.56\textwidth]{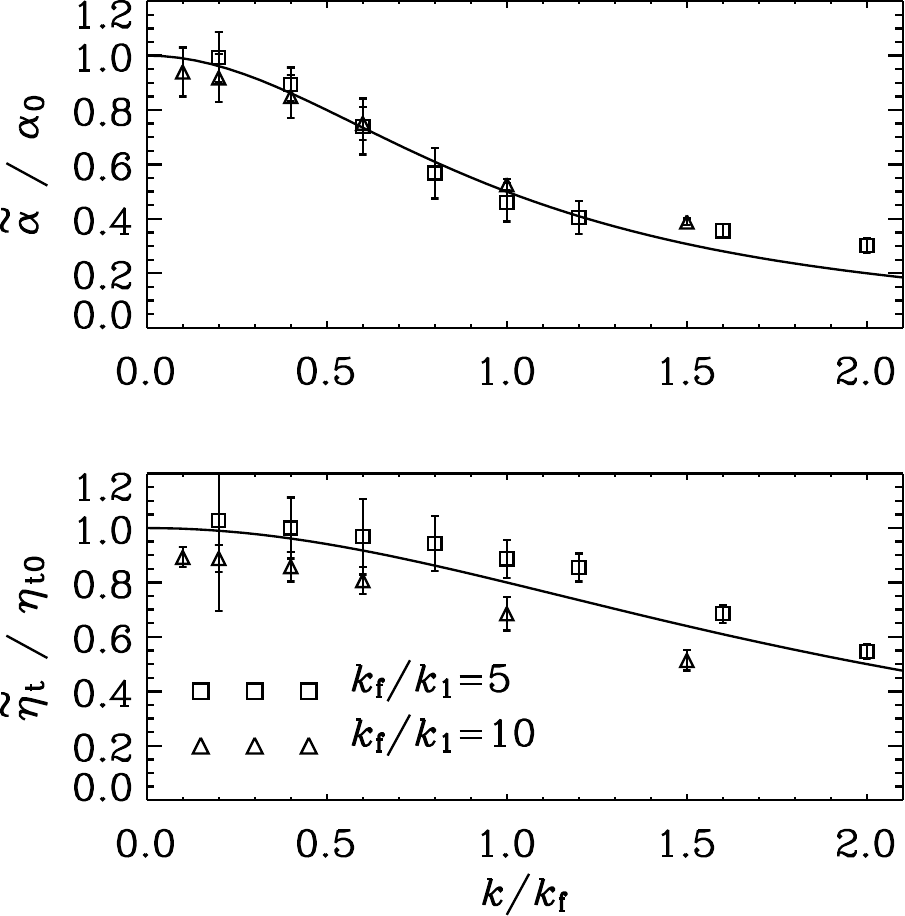}\hspace{.25cm}\includegraphics[width=.39\textwidth]{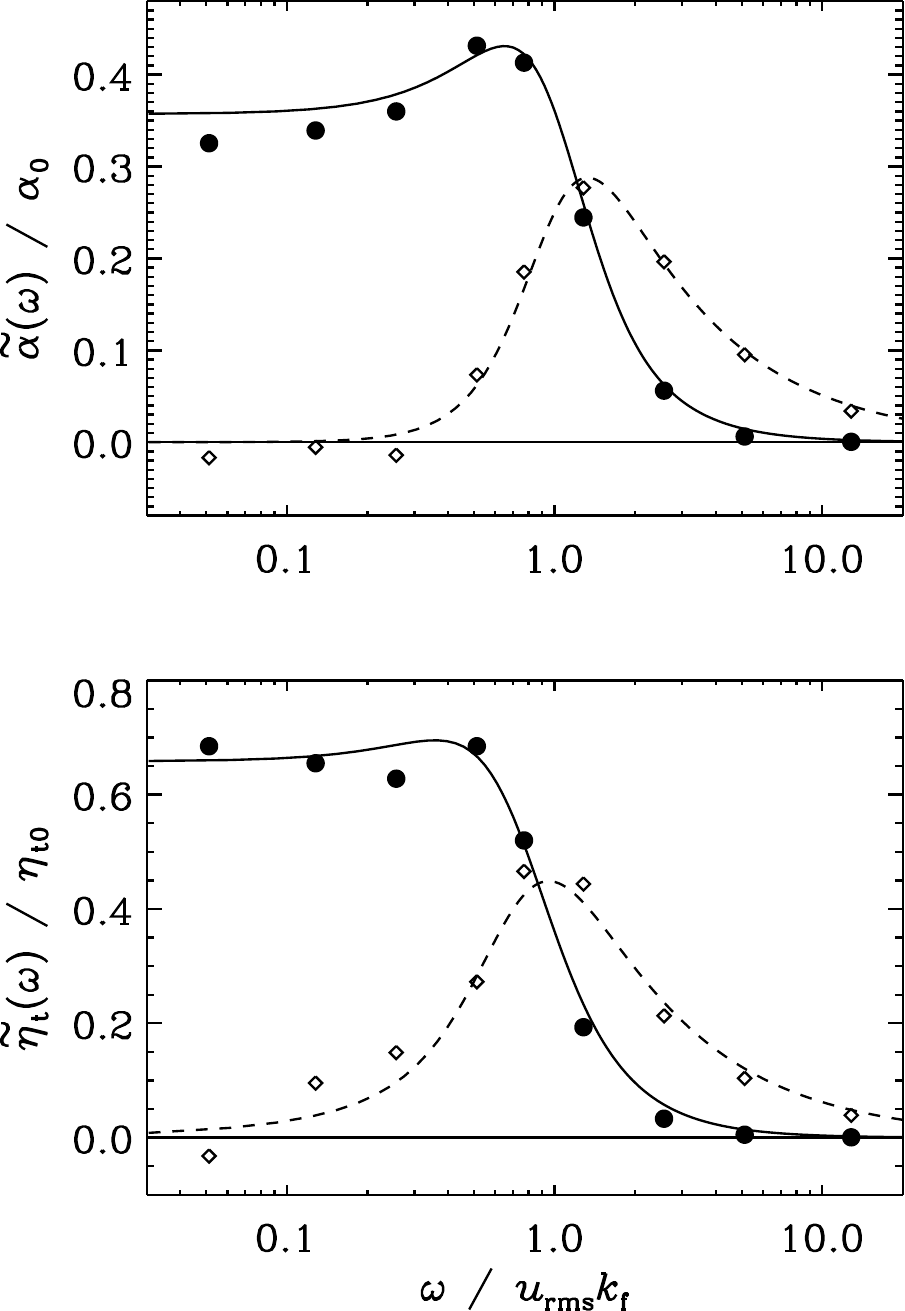}
\end{center}
\caption[]{Left: $\tilde{\alpha}(k)$ (top) and $\tilde{\eta}_{\rm
    t}(k)$ (bottom) normalized by the FOSA estimates from
  isotropically forced helical turbulence as a function of the
  wavenumber $k$ of the test fields. Adapted from \cite{BRS08}. Right:
  $\tilde{\alpha}(\omega)$ (top) and $\tilde{\eta}_{\rm t}(\omega)$
  (bottom) normalized by the FOSA estimates from isotropically forced
  helical turbulence as a function of the frequency $\omega$ of the
  test fields. Adapted from \cite{2009ApJ...706..712H}.}
\label{fig:BRS08_alpeta}
\end{figure}

The scale dependence of $\alpha$ and $\etat$ for the Roberts flow and
for isotropically forced homogeneous turbulence were studied in
\cite{BRS08}. The dependence of the coefficients on spatial
scale was found to be approximately Lorentzian:
\begin{eqnarray}
\tilde{\alpha}(k) = \frac{\alpha_0}{1+(k/\kf)^2},\ \ \tilde{\eta}_{\rm t}(k) = \frac{\etatz}{1+(k/2\kf)^2}.
\end{eqnarray}
The results are shown in the left panels of
\Figa{fig:BRS08_alpeta}. Such non-local contributions to $\alpha$ and
$\etat$ lead to a modification of the growth rate of the
dynamo. However, \cite{BRS08} concluded that the non-local effects
become important only when the scale of the mean field is comparable
to the dominant scale of turbulence. In a subsequent study,
\cite{2009ApJ...706..712H} investigated temporal non-locality or
memory effects for passive scalar transport and dynamos in Roberts
flow and in turbulence. In analogy to the spatial non-locality, they
found that memory effects become important when the large-scale field
varies on a similar timescale as the flow itself, translating to a
Strouhal number of the order of unity or larger; see the right panels
of \Figa{fig:BRS08_alpeta}. Finally, \cite{2012AN....333...71R}
considered the case where both spatial and temporal scale separations
are poor and suggested to represent the non-local EMF as
\citep[see also][]{Pipin_2023_MNRAS_522_2919}
\begin{eqnarray}
\left(1 + \tau\frac{\pd}{\pd t} - \ell^2 \frac{\pd^2}{\pd z^2}\right)\EMFi = \alpij^{(0)}\mBBj + \etaijk^{(0)}\mean{B}_{j,k},\label{equ:nonlocalEMF}
\end{eqnarray}
where $\tau$ and $\ell$ are a temporal and spatial scale, and where
the superscript zero refers to the transport coefficients for
$k\rightarrow 0$ and $\omega \rightarrow
0$. \cite{2012AN....333...71R} admitted that higher order terms are
likely needed to fully capture the non-locality but that this
expression nevertheless gives a flavor of the issue. A qualitative
change of behavior in comparison to the purely local description is
that the excitation threshold for the dynamo is lowered if the dynamo
is oscillatory \citep[e.g.][]{2012AN....333...71R}. A similar
conclusion was reached by \cite{BC18} who implemented
\Eq{equ:nonlocalEMF} in a spherical mean-field dynamo model
representative of the Sun. Furthermore, \cite{2014MNRAS.441..116R}
showed that the dynamos in two flavors of non-helical Roberts flow
\citep{1972RSPTA.271..411R} are driven by off-diagonal components of
the $a_{ij}$ tensor when memory effects are retained.

Quenching of turbulent transport coefficients was studied by
\cite{2014ApJ...795...16K} by means of the test field method from
simulations of the Roberts flow, forced turbulence, and convection
where a large-scale external field was imposed. Quenching formula
proportional to $[1+p_i(\mBBB/\Beq)^{q_i}]^{-1}$ was assumed, where
$p_i$ and $q_i$ were fit parameters. The quenching exponent $q_i$
depends on the type of the flow and whether $\Beq$ is estimated from
the unquenched flow or not. In the latter case the exponent $q_\alpha$
for the $\alpha$ effect was 2 for turbulent convection and 3 for
isotropically forced homogeneous turbulence whereas the $q_{\eta_{\rm
    t}}$ for the turbulent diffusivity ranged from 1.1 to
1.3. Strongly anisotropic quenching has been assumed in some solar
dynamo models in order to assure that turbulent diffusion remains
subdominant in determining the cycle period
\citep[e.g.][]{CNC04,Karak_Choudhuri_2011_MNRAS_410_1503}. Curiously,
\cite{2014ApJ...795...16K} found no evidence of such strong anisotropy
even in cases where the large-scale field reached $10^2$ times
equipartition.

\begin{figure}[ht]
\begin{center}
\includegraphics[width=0.9\textwidth]{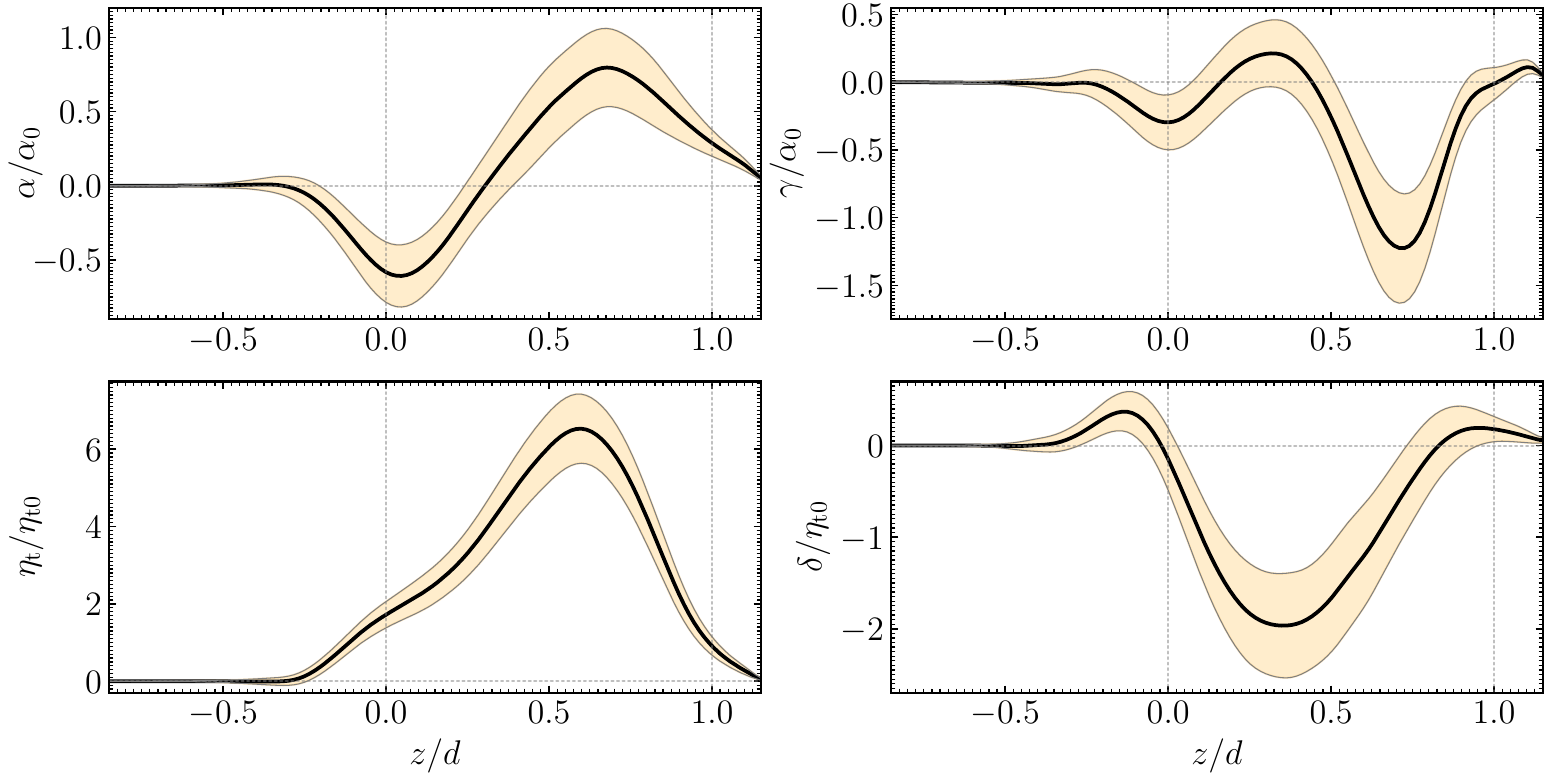}
\end{center}
\caption[]{Turbulent transport coefficients $\alpha(z)$ (top left),
  $\gamma(z)$ (top right), $\etat(z)$ (bottom left), and $\delta(z)$
  (bottom right) from Run~B of \cite{KKB09a} with $\Rey=35$ and
  $\Co=0.36$. The convection convection zone is situated between
  $z/d=0$ and $z/d=1$.}
\label{fig:alpeta_KKB09}
\end{figure}

\begin{figure}[ht]
\begin{center}
\includegraphics[width=.9\textwidth]{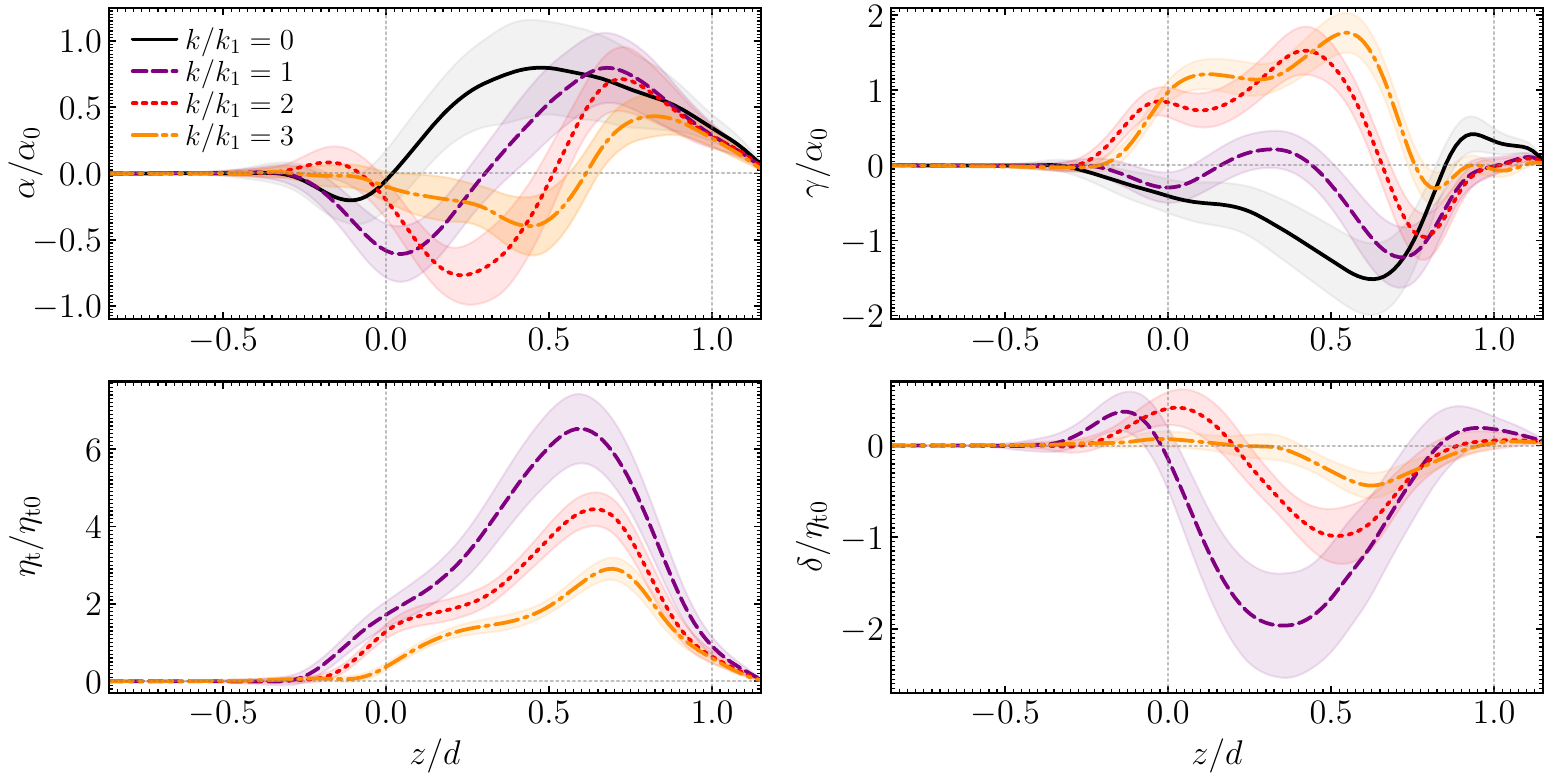}
\end{center}
\caption[]{Effects of non-locality on the turbulent transport
  coefficients $\alpha$, $\gamma$, $\etat$, and $\delta$ from
  stratified convection in Cartesian geometry. Data from Runs~B11-B14
  of \cite{KKB09a}. The legend indicates the normalized wavenumber of
  the test fields.}
\label{fig:KKB09_nonloc}
\end{figure}

The test field method as presented above can also be applied to
inhomogeneous cases such as convection in a Cartesian box. An example
is shown in \Figu{fig:alpeta_KKB09} where horizontally averaged
kinetic helicity and the coefficients according to
\Eqs{equ:zdepco1}{equ:zdepco2} are shown from a rotating density
stratified convection simulation in Cartesian coordinates
\citep{KKB09a}. The profiles of $\alpha$ and $\gamma$ are similar to
those obtained with the imposed field method \citep[][]{OSB01}, which
also roughly agree with theoretical (FOSA) predictions. The vertical
pumping effect $\gamma$ is downward only in the upper convection zone,
whereas $\etat$ is positive everywhere. Stratified convection is
highly non-local such that flows can traverse long distances and even
penetrate the whole convection zone. Therefore the assumption of a
local and instantaneous connection between $\mEMF$ and $\mBBB$ is
\emph{a priori} not well justified. \Figu{fig:KKB09_nonloc} shows the
same turbulent transport coefficients as in \Figa{fig:alpeta_KKB09}
but computed with test fields with different vertical wavenumber $k$
ranging between $0\ldots 3$, where $k=0$ corresponds to an imposed
field. These results show that some of the coefficients, such as
$\alpha$ and $\gamma$ can even change sign as a function of spatial
scale, whereas the amplitudes of $\etat$ and $\delta$ are reduced for
higher values of $k$\footnote{In \cite{KKB09a} the coefficient
$\delta$ had a sign error which is corrected in
\Figa{fig:KKB09_nonloc}.}.

\begin{figure}[ht]
\begin{center}
\includegraphics[width=.5\textwidth]{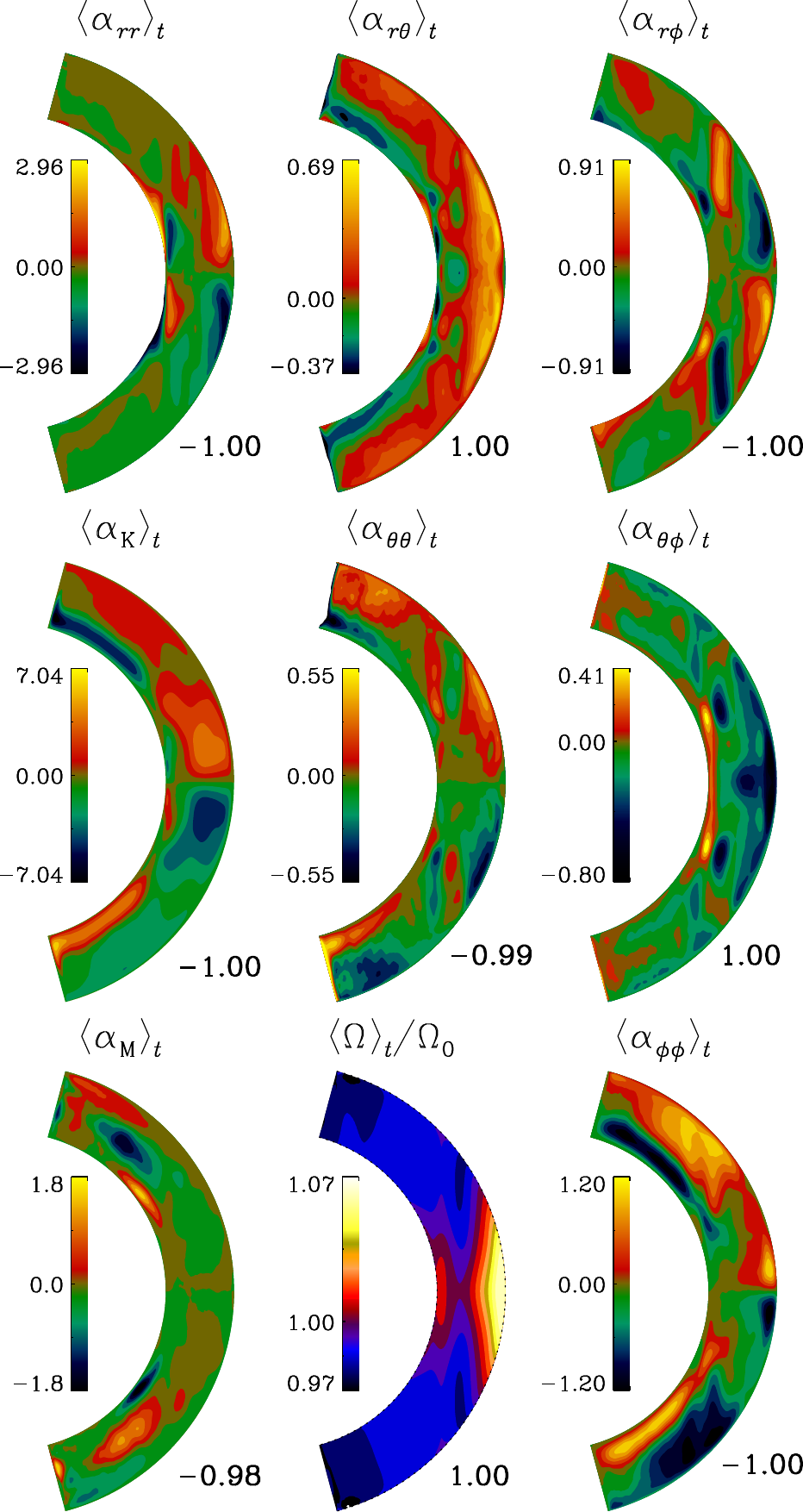}\includegraphics[width=.5\textwidth]{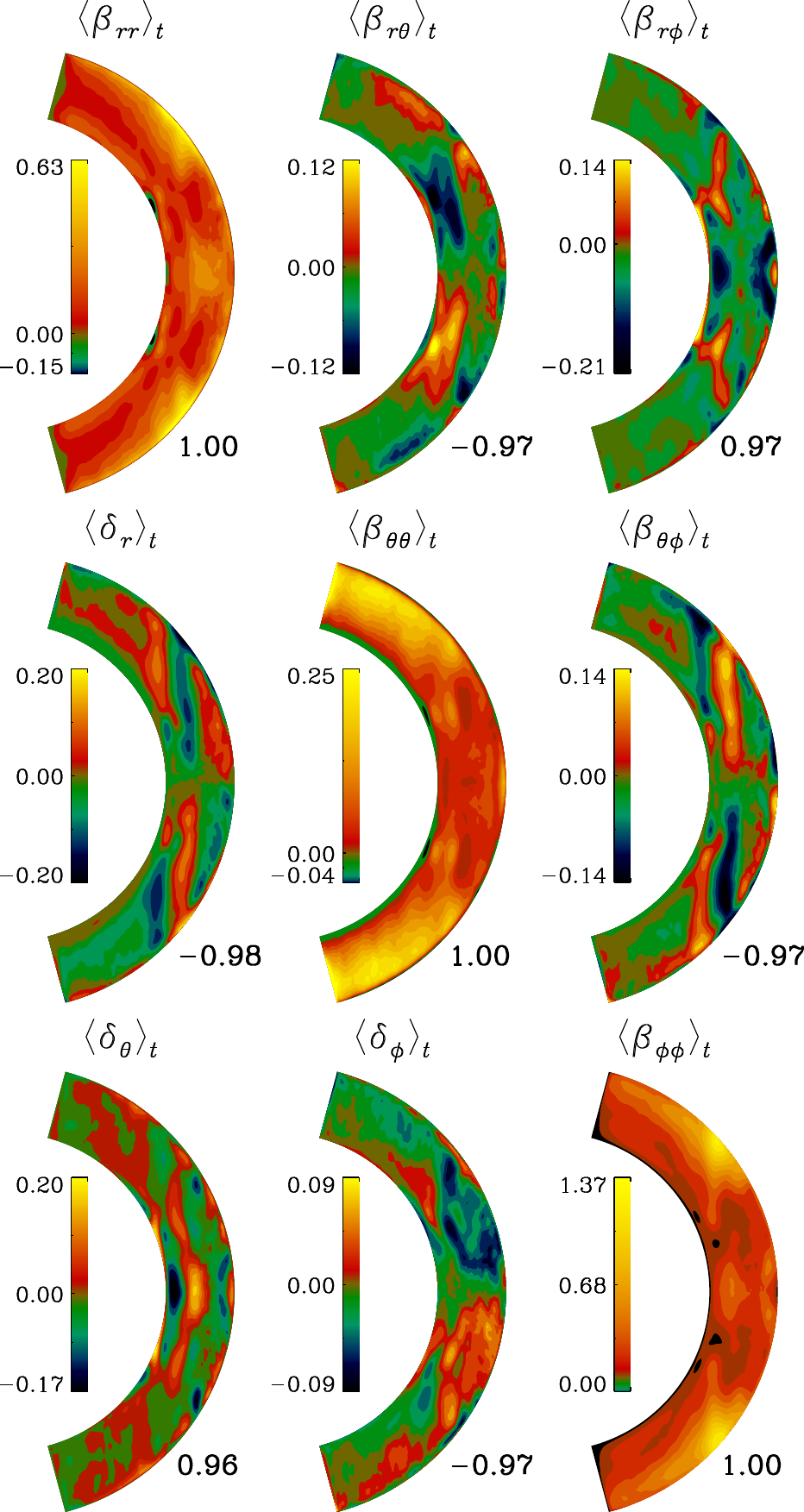}
\end{center}
\caption[]{Nine panels on the left: time averaged components of the
  $\alpha_{ij}$ tensor, $\alphaK = - \onethird \tau \mkinhel$,
  $\alphaM = \onethird \tau \mcurhel$, and $\mOm = \mUUp/r \sin\theta
  + \Omega_0$ from a convection simulation in a spherical wedge
  \citep{2018A&A...609A..51W}. Nine panels on the right: Independent
  components of $\bm\beta$ and $\bm\delta$ from the same simulation.}
\label{fig:alpbeta_Warnecke2018}
\end{figure}

Considering stars such as the Sun, the method has to operate spherical
coordinates. Assuming axisymmetry and that non-local effects can be
omitted, a total of 27 independent $a_{ij}$ and $b_{ijk}$ coefficients
are needed; see \cite{SRSRC05,SRSRC07}. In these, and in subsequent
works
\citep[e.g.][]{Sch11,SPD11,SPD12,2013MNRAS.431L..78S,2018A&A...609A..51W,2019ApJ...886...21V},
the test fields were chosen to be
\begin{eqnarray}
(\mBBr,\mBBt,\mBBp)\!=\!(1,1,1),\ (\mBBr,\mBBt,\mBBp)\!=\!(r,r,r),\ (\mBBr,\mBBt,\mBBp)\!=\!(\theta,\theta,\theta).\label{equ:BBtestsph}
\end{eqnarray}
The EMF is given by
\begin{eqnarray}
\EMFi = \tilde{a}_{ij}\mBBj + \tilde{b}_{ijr}\pd_r \mean{B}_j + \tilde{b}_{ij\theta}\pd_\theta \mean{B}_j,\ \  i,j = r,\theta,\phi.\label{equ:EMFisph}
\end{eqnarray}
The coefficients can be worked out from \Eq{equ:EMFisph} using
\Eq{equ:R80EMF}, \citep[see e.g.\ Appendix A
  of][]{2019ApJ...886...21V}. \Figu{fig:alpbeta_Warnecke2018} shows
representative results for the components of $\bm\alpha$ and
$\bm\beta$ from \cite{2018A&A...609A..51W}. The diagonal components of
$\alpha_{ij}$ and the estimate $\alpha_{\rm K} = -\onethird \tau
\mean{\bm\omega\bm\cdot\uuu}$, where $\tau = (\urms k_1)^{-1}$ is an
estimate of the convective turnover time, are in rough qualitative
agreement. This entails a positive $\alpha$ effect in the bulk of the
convection zone in the northern hemisphere. However, the $\alpha$
effect in density stratified rotating convection is necessarily
anisotropic and the $\alphaK$ estimate cannot be expected to be
accurate in detail
\citep[e.g.][]{Kleeorin_Rogachevskii_2003_PhRvE_67_026321}.
Furthermore, the diagonal components of $\beta_{ij}$ are predominantly
positive. Non-positivity of turbulent diffusion is most likely
unphysical but it can also be related to insufficient data or
non-local effects which were not considered.

\begin{figure}[ht]
\begin{center}
\includegraphics[width=\textwidth]{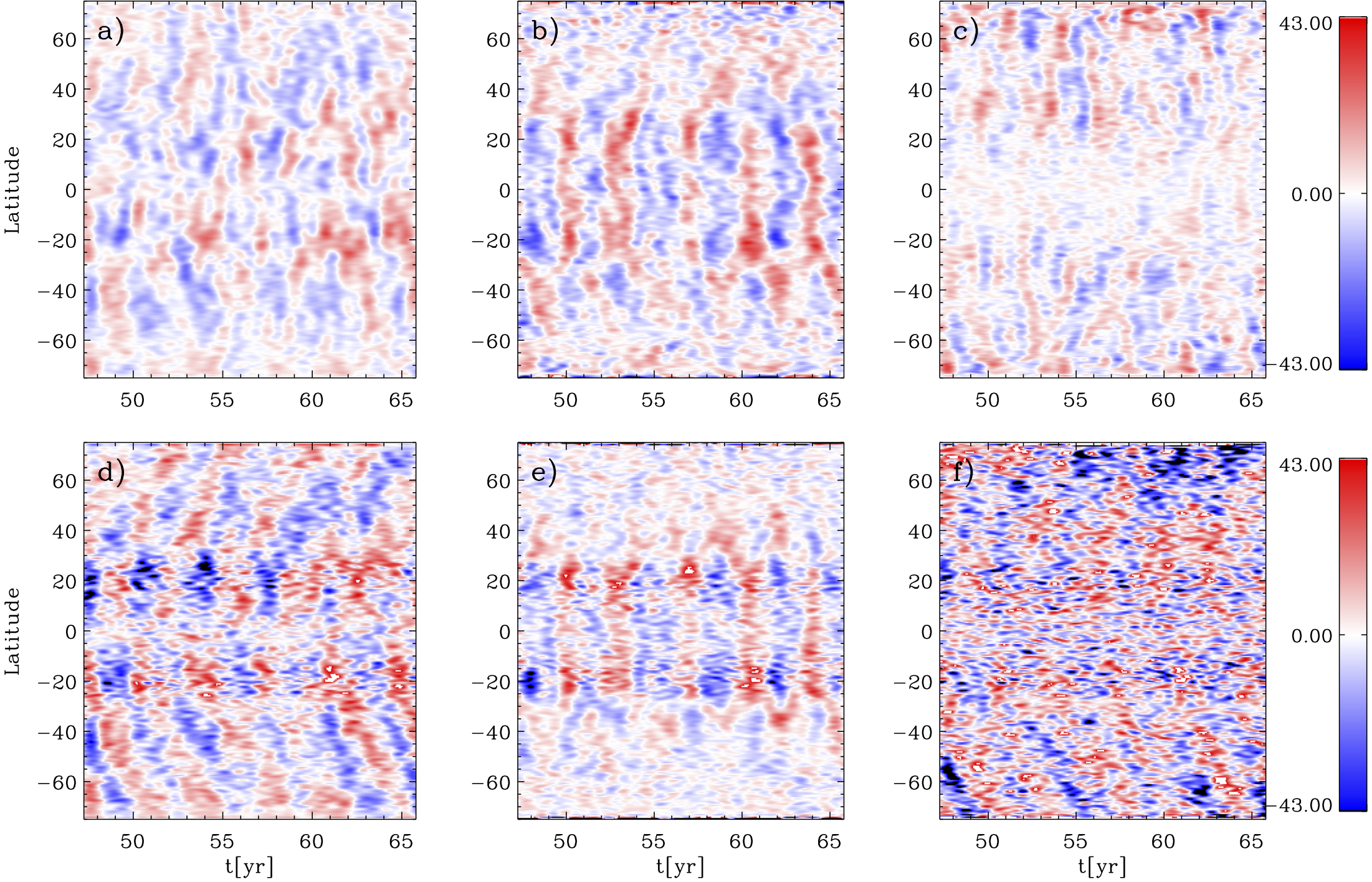}
\end{center}
\caption[]{Top row: The actual radial, latitudinal and longitudinal
  components of the $\mEMF$ from the simulation presented in
  \cite{2019ApJ...886...21V}. Bottom row: the same quantities but
  reconstructed using \Eq{equ:R80EMF} with the measured turbulent
  transport coefficients and mean magnetic field $\mBBB$ in the same
  simulation.}
\label{fig:EMFreconst_Viviani2019}
\end{figure}

A check of the validity of the computed transport coefficients is that
they are used to reconstruct the actual EMF of the main run. An
example is shown in \Figa{fig:EMFreconst_Viviani2019} for the
simulation presented in \cite{2019ApJ...886...21V}. While reasonable
qualitative agreement can be found especially at low latitudes, the
magnitude of the reconstructed EMF is higher than the actual EMF by a
factor of two to three; see \cite{2019ApJ...886...21V}. A possible
explanation is that the test fields in \Eq{equ:BBtestsph}, vary on
large scales that are comparable to coherent structures in convective
flows in which case non-local effects can become important. Therefore
the assumption of a local and instantaneous correspondence between
$\mBBB$ and $\mEMF$ can be questioned in the case of convection. On
the other hand, correspondence between a DNS and a mean-field model
was found to be much better for a forced turbulence simulation with a
significantly higher scale separation and $\ReM\approx 1$ where the
effects of non-locality are likely to be less important
\citep{2018A&A...609A..51W}.

\subsubsection{Nonlinear test field methods}
\label{sec:nltf}

The quasi-kinematic test field method is formally applicable to
situations where also the small scale fields owe their existence to
$\mBBB$. This is a rather restrictive condition from astrophysical
perspective where small-scale dynamos are likely ubiquitous
\citep[e.g.][]{2023SSRv..219...36R}. A succession of methods to
generalize the quasi-kinematic test field method have been developed
to accommodate this
\citep{2010A&A...520A..28R,2020ApJ...905..179K,2022ApJ...932....8K}.
These methods have gradually included more of the terms in the
Navier-Stokes equation that are now needed for the calculation of the
test field $\mEMF^{(p)}$. Here only the most general case, the
compressible test field method, is discussed
\citep{2022ApJ...932....8K}. In this method the full set of MHD
equations are solved for the main run (mr) where a self-consistent
small-scale and large-scale dynamos can operate, a zero run (0) where
the mean magnetic field is zero, and for each of the imposed test
fields ($B$). For example, for isotropically forced homogeneous
turbulence this corresponds to evolving $\uuu$, $\bbb$, and $h = \cst
\ln\rho$ for each of these cases. A full mean-field representation of
this system requires that the mean electromotive force $\mEMF^{(B)}$,
the ponderomotive force
\begin{eqnarray}
\mcbm{F}^{(B)}=(\mean{\jjj\times \bbb)/\rhoref -
  \uuu\bm\cdot\bm\nabla\uuu + 2\nu \bm{\mathsf{s}}\bm\cdot\bm\nabla h})^{(B)},
\end{eqnarray}
and the mean mass source
\begin{eqnarray}
\mcbm{Q}^{(B)} = -(\mean{\uuu\bm\cdot\bm\nabla h})^{(B)},
\end{eqnarray}
where $\rhoref$ is a constant reference density and $\bm{\mathsf{s}}$
is the fluctuating rate-of-strain tensor, are computed. However, all
of these correlations contain terms that are ultimately non-linear in
$\mBBB$. Furthermore, there is no direct connection between the main
run and the test field equations. In \cite{2022ApJ...932....8K} this
is circumvented by assuming that $\bbb^{(\rm mr)} \approx \bbb =
\bbb^{(0)} + \bbb^{(B)}$. This is not fully rigorous but it is assumed
to be sufficiently accurate if the actual mean field in the main run
and the test fields are similar. This leads to freedom in choosing
which combinations of the fluctuating quantities from the main run,
zero run, and test field runs are used to construct the turbulent
correlations that are non-linear, such as $(\uuu\times\bbb)'^{(B)}$,
in the mean field. In the case studied in \cite{2022ApJ...932....8K}
this leads to a total of 32 possible flavors of the compressible test
field method, four of which \citep[see also][]{2010A&A...520A..28R}
were studied in more detail.

So far the compressible test field method has been applied to study
the shear dynamo problem along with simpler configurations involving
the Roberts flow. For the shear dynamo the results from compressible
test field method are in agreement with those of the quasi-kinematic
test field method in that no evidence of a coherent shear-current or
R\"adler effects were found, even in a regime where a strong
small-scale dynamo was present. Currently the usefulness of the
compressible test field method is limited to relatively low $\ReM$
because the linear test field solutions are prone to a small-scale
dynamo-like instability. This issue is alleviated to some degree by
resetting similarly as in the case of the quasi-kinematic test field
method \citep[e.g.][]{2009MNRAS.398.1891H}.

\section{Comparisons of 3D dynamo simulations with mean-field theory and models}
\label{sec:comp}

\subsection{Forced turbulence simulations with and without shear (Class 1)}

\subsubsection{Helically forced turbulence without shear ($\alpha^2$ dynamo)}

Forced turbulence simulations in periodic cubes offer the simplest
point of comparison between DNS and mean-field models. The helically
forced case with no shear constitutes the minimal ingredients of an
$\alpha^2$ dynamo in the parlance of mean-field theory. The dispersion
relation for such a dynamo in the kinematic regime is given by
\citep[e.g.][]{M78,KR80}
\begin{eqnarray}
\lambda = |\alpha| k - \etaT k^2,\label{equ:a2disp}
\end{eqnarray}
where $k$ is the wavenumber of the magnetic field and $\etaT = \etat +
\eta$. The maximum growth rate is obtained at $\kmax =
|\alpha|/(2\etaT)$, where $\kmax$ is the wavenumber corresponding to
the fastest growing mode. Assuming fully helical turbulence forced at
wavenumber $\kf$, such that $\alpha = \onethird \urms$ and $\etat =
\onethird \urms \kf^{-1}$, the wavenumber of the fastest growing for
sufficiently large $\ReM$ with $\etat \gg \eta$ is $\kmax = \onehalf
\kf$ \citep{2002AN....323...99B}. In the general case the flow is not
fully helical and the molecular diffusivity $\eta$ cannot be
neglected. Then, $\alpha = \onethird \epsilon_f \urms$, where
$\epsilon_f = \mkinhel/(\urms\orms)$ is the fractional helicity, and
the expression for the wavenumber of the fastest growing mode is
\begin{eqnarray}
\kmax = \frac{\epsilon_f k_f}{2(1+\ReM^{-1})}.\label{equ:kmax}
\end{eqnarray}
This expression was obtained by \cite{2002AN....323...99B} who also
found from simulations that $\kmax$ increases for increasing $\ReM$ in
accordance with \Eq{equ:kmax}.

\begin{figure}[ht]
\begin{center}
\includegraphics[width=.39\textwidth]{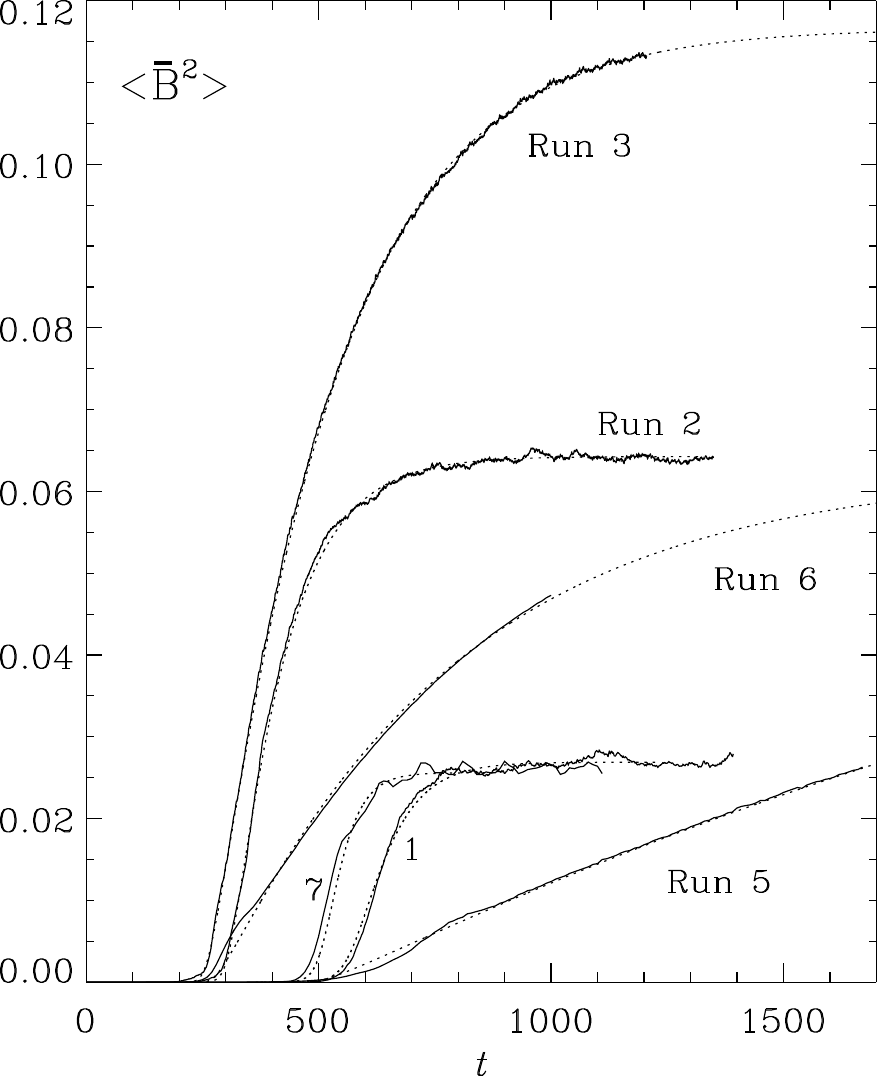}\hspace{.25cm}\includegraphics[width=.58\textwidth]{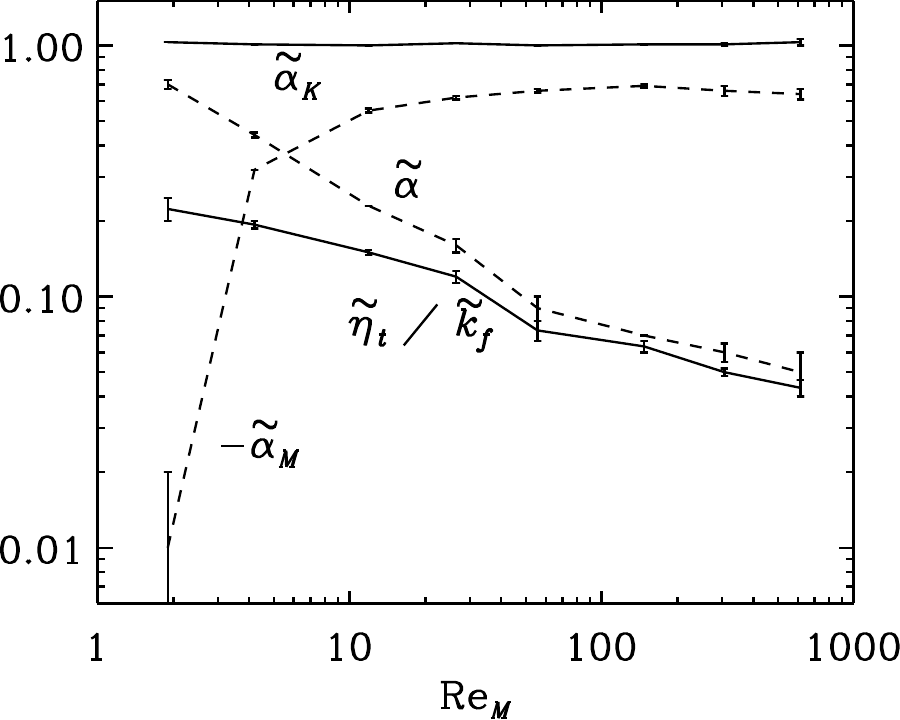}
\end{center}
\caption[]{Left: Average energy of the mean magnetic field as a
  function of time (solid lines) along with predictions based on
  magnetic helicity conservation (dotted lines). Adapted from
  \cite{B01}. Right: quenching of $\alpha$ and $\etat$ from dynamo
  simulations driven by helical turbulence
  \citep{2008ApJ...687L..49B}}
\label{fig:helicalturbdyn}
\end{figure}

The wavenumber of the fastest growing mode corresponds to a maximum
growth rate is $\lammax = \alpha^2/(4\etaT)$, suggesting that the
growth rate of the magnetic field is quadratic in the kinetic helicity
due to $\alpha \propto \mkinhel$. Simulations provide support for such
scaling; see Fig.~8 of \cite{SB14}. However, in the same study
$\lammax$ was found to be about twice larger than the actual growth
rate $\lambda$ in the 3D simulation. The reason for this discrepancy
is currently unclear but it is known that an additional correction to
the dispersion relation arises due to a kinetic helicity dependence of
turbulent diffusivity
\citep{1988GApFD..43..149N,Mizerski_2023_PRE_107_055205,2025ApJ...985...18R}
which was reported from simulations by
\cite{2017AN....338..790B,2025ApJ...984...88B}. Furthermore, the
dispersion relation (\ref{equ:a2disp}) is also modified if a memory
effect, or temporal non-locality, is present such that $\alpha =
\alpha(\lambda)$ and $\etat = \etat(\lambda)$
\citep{2009ApJ...706..712H}. While this effect was found to be
important for the Roberts flow in \cite{2009ApJ...706..712H}, it has
yet to be demonstrated for turbulence.

The non-linear state of a fully periodic helical dynamo is of
mean-field theoretical interest because it can be used to study the
effects of magnetic helicity conservation. In this context, \cite{B01}
studied the non-linear evolution of large-scale dynamos using
helically forced turbulence without shear. Such simulations produce
large-scale fields are of Beltrami type for which $\mJJJ = a
\mBBB$. Furthermore, the saturation of the mean field in these
simulations was found to occur on a resistive timescale in accordance
with arguments arising from mean-field theory and magnetic helicity
conservation; see the left panel of
\Figa{fig:helicalturbdyn}. Furthermore, quenching of $\alpha$ and
$\etat$ as functions of $\ReM$ was studied using the quasi-kinematic
test field method in \cite{2008ApJ...687L..49B}. The magnetic
diffusivity was shown to be decreased by a factor of roughly 5 in the
range $\ReM = 2\ldots 600$, while the total $\alpha$ effect defined
via \Eq{equ:MTAcoefs} was reduced by a factor of 14; see right panel
of \Figa{fig:helicalturbdyn}. The reduction of $\alpha$ was attributed
to be almost solely due to the increasing $\alphaM$ with $\ReM$.

\subsubsection{Helically forced turbulence with shear ($\alpha\Omega$ dynamo)}
\label{sec:helforce}

\begin{figure}[ht]
\begin{center}
\includegraphics[width=.65\textwidth]{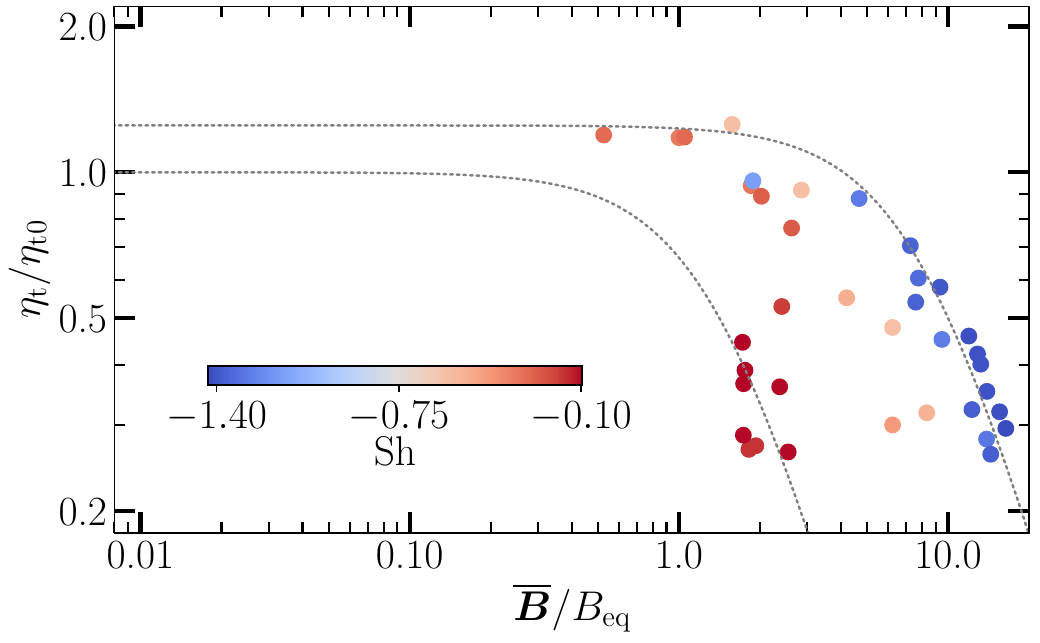}
\end{center}
\caption[]{Normalized turbulent magnetic diffusivity $\tetat =
  \etat/\etatz$, where $\etatz=\urms/(3\kf)$, where $\etat$ is
  computed from \Eq{equ:omcycashear} as a function of the mean
  magnetic field in units of $\Beq$. The curves show quenching
  functions proportional to $\tetat/[1 + p_{\eta_{\rm
        t}}(\mBBB/\Beq)^2]$ with $\tetat=1.25$ ($1.0$) and
  $p_{\eta_{\rm t}}=0.015$ ($0.5$) for the upper (lower) curve. Based
  on data from Fig.~10 of \cite{KB09}.}
\label{fig:Kapyla_Brandenburg_2009_etaquench}
\end{figure}

Another example of a Class 1 dynamo is the $\alpha$-shear dynamo,
which is driven by imposed shear flow and helical turbulence
\citep[e.g.][]{KB09,HRB11,2013Natur.497..463T,2014ApJ...789...70C,2016ApJ...825...23P}. Such
a setup corresponds to classical $\alpha\Omega$ or $\alpha^2\Omega$
dynamos depending on the strength of the shear. In shearing box
simulations with uniform imposed shear $\mUUU=(0,xS,0)$ this is
denoted by the shear number $\Sh = S/(\urms \kf)$. Such dynamos
produce cyclic magnetic fields where the propagation direction is
consistent with the Parker-Yoshimura rule
\citep[][]{2007NJPh....9..305B}. Furthermore, assuming a stationary
saturated state of the dynamo, the cycle frequency $\omcyc$ is given
by
\begin{eqnarray}
\omcyc = \etaT k_m^2,\label{equ:omcycashear}
\end{eqnarray}
where $k_m$ is the wavenumber of the large-scale mean magnetic
field. Assuming \Eq{equ:omcycashear} to hold also in the nonlinear
regime where $\etaT=\etaT(\mBBB)$, the relation $\ocyc(\mBBB)$ can be
interpreted as a proxy of the quenching of the magnetic
diffusivity. This was done in \cite{KB09}; see
\Figa{fig:Kapyla_Brandenburg_2009_etaquench}. The results suggest that
quenching becomes significant for $\mBBB/\Beq \gtrsim 1$ and that the
onset of quenching depends on the strength of the shear measured by
$\Sh$.

An important question to consider is the influence of a concurrent
small-scale dynamo on the large-scale field generation. This was
studied by \cite{2016ApJ...816...28K} in a series of simulations where
the relative importance of the small-scale dynamo was varied. They
found that when the small-scale dynamo was absent, the small-scale
magnetic fields were positively correlated with the cyclic large-scale
field. When the small-scale dynamo was also excited, two scenarios
appeared. First, if the large-scale field was weaker than the
equipartition value, the small-scale fields were almost independent of
the large-scale field. Second, if the large-scale field was stronger
than equipartition, the small-scale field was anti-correlated with the
cycle, interpreted as suppression of the SSD. In the Sun the
small-scale magnetic fields are independent of -- or weakly
anticorrelated with -- the cycle, suggesting that both a small-scale
and a large-scale dynamo are operating. Furthermore,
\cite{2016ApJ...816...28K} found decent agreement between their
simulation results and analytic theory of tangling of small-scale
fields in the absence of a small-scale dynamo \citep{RK07}.

\cite{HRB11} studied helically forced dynamos with shear in a
parameter regime where both $\alpha^2$ and $\alpha^2\Omega$ dynamos
were possible. While an oscillatory $\alpha^2\Omega$ mode was found to
have the fastest growth rate, it was often overwhelmed by a
non-oscillatory $\alpha^2$ mode in the nonlinear regime. Sometimes
such transition was found to occur long after the dynamo had reached a
saturated state. This was mentioned as a challenge for mean-field
models and calls for methods to extract turbulent transport
coefficients in the nonlinear regime. \cite{HRB11} connected these
transitions to the Sun and conjectured that a transition between
$\alpha^2\Omega$ and $\alpha^2$ modes could explain, for example, the
Maunder minimum. However, \cite{HRB11} found only a unidirectional
transition $\alpha^2\Omega \rightarrow \alpha^2$ such that the
original oscillatory mode never recovered.

\subsubsection{Nonhelically forced turbulence with shear}
\label{sec:nonhelforced}

\begin{figure}[ht]
\begin{center}
\includegraphics[width=.64\textwidth]{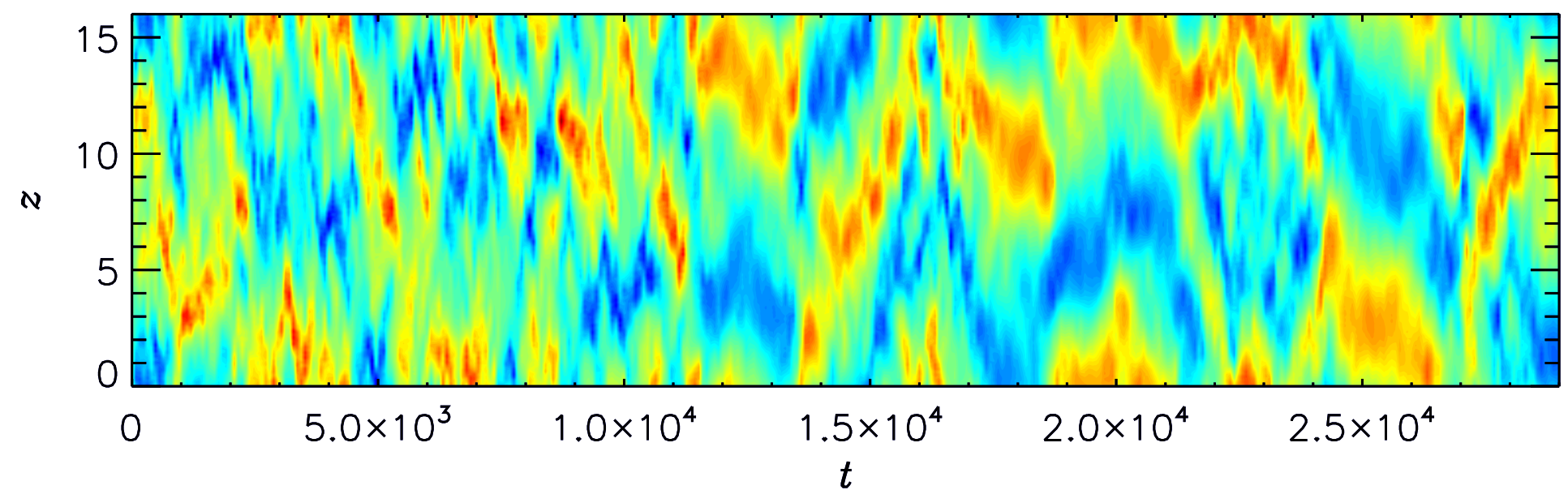}\hspace{0.25cm}\includegraphics[width=.33\textwidth]{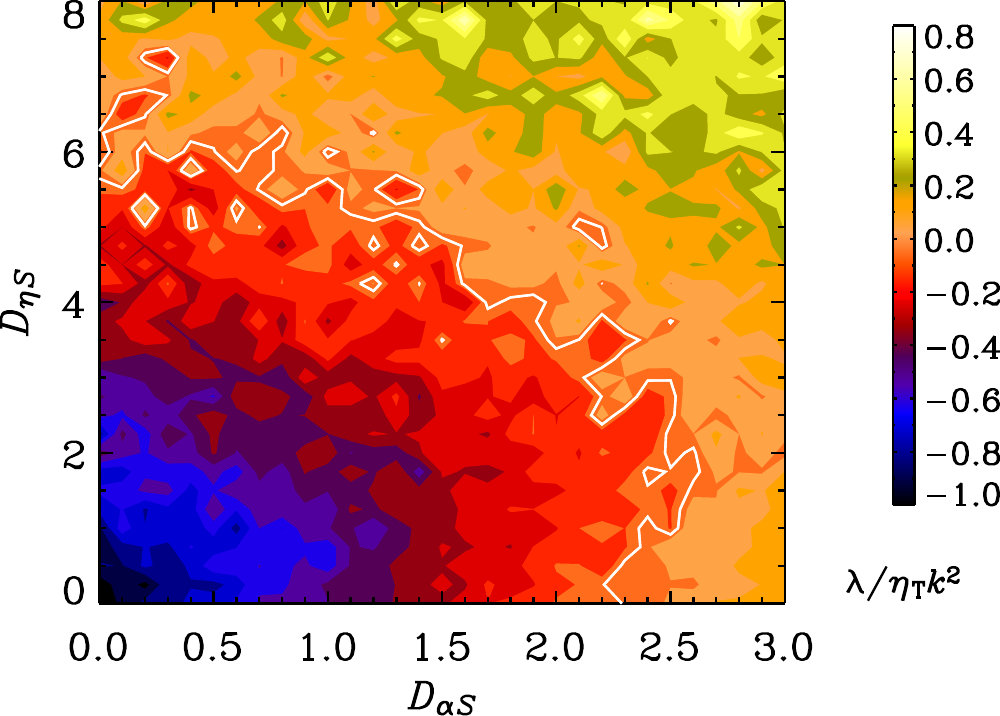}
\end{center}
\caption[]{Left: Horizontally averaged stream-wise magnetic field
  $B_y$ from a simulation of nonhelical shear dynamo. Adapted from
  \cite{Teed_Proctor_2017_MNRAS_467_4858}. Right: Dynamo growth rate
  from mean-field models as functions of the strengths of the
  incoherent $\alpha$ and shear-current effects quantified by the
  dynamo numbers $D_{\alpha S}$ and $D_{\eta S}$. Adapted from
  \cite{BRRK08}.}
\label{fig:sheardyn}
\end{figure}

While the theoretical interpretation of helically forced dynamos with
shear is relatively straightforward, the situation is much less
obvious when the turbulence is non-helical. Mean-field effects leading
to such shear dynamo were derived theoretically much before a
numerical demonstration. These include the $\mvOm\times\mJJJ$ or
R\"adler effect \citep{1969VeGG...13..131R} and the shear-current
effect \citep{2003PhRvE..68c6301R,2004PhRvE..70d6310R}, which rely on
the off-diagonal component $\eta_{yx}$ of the diffusivity tensor for
driving the dynamo. More specifically, a necessary condition is that
$\eta_{yx}$ and $S$ have opposite signs \citep[e.g.][]{BS05}. Shear
dynamo action was demonstrated numerically by
\cite{YHSKRICM08,YHRSKRCM08}, and \cite{BRRK08}, showing that the
growth rate of the dynamo is proportional to the shear rate $S$ for
weak shear. The large-scale magnetic fields produced by such dynamos
tend to be quasi-steady or show random polarity reversals
\citep[e.g.][]{BRRK08,Teed_Proctor_2017_MNRAS_467_4858}; see the left
panel of \Figa{fig:sheardyn}.

However, test field simulations of \cite{BRRK08} yielded a non-zero
$\eta_{yx}$, but the sign was not conducive to dynamo
action. Subsequent studies have either found consistency with zero or
the wrong sign for $\eta_{yx}$ \citep[e.g.][]{MKTB09}. More recently,
in a series of papers Squire \& Bhattacharjee introduced a magnetic
shear-current effect that is analogous to its kinematic counterpart,
but it is driven by small-scale magnetism
\citep{2015ApJ...813...52S,2016JPlPh..82b5301S}. However, results from
nonlinear test field method have not confirmed a sign change between
the kinematic and small-scale dynamo cases
\citep{2020ApJ...905..179K,2022ApJ...932....8K}. Therefore the
existence of coherent shear-current effect -- in its kinematic and
magnetic incarnations -- remains inconclusive. The most likely cause
for the nonhelical shear dynamo is due to incoherent dynamo effects
such as the cumulative effects of fluctuations of the on average zero
$\alpha$ effect or kinetic helicity with the shear flow
\citep[e.g.][]{BRRK08,2022ApJ...932....8K}. This is demonstrated in
the right panel of \Figa{fig:sheardyn} where the dynamo growth rate is
shown as a function of dynamo numbers $D_{\alpha S}= \alpha_{\rm
  rms}|S|/(\etaT^2 k^3)$ and $D_{\eta S}= \eta_{xy}^{\rm
  rms}|S|/(\etaT^2 k^2)$ describing the incoherent $\alpha$ and
shear-current effects. The simulations of \cite{BRRK08} typically had
$D_{\alpha S}\approx D_{\eta S}\approx 4$ indicating that the
incoherent $\alpha$ effect was the driver of the dynamo whereas even
the incoherent shear-current effect was subcritical.

\subsection{Convective dynamos in local boxes (Class 2)}

\begin{figure}
\begin{center}
\includegraphics[width=.45\textwidth]{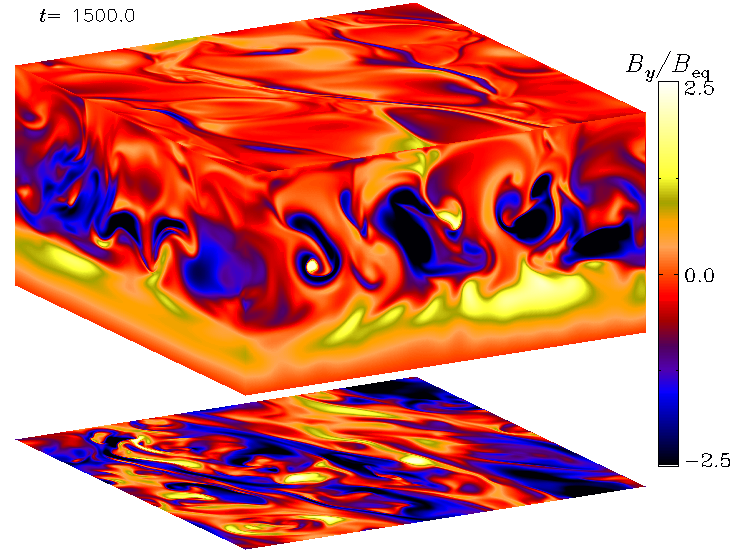}\includegraphics[width=.55\textwidth]{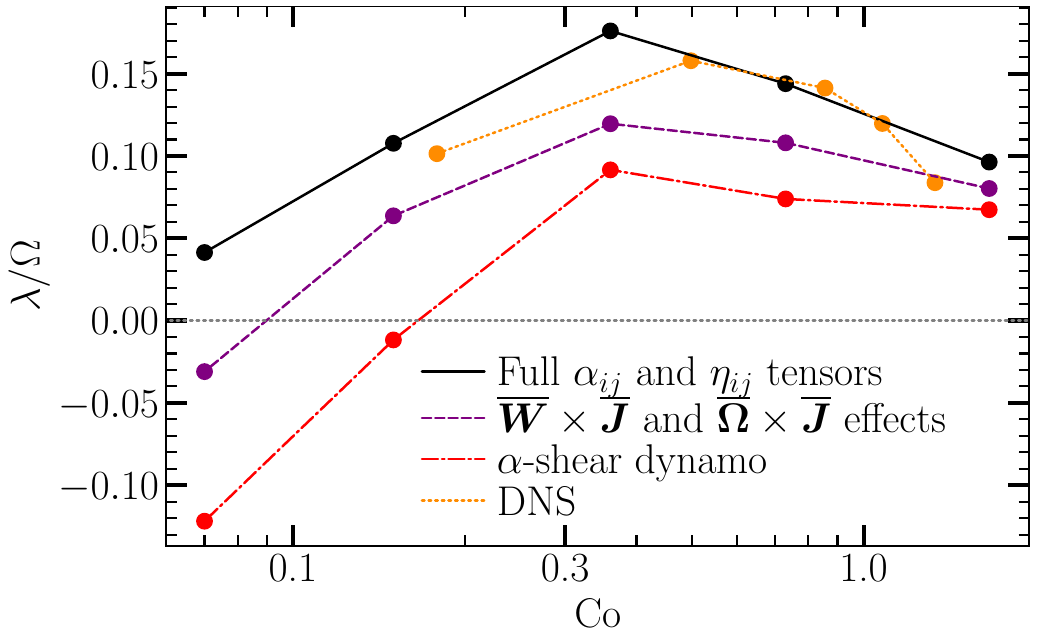}
\end{center}
\caption[]{Left: Stream-wise magnetic field component $B_y$ at the
  periphery of the simulation domain in a convection simulation with
  linear shear and rotation from \cite{KKB08}. Right: Growth rate
  $\lambda$ of the magnetic field, normalized by the rotation rate
  $\Omega$, from Cartesian DNS (orange dotted line), and
  one-dimensional mean-field models using the full $\alpha_{ij}$ and
  $\eta_{ij}$ tensors (black solid), only the shear-current and
  R\"adler effects (purple dashed), and a pure $\alpha$-shear dynamo
  (red dash-dotted). The turbulent transport coefficients were
  acquired from corresponding DNS using the test field method
  \citep{KKB09a}.}
\label{fig:LSD_gr_KKB08}
\end{figure}

A step forward from forced turbulence simulations is to consider
thermally driven convection under the influence of rotation and/or
shear in Cartesian geometry; see the left panel of
\Figa{fig:LSD_gr_KKB08}. Inhomogeneous convection with rotation and/or
shear leads to kinetic helicity production and an $\alpha$ effect
\citep[e.g.][]{2006PhRvE..73e6311R}, and the interaction between shear
and turbulent flows enables the shear dynamo via the R\"adler and
incoherent dynamo effects. Furthermore, the magnetorotational
instability \citep[][]{Veli59,1991ApJ...376..214B} can also operate
given suitable signs of shear and rotation \citep[e.g.][]{KMB13}. The
mean fields in this type of simulations are one- \citep[e.g.][]{KKB08}
or two-dimensional \citep[e.g.][]{HP09,KKB10b} depending, e.g., on the
spatial structure of the imposed shear flow.

The first successful large-scale dynamos were obtained from local
convection simulations where a large-scale horizontal shear was
additionally imposed \citep{KKB08,HP09}. These setups include all the
ingredients of classical $\alpha\Omega$ dynamo similarly to the
$\alpha$-shear dynamos discussed in \Seca{sec:helforce}. For weak
shear the growth rate $\lambda$ of the large-scale magnetic field was
found to be roughly proportional to the shear rate $S$ in both
studies, similarly to the nonhelical shear dynamo simulations of
\cite{YHSKRICM08}. In comparison, in a classical $\alpha\Omega$
dynamo, albeit with spatially uniform $\alpha$ and $\etat$, the growth
rate scales as $S^{1/2}$ \citep[e.g.][]{BS05}. Another difference to
classical $\alpha\Omega$ dynamos is that the large-scale fields in
this type of simulations are almost always non-oscillatory \citep[see,
  however][]{KMB13}. In \cite{KKB09a} the turbulent transport
coefficients $\alpij(z)$ and $\etaij(z)$ were computed using the
quasi-kinematic test field method for a set of simulations similar to
those in \cite{KKB08}. Kinematic growth rates realized in the DNS were
compared to one-dimensional mean-field models where the test field
coefficients from corresponding DNS were used. The results are shown
in the right panel of \Figa{fig:LSD_gr_KKB08}. The growth rates from
DNS match well with the mean-field models when all of the components
of $\alpha_{ij}$ and $\eta_{ij}$ were retained. These results also
suggest that the R\"adler/shear-current effects contribute to the
dynamo but that it is subdominant in comparison to the contribution
from the $\alpha$ effect. These simulations were made with modest
$\ReM$ of around 35, such that no small-scale dynamo was
present. Important caveats include the omission of incoherent dynamo
effects \citep[e.g.][]{BRRK08} and the effects of nonlocality
\citep[][]{KKB09a}.

\begin{figure}[ht]
\begin{center}
\includegraphics[width=.75\textwidth]{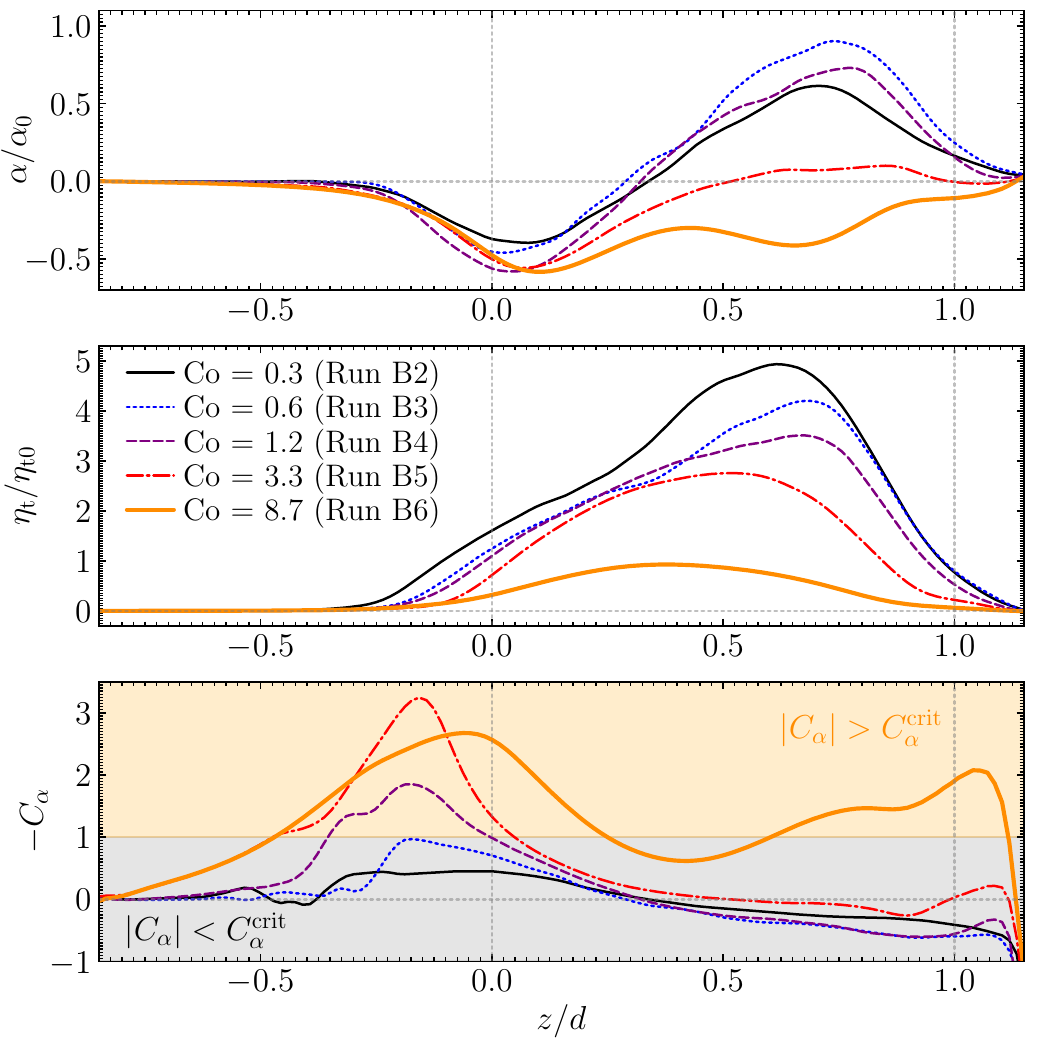}
\end{center}
\caption[]{Top and middle panels: $\alpha$ and $\etat$, respectively,
  as functions of height from a set of Cartesian convection
  simulations from \cite{KKB09b}. The Coriolis number is indicated by
  the legend in the middle panel. Bottom: Dynamo number $C_\alpha$
  corresponding to the simulations on the upper panels. Gray (orange)
  shading indicates $|\Calp|<\Calpcrit$ ($>\Calpcrit$). The dotted
  vertical lines indicate the extent of the initially convectively
  unstable layer.}
\label{fig:rotconv_KKB09}
\end{figure}

Simulations of rigidly rotating convection without shear flows, which
have all the hallmarks of a classical $\alpha^2$ dynamo, routinely
produced dynamos but most often no appreciable large-scale magnetic
fields were reported
\citep[e.g.][]{2000JFM...404..311J,CH06,HC08,2008ApJ...685..596T,KKB08}.
This happened despite the presence of an $\alpha$ effect in such
simulations \citep[e.g.][]{BNPST90,OSB01,OSBG02,KKOS06}. The test
field calculations of \cite{KKB09a} showed that the $\alpha$ effect
increases and turbulent diffusivity $\etat$ decreases as a function of
rotation. For a one-dimensional $z$-dependent $\alpha^2$ dynamo, the
dynamo number $C_\alpha(z) = \alpha(z)/[\etat(z) k_1]$, where
$k_1=2\pi/L_z$ and $L_z$ is the vertical size of the system, has to
exceed the critical value $\Calpcrit=1$ for dynamo action
\citep[e.g.][]{BS05}. This condition was fulfilled for sufficiently
large $\Co$ in \cite{KKB09b}; see \Figa{fig:rotconv_KKB09}\footnote{In
\cite{KKB09b}, $c_\alpha = \alpha(z)/(\etat(z) \kf)$, where $\kf =
2\pi/d$, where $d$ is the depth of the initially convectively unstable
layer, was used to characterize the dynamo. With $d = L_z/2$ and
$\kf=2k_1$, the critical value in terms of $c_\alpha$ corresponds to
$|c_\alpha^{\rm crit}| > 0.5$.}. This was confirmed by a corresponding
dynamo simulations. These results suggest that mean-field theory has
predictive power at least in a limited sense considered here. However,
in \cite{KKB09b} large-scale dynamo action was obtained only in cases
where large-scale hydrodynamic vortices were excited in the kinematic
regime of the simulations. The large-scale vorticity generation occurs
when $\Co$ and $\Rey$ exceed certain threshold values; see
e.g.\ \cite{Chan07}, \cite{2011ApJ...742...34K}, and \cite{GHJ14}. The
origin of the vortices is often attributed to two-dimensionalization
of turbulence but the observed threshold behavior with respect to
Reynolds and Coriolis numbers could also be signs of an
instability. Such large-scale vortices aid the dynamo in the kinematic
regime, whereas once the magnetic fields become dynamically important,
the vortices are quenched
\citep[e.g.][]{KMB13,2014ApJ...794L...6M,GHJ15,2017JFM...815..333G,2018A&A...612A..97B}. The
non-linear state of such dynamos is cyclic; see an example in
\Figa{fig:locosims}. There is considerable uncertainty as to how this
dynamo is generated, although an oscillatory $\alpha^2$ dynamo is a
possible candidate \citep[e.g.][]{BS87,Ra87}.

\begin{figure}[ht]
\begin{center}
\includegraphics[width=\textwidth]{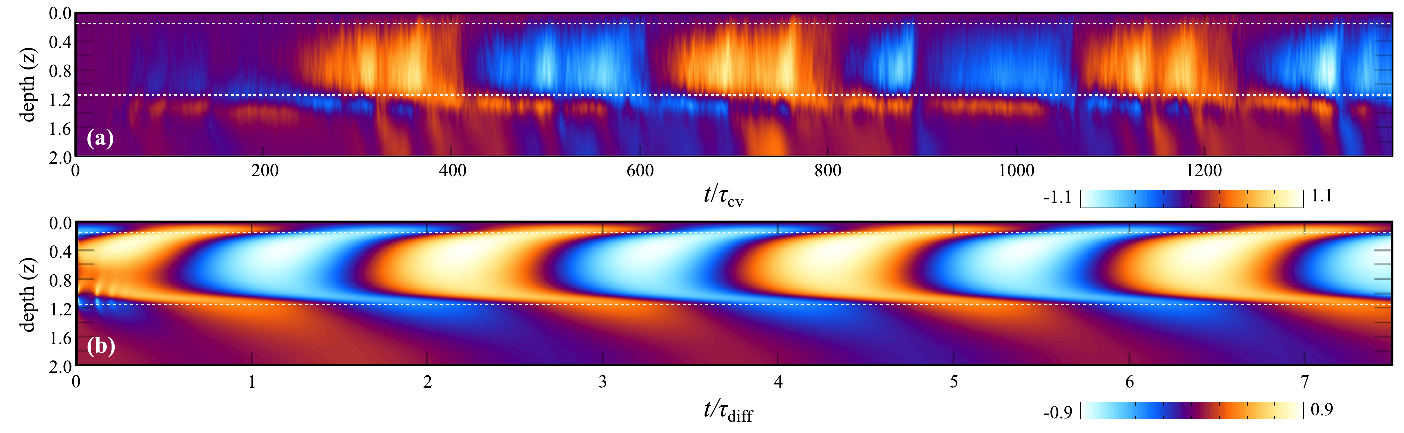}
\end{center}
\caption[]{Time-depth diagram of the horizontally averaged horizontal
  magnetic field component $\mBBx(z,t)$ from a DNS (upper panel) and
  from a corresponding mean-field model (lower panel). Adapted from
  \cite{2014ApJ...794L...6M}.}
\label{fig:butters_MS14}
\end{figure}

\cite{2014ApJ...794L...6M} made a comparisons of DNS and mean-field
models of this type of systems and found that the mean-field model
reproduced the cyclic behavior of the mean field of the DNS. Magnetic
$\alpha$ effect was invoked as the non-linearity whereas turbulent
pumping and diffusivity were assumed to be catastrophically quenched.
However, the mean-field models used isotropic FOSA expressions for the
kinematic $\alpha$ and $\etat$. Nevertheless, the correspondence of
the DNS and mean-field models is remarkably good; see
\Figa{fig:butters_MS14}. A similar analysis was made of simulations
with significantly larger density stratification in
\cite{2022arXiv220606566M}, where similar cyclic solutions were found
in an earlier study \citep{MS16}.

\subsection{Global simulations of convection in spherical shells (Class 3)}
\label{subsec:globalco}

\subsubsection{Interpretation in terms of $\alpha\Omega$ dynamos}

\begin{figure}[ht]
\begin{center}
\includegraphics[width=.65\textwidth]{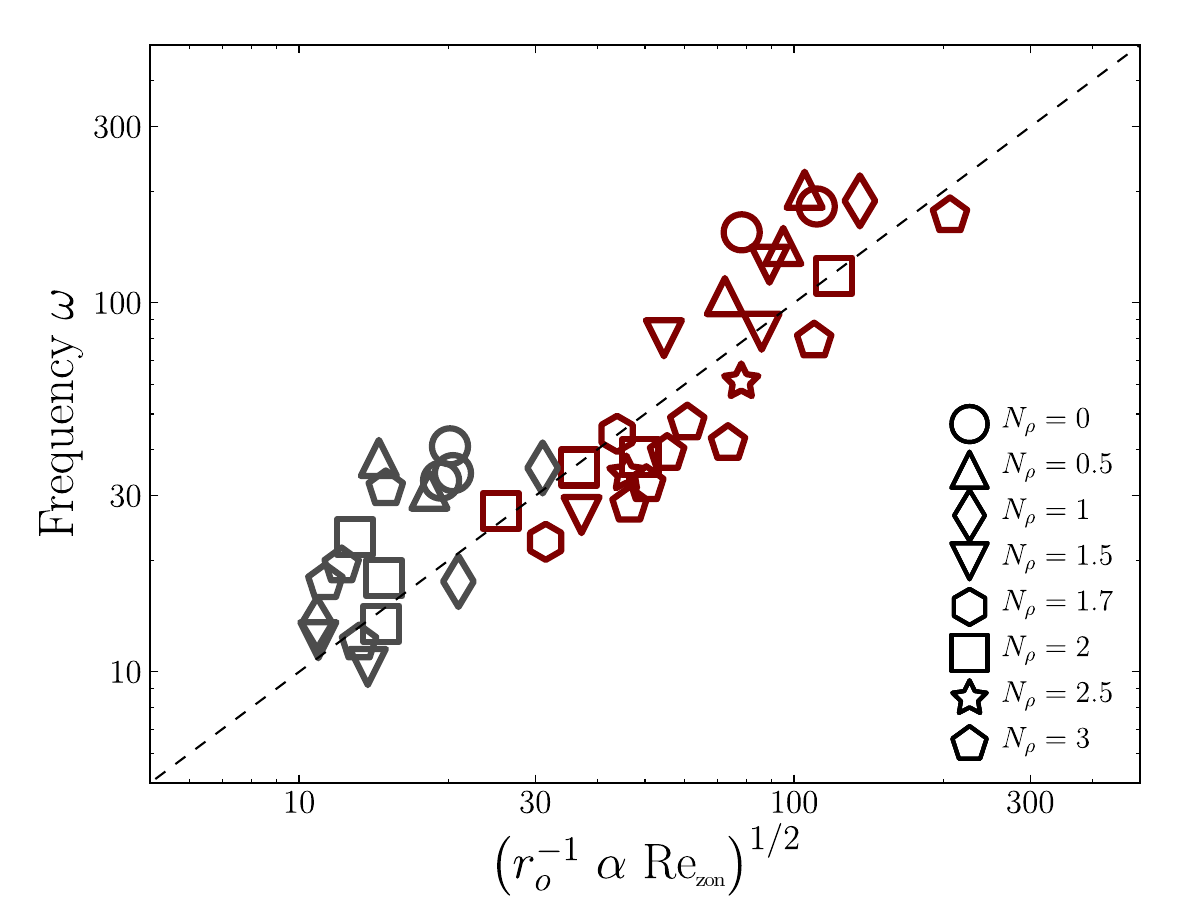}
\end{center}
\caption[]{Frequency of the dynamo wave versus a proxy from mean-field
  $\alpha\Omega$ theory. In comparison to \Eq{equ:ocycaO}, $r_o\propto
  k_y^{-1}$ is the outer radius of the shell and $\Rey_{\rm zon}
  \propto \pd\mOm/\pd r$ is the Reynolds number related to the mean
  zonal flow. The symbols refer to different density stratifications
  whereas the red (gray) symbols denote thick (thin) shells. Adapted
  from \cite{GDW12}.}
\label{fig:GastineParYos}
\end{figure}

Early successful dynamos in global geometry by \cite{Gi83} and
\cite{Gl85} produced cyclic large-scale magnetic fields that migrated
poleward. Given that in these simulations $\pd\mOm/\pd r >0$, and
$\mean{\ooo\bm\cdot\uuu} <0$ in the northern hemisphere, poleward
propagation of the dynamo wave is consistent with the Parker-Yoshimura
rule \citep{Pa55b,Yo75}. Similar poleward propagating dynamos have
been found in numerous other studies in Boussinesq
\citep[e.g.][]{BS06}, anelastic \citep[e.g.][]{BMBBT11,GDW12}, as well
as fully compressible simulations
\citep[e.g.][]{KKBMT10,WKKB14,MMK15}, where they have often been
interpreted in terms of mean-field $\alpha\Omega$ dynamos. In
\cite{2018A&A...616A..72W} the cycle frequency was shown to correspond
to that obtained from dispersion relation of the $\alpha\Omega$ dynamo
\begin{eqnarray}
\ocyc = \left(\frac{\alpha k_\theta}{2} r \cos \theta \frac{\pd \mOm}{\pd r} \right)^{1/2},\label{equ:ocycaO}
\end{eqnarray}
where $k_\theta$ is a latitudinal wavenumber and $\alpha$ corresponds
to the isotropic FOSA expression (\ref{equ:coefsfinalFOSA}). In
reality, the periods of dynamo cycles are independent of $r$ and
$\theta$ and therefore typical values of $\alpha$ and $\pd\mOm/\pd r$
from the dynamo region are used in \Eq{equ:ocycaO}. This relation was
found to be consistent with a variety of simulations with different
density stratifications and shell thicknesses in \cite{GDW12}; see
\Figa{fig:GastineParYos}. \cite{2018A&A...616A..72W} studied the
dependence of simulated cycles on rotation. The simulation results
were shown to be consistent with \Eq{equ:ocycaO} in the cases where
clear cycles were present by using simple estimates of $\alpha$ and
$\etat$ from \Eq{equ:MTAcoefs}. Furthermore, assuming a
quasi-stationary saturated state of the dynamo, \Eq{equ:omcycashear}
can be used to estimate the turbulent diffusivity $\etat$, which leads
to $\etat = \Delta r R^2/(2 \Pcyc)$, where $\Delta r$ is the depth of
the convection zone and $\Pcyc = 2\pi/\ocyc$ \citep{RS72}. This was
shown to coincide with FOSA estimate, \Eq{equ:coefsfinalFOSA}, and the
cycle period was interpreted to be controlled by the turbulent
diffusivity.

As mentioned in \Seca{sec:dynsims} since the pioneering efforts of the
1980s, there have been several simulations that produce solar-like
equatorward migration
\citep[e.g.][]{KMB12,ABMT15,DWBG16,SBCBN17,2018ApJ...863...35S,2018A&A...616A..72W,2020ApJ...892..106M,2022ApJ...931L..17K,2022ApJ...926...21B}. The
results of many of these simulations can also be understood in terms
of $\alpha\Omega$ dynamos, although with a twist that makes them
incompatible with the solar dynamo. This was shown in \cite{WKKB14}
who studied the cause of equatorward migration of dynamo waves in
simulations similar to those presented in \cite{KMB12} and
\cite{KMCWB13}, again employing the simple FOSA expressions for
$\alpha$ and $\etat$. The main conclusion of this study was that the
oscillatory solutions can be qualitatively explained by the
Parker-Yoshimura rule and that equatorward migration is due to a
mid-latitude minimum in the angular velocity, that leads to a
localized region where $\pd\mOm/\pd r < 0$. An otherwise similar
simulation but where the mid-latitude minimum in the angular velocity
is absent shows poleward migration, again in accordance with the
Parker-Yoshimura rule. A similar analysis was made in
\cite{2018A&A...609A..51W}, but using turbulent transport coefficients
computed using the test field method. Qualitative agreement with the
Parker-Yoshimura rule is found also in this case. A similar local
mid-latitude minimum of $\mOm$ is also present in the equatorward
migrating solutions in \cite{ABMT15}.

Another mechanism to achieve equatorward migration was introduced in
\cite{DWBG16} where the sign of the kinetic helicity in much of the
convective layer is reversed. This is commonly encountered in the
overshoot region below the convection zone but in the simulations of
\cite{DWBG16} the reversal extended to much of the convection zone. It
is unclear why exactly the helicity reversal occurs and how robust it
is. \cite{DWBG16} attribute it to a combination of low $\Pra$, weak
stratification in the layer with the reversal, distributed internal
heat sources, and fixed flux boundary conditions. The helicity
reversal leads to a negative $\alpha$ effect in the northern
hemisphere and equatorward migration of dynamo waves with a solar-like
differential rotation in accordance with the Parker-Yoshimura
rule. However, the simulation setup in these models significantly
differs from the Sun in that much deeper convective shells were
studied and in some cases convection transported only a small fraction
of the total flux. Further studies are needed to ascertain if this
mechanism can operate in the Sun.

\subsubsection{Magnetic driving of dynamos?}

The analyses discussed in the previous section have assumed that the
non-linear evolution of the large-scale dynamos can be represented in
terms of the standard mean-field dynamo theory where the influence of
the magnetic field is taken into account in the turbulent transport
coefficients. However, it is also possible that MHD instabilities that
require a finite-amplitude magnetic field to begin with can drive
dynamo action.

For example, \cite{2019ApJ...880....6G} analyzed simulations with and
without tachoclines employing the isotropic MTA expressions for
$\alpha$ and $\etat$ where the magnetic contribution to $\alpha$ was
retained; see \Eq{equ:MTAcoefs}. Their results suggest that dynamos in
their simulations are of $\alpha^2\Omega$ type with the $\alpha$
effect being dominated by the magnetic contribution in the stably
stratified regions below the convection zone. The source of the
magnetic $\alpha$ effect in the layers below the convection zone was
conjectured to be the buoyancy \citep{Pa55a} or Tayler instability
\citep{1973MNRAS.161..365T}. The former has been shown to lead to an
$\alpha$ effect \citep[e.g.][]{2011A&A...534A..46C} while the latter
has been demonstrated to lead to dynamos recently
\citep[e.g.][]{2023Sci...379..300P}. Furthermore,
\cite{2018ApJ...863...35S} argued that the cycles in their simulations
arise from the non-linear influence the magnetic field exerts on the
differential rotation. However, more refined analyses, e.g., with
nonlinear test field methods, are required to definitively establish
the cause of dynamo action in these cases

\subsubsection{Relative strength of dynamo effects as function of rotation}

Magnetic activity in stars is a strong function of stellar rotation
\citep[e.g.][]{2017LRSP...14....4B}. Numerical simulations of
convection-driven dynamos have been used to study the
rotation-activity relation by several groups
\citep[e.g.][]{2018A&A...616A.160V,2018ApJ...863...35S,2018A&A...616A..72W,2020A&A...642A..66W,2022ApJ...926...21B}. From
the dynamo-theoretical point of view the interest lies in the rotation
dependence of various dynamo effects. This was studied by
\cite{2020A&A...642A..66W} who computed the turbulent transport
coefficients using the spherical quasi-kinematic test field method
from a similar set of dynamo simulations as in
\cite{2018A&A...616A..72W}. The contributions of individual
mean-fields effects were compared over a large range of Coriolis
numbers. This entails comparing the volume averaged rms values of
terms such as $\bm\nabla \times (\bm\alpha\bm\cdot\mBBB)$, $\bm\nabla
\times (\bm\beta\bm\cdot \bm\nabla \times \mBBB)$, etc., as functions
of rotation.

\cite{2020A&A...642A..66W} concluded that for slow rotation with
strong anti-solar differential rotation the dominant effects besides
the differential rotation are the $\alpha$ and R\"adler effects. In
the intermediate rotation regime where predominantly axisymmetric
oscillatory dynamos exist, the $\alpha$ effect contributes both to
poloidal and toroidal fields whereas the $\Omega$ effect is also
important for the latter. Thus these dynamos can be classified to be
of $\alpha^2\Omega$ type. In the rapid rotation regime differential
rotation is quenched and the highly anisotropic $\alpha$ effect
dominates, with these dynamos being classified as
$\alpha^2$. \cite{2022ApJ...926...21B} concluded that most of their
simulations are more likely $\alpha\Omega$ type based on an SVD
analysis of the $\alpij$ coefficients and the strength of differential
rotation.

\subsubsection{Mean-field modeling}
\label{sec:MFmodeling}

The most rigorous test of the turbulent transport coefficients
extracted from simulations is to use them in a mean-field model
corresponding to the simulation where they were extracted from.
Several levels of rigor how this is done can be distinguished. In the
simplest cases only a fraction of the possible turbulent transport
coefficients are extracted while the rest are either omitted, replaced
by physically plausible parameterized alternatives, or used as free
parameters. Similar arguments apply to the mean flows that go into the
mean-field models. Here the starting point is in cases with most free
parameters followed by more constrained and rigorous comparisons.

However, a related complementary approach combines local simulations
and global mean-field models \citep[e.g.][]{KKOS06,KKT06}. In
\cite{KKOS06}, $\bm\alpha$ and $\bm\gamma$ were computed using the
imposed field method from local simulations at different latitudes and
Coriolis numbers. The latter was interpreted to correspond to
different depths in the solar convection zone where $\Co$ ranges
between $10^{-4}$ near the surface to about 10 at the base
\citep[e.g.][]{O03}. The data from 3D simulations was combined to
meridional profiles of $\alpha_{ij}(r,\theta)$ and
$\gamma_i(r,\theta)$ and used in mean-field dynamo models of the Sun
\citep{KKT06}. Turbulent diffusivity was assumed to be isotropic and
uniform with a value of the order of
$10^8$~m$^2$~s$^{-1}$. Furthermore, the helioseismic rotation profile
of the Sun and a plausible single-cell counterclockwise meridional
flow were used. These models yield solar-like equatorward migration
especially when equatorward pumping $\gamma_\theta$ and meridional
flow were included. The near-surface shear layer also contributes to
this but its effect is subdominant. This case is rather more
illustrative than predictive but this approach shows some promise in
probing turbulent phenomena locally.

\begin{figure}[ht]
\begin{center}
\includegraphics[width=.65\textwidth]{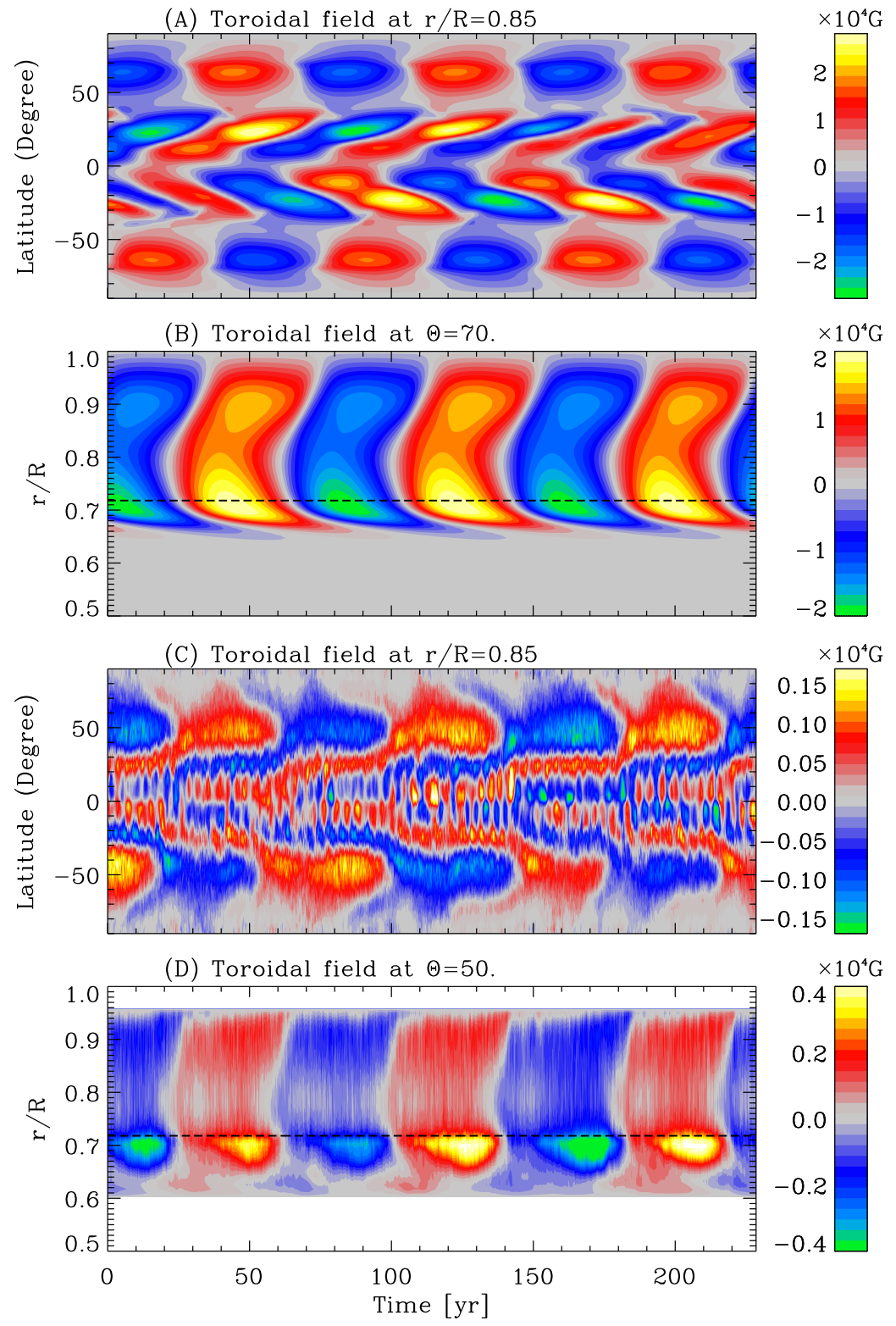}
\end{center}
\caption[]{Top panels: Mean toroidal field $\mBBp(\theta,t)$ at the
  middle of the convection zone at $r/R=0.85$ and $\mBBp(r,t)$ at
  latitude $\theta=70^\circ$ in an $\alpha^2\Omega$ mean-field model
  from \cite{SCB13}. Bottom panels: corresponding azimuthally averaged
  magnetic fields from the 3D convective dynamo simulation of
  \cite{GCS10}.}
\label{fig:Simard_et_al_2013}
\end{figure}

One of the first efforts to explain global 3D dynamo simulation with
mean-field models was presented by \cite{2013ApJ...775...69D}, who
extracted the differential rotation, and $\alpha$ and $\etat$ using
FOSA estimates from a suite of global 3D simulations similar to that
of \cite{GCS10}. Meridional flows in the simulations were weak and
therefore assumed to vanish in the mean-field models. Despite these
simplifications, the mean-field models with these ingredients were
able to recover the large-scale magnetic fields of the 3D simulations
remarkably well in some cases. The next step in rigor came with the
study of \cite{RCGBS11}, who extracted $\bm\alpha$ from the simulation
presented in \cite{GCS10} using the SVD method; see
\Seca{sec:SVDmethod}. \cite{SCB13} used these coefficients to drive a
kinematic $\alpha\Omega$, $\alpha^2$, and $\alpha^2\Omega$ mean-field
dynamo models with the rotation profile from a hydrodynamic simulation
and a generic single cell per hemisphere meridional flow circulation
pattern. The magnetic diffusivity was assumed to be a scalar with
uniform value in the convection zone. The best agreement between the
simulation of \cite{GCS10} and the mean-field models was obtained when
the full $\alpha$ tensor was included. Despite the differences to the
original simulation, remarkably good correspondence between the
simulation and the mean-field model was found; see
\Figa{fig:Simard_et_al_2013}.

In a related study, \cite{2016AdSpR..58.1522S} used the time series
encompassing about 40 magnetic cycles from the simulation of
\cite{PC14} to extract the $a_{ij}$ and $b_{ijk}$ coefficients using
the SVD method. Mean-field modeling applying these coefficients was
done by \cite{2016ApJ...826..138B} with similar simplifications as in
\cite{SCB13}. The main motive of \cite{2016ApJ...826..138B} was to
study the origin of seemingly two coexisting dynamo modes in the
simulation of \cite{PC14} that resemble the quasi-biennial
oscillations observed in the Sun. The mean-field models indeed suggest
two spatially separated dynamos. However, here the correspondence
between the simulation and the mean-field models was significantly
worse than in \cite{SCB13}. Finally, \cite{2020JSWSC..10....9S}
followed a very similar approach but allowed for a back-reaction of
the magnetic field on the differential rotation via the large-scale
Lorentz force, and found that this leads to long-term modulation of
cycles and Maunder minimum-like events.

The use of isotropic turbulent diffusivity and generic mean flow
profiles in mean-field modeling is problematic because these are
integral parts of the dynamo solutions of 3D simulations. The first
studies that took into account also the tensorial turbulent
diffusivity $b_{ijk}$ were done with the test field method by
\cite{SRSRC05,SRSRC07}. They computed the turbulent transport
coefficients from Boussinesq magnetoconvection and dynamo simulations
in spherical shells. The actual and reconstructed EMFs were compared
and mean-field dynamo models of the dynamo simulations were made. The
correspondence between the DNS and mean-field models was found to be
satisfactory in cases where convection and dynamos were close to
marginal and time-independent.  Clearer differences appeared more
supercritical time-dependent quasi-stationary dynamos. \cite{SRSRC07}
noted that the ansatz \Eq{equ:R80EMF} was not sufficient to reproduce
the EMF and that either higher order terms in the expansion or the
effects of non-locality are a possible causes for this. \cite{Sch11}
found that the issues related to poor scale separation could be
alleviated by simultaneous spatial and temporal
averaging. Nevertheless, although the EMF and the resulting DNS and
mean-field solutions differed in details, the large-scale dynamo mode
was correctly captured in all of the cases \citep[see also][]{SPD11}.

Finally, \cite{2021ApJ...919L..13W} did mean-field modeling using the
turbulence transport coefficients measured with the test field method
in \cite{2018A&A...609A..51W} from a density-stratified simulation of
rotating convection in spherical wedge geometry. This simulation was
targeted to model stellar dynamos and it includes higher density
stratification and it is more supercritical than the earlier models
of, for example, \cite{Sch11}. In this simulation a solar-like cyclic
large-scale magnetic field arises showing a dominant dynamo wave that
propagates equatorward (poleward) at low (high) latitudes with a long
cycle and a secondary poleward cycle near the equator with a much
shorter cycle; see also \cite{KMB12,KKOBWKP16}. Time-averaged
turbulent transport coefficients and mean flows were extracted and
post-processed to remove rapid spatial and temporal variations.

\begin{figure}[ht]
\begin{center}
\includegraphics[width=\textwidth]{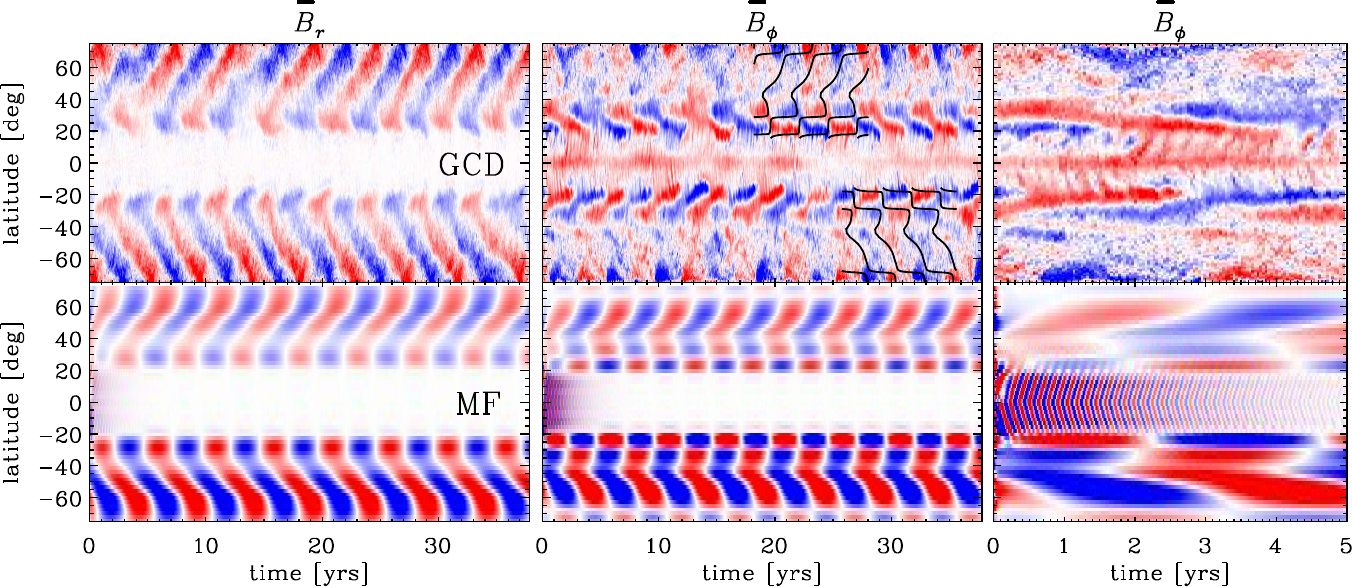}
\end{center}
\caption[]{Butterfly diagrams of radial (left column) and azimuthal
  (middle and right columns) magnetic fields from a global 3D
  simulation (top row), and a mean-field model using turbulent
  transport coefficients from the test field method (lower
  row). Adapted from \cite{2021ApJ...919L..13W}.}
\label{fig:Warnecke_et_al_2021}
\end{figure}

Linear mean-field models were used which is compatible with the fact
that the coefficients were obtained from a simulation where the
magnetic field is already saturated. The mean-field model reproduces
large-scale features such as the cycle period and both the poleward
and equatorward migration of the magnetic field in the direct
simulation when the magnitude of $\alpha$ tensor was scaled up by a
factor that varies between $1.40$ and $1.525$; see
\Figa{fig:Warnecke_et_al_2021}. By systematically turning on and off
various effects, \cite{2021ApJ...919L..13W} concluded that almost all
of the turbulent transport coefficients are needed to reproduce the
large-scale magnetic field solution, and that the meridional flow has
a negligible effect of the solutions in these simulations. The authors
further concluded that the dynamo in the simulation in question is of
$\alpha^2\Omega$ type and that effects of non-locality are not needed
to reproduce the evolution of the large-scale magnetic
field. Remarkably also the short secondary cycle is reproduced with
the mean-field models and conjectured to be of $\alpha^2$ type.  This
can have signifigance for the interpretation of stellar cycles where
co-existing long and short cycles have been pointed out by
\cite{2017ApJ...845...79B} using data from the Mount Wilson Survey.

To conclude, detailed mean-field modeling consistently reproduces the
large-scale dynamo mode of the DNS at least qualitatively. This is
despite the fact that the reconstructed EMF is typically significantly
less well reproduced \citep[e.g.][]{2019ApJ...886...21V}, such that
the amplitudes of the reconstructed and actual $\mEMF$ can differ by a
factor of two. Furthermore, it is puzzling that the studies that use
only some of the coefficients from simulations or which apply an
incomplete ansatz for the electromotive force also tend to often
capture the large-scale field evolution rather well
\citep[e.g.][]{SCB13}. This can mean that the large-scale dynamo mode
established in the simulations is quite insensitive to the choice of
the turbulent transport coefficients. None of the studies discussed
above consider non-locality or incoherent dynamo effects that have
been suggested as drivers of dynamos in simpler settings
\citep[e.g.][]{2014MNRAS.441..116R,BRRK08}.

\subsection{Magnetic helicity conservation and saturation of large-scale dynamos}

The relation of magnetic helicity conservation to the non-linear
saturation of large-scale dynamos is a topic that has been actively
studied using numerical simulations. The simplest way of testing this
is to change the magnetic boundary conditions of the system such that
they either allow or prevent the flux of magnetic helicity across the
boundary. Such experiments were conducted by \cite{KKB10b}, with local
simulations of convection with shear and rotation. In such cases a
shear-mediated magnetic helicity flux has been suggested to
alleviate catastrophic quenching
\citep[e.g.][]{2001ApJ...550..752V}. The results of \cite{KKB10b} show
that the dynamos in cases with open and closed boundaries show
dramatically different behavior especially at high values of
$\ReM$. The saturation amplitude of the total magnetic energy is
independent of $\ReM$ for open boundaries and proportional to
$\ReM^{-1}$ for closed (perfectly conducting) boundaries, the latter
coinciding with catastrophic quenching. However, the mean magnetic
field was proportional to $\ReM^{-0.25}$ ($\ReM^{-1.6}$) for open
(closed) boundaries and for closed boundaries. Both of these trends
are steeper than that expected from efficient small-scale magnetic
helicity fluxes out of the system and catastrophic quenching,
respectively. The magnetic Reynolds numbers in this study were still
very modest ($\ReM\approx 200$) compared to astrophysically relevant
regimes and it is plausible that significantly higher $\ReM$ are
needed to reach a regime where asymptotic scaling manifests. A similar
conclusion was drawn earlier from mean-field models applying the
dynamical quenching formalism by \cite{BCC09} who found that an
asymptotic regime is reached for $\ReM$ of the order of $10^4$.

\begin{figure}[ht]
\begin{center}
  \includegraphics[width=0.65\textwidth]{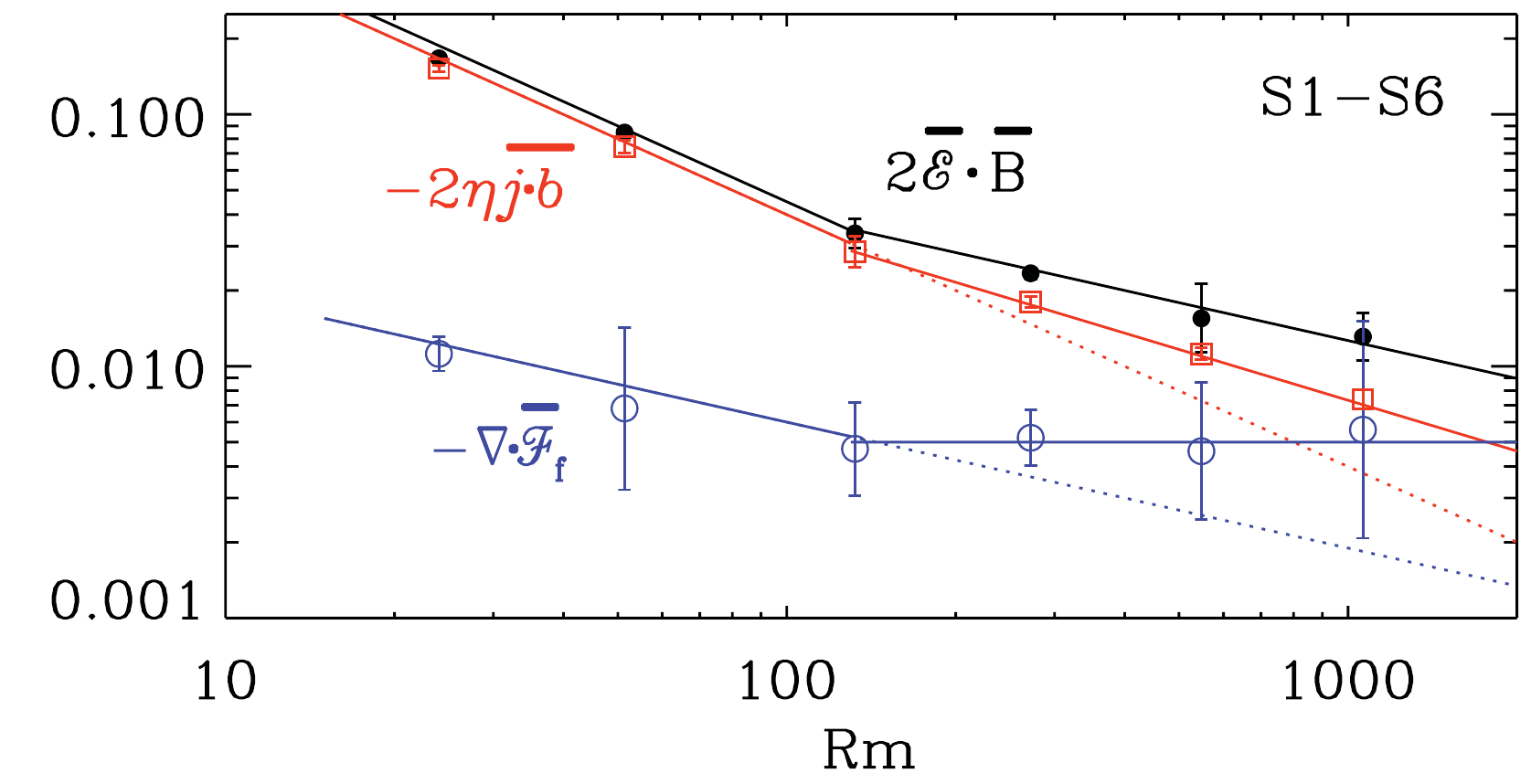}
  \includegraphics[width=0.65\textwidth]{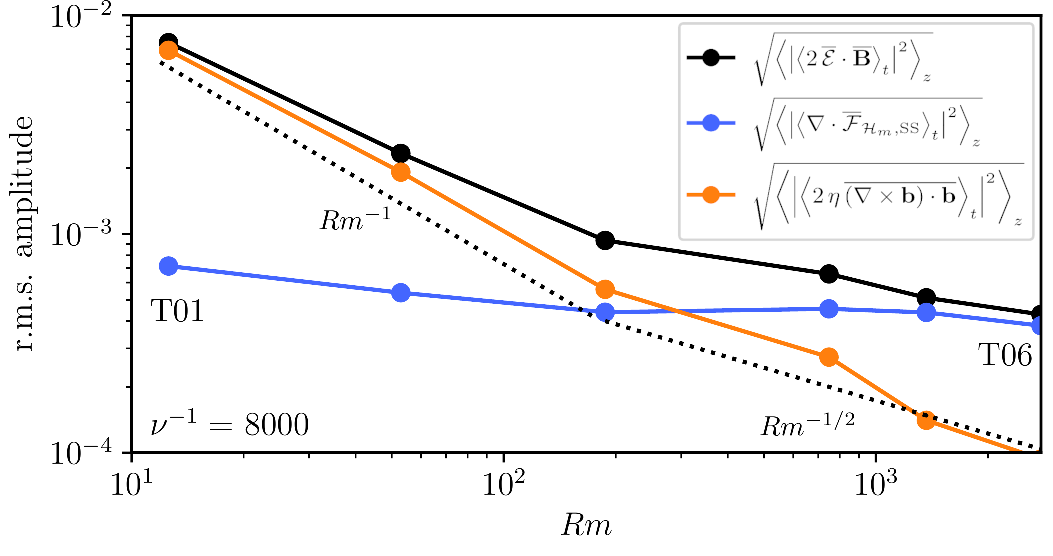}
  \includegraphics[width=0.65\textwidth]{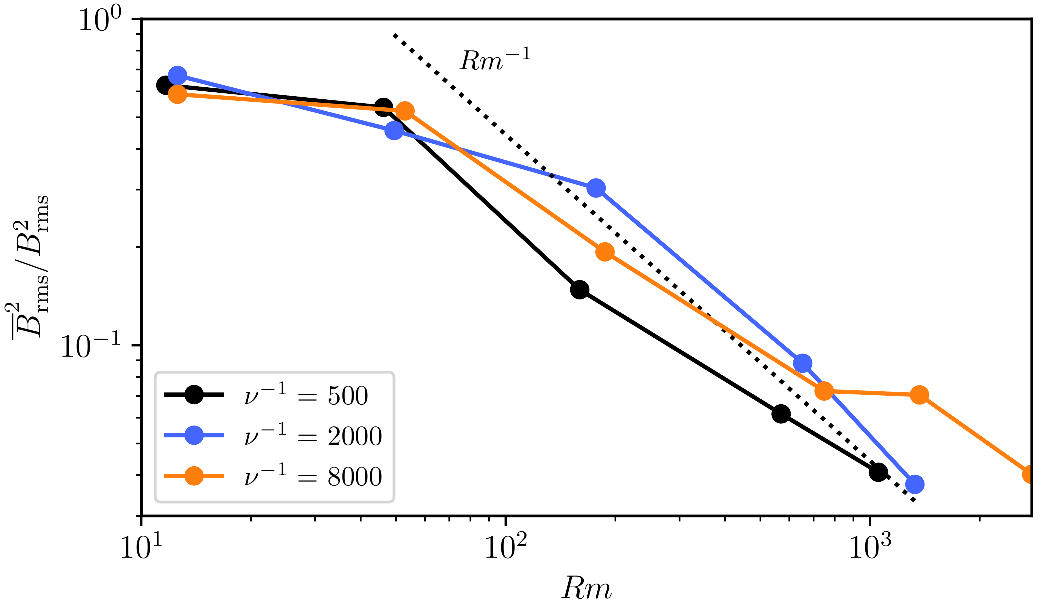}
\end{center}
\caption[]{Top: Terms in the conservation equation of small-scale
  magnetic helicity with $2\mEMF\bm\cdot\mBBB$,
  $-2\eta\mean{\jjj\bm\cdot\bbb}$, and
  $-\bm\nabla\bm\cdot\mean{\cal{F}}_{\rm f}$ corresponding to magnetic
  helicity production, resistive losses, and divergence of advective
  flux due to wind, respectively. Adapted from \cite{DSGB13}. Middle:
  Rms amplitudes of the terms in the small-scale current helicity
  evolution equation as functions of $Rm$ ($=\ReM$) from a set of 3D
  simulations of a Galloway-Proctor-like helical flow with sinusoidal
  variation of kinetic helicity from \cite{2021PhRvF...6l1701R}.
  Bottom: Mean magnetic field energy as a function of $\ReM$ from an
  extended set of runs from the same study.}
\label{fig:Rincon_2021}
\end{figure}

\cite{Hubbard_Brandenburg_2012_ApJ_748_51} distinguished two types of
catastrophic quenching which refer to low saturated field amplitude
(Type I) and to slow saturation in resistive timescale (Type
II). Although the magnetic Reynolds numbers were rather modest, no
evidence of Type I quenching was found from helically forced
$\alpha^2$ dynamos. \cite{Hubbard_Brandenburg_2012_ApJ_748_51} also
concluded that if shear is present in the system, dynamo-generated
large-scale fields are typically weakly helical, and that such fields
can grow to be dynamically significant already in the kinematic phase
of the dynamo without yet being affected by the magnetic helicity
constraint. Furthermore, open boundaries help to avoid also Type II
catastrophic quenching. This was explored in \cite{DSGB13} with
helically forced turbulence simulations where a wind out of the dynamo
region was introduced. The top panel of \Figa{fig:Rincon_2021} shows
that the contribution by the wind to the negative divergence of the
magnetic helicity flux ($-\bm\nabla\bm\cdot\mean{\cal{F}}_{\rm f}$)
becomes independent of $\ReM$ around $\ReM=170$ and whereas the
resistive losses ($-2\eta\mean{\jjj\bm\cdot\bbb}$) decrease as
$\ReM^{-2/3}$. Furthermore, the advective term reaches the magnitude
of the resistive term around $\ReM=10^3$ such that it alleviates
otherwise catastrophic quenching.

As discussed in \Seca{sec:maghel}, catastrophic quenching can also be
alleviated by internal magnetic helicity fluxes even if the domain is
closed \citep[e.g.][]{2010AN....331..130M}. High resolution 3D
simulations by \cite{2021PhRvF...6l1701R} used a setup where the mean
kinetic helicity had a sinusoidal variation with a sign change at an
equator, similar to that expected in the
Sun. \cite{2021PhRvF...6l1701R} found that the small-scale magnetic
helicity flux overcomes the resistive contribution and compensates for
the transfer term proportional to $\mEMF\bm\cdot\mBBB$ at $\ReM
\gtrsim 10^3$; see the top panel of \Figa{fig:Rincon_2021}. This can be
attributed to a flux mediated by turbulent diffusion. These results
corroborate the findings of \cite{BCC09} and suggest that
non-diffusive internal magnetic helicity fluxes become effective only
near the highest currently numerically achievable $\ReM$ in excess of
$10^3$. However, even in the cases with the highest $\ReM$ the
energy of the mean magnetic field is declining nearly proportional to
$\ReM^{-1}$; see the bottom panel of \Figa{fig:Rincon_2021}.

\subsection{Active region formation via negative effective magnetic pressure}

\begin{figure}[ht]
\begin{center}
  \includegraphics[width=0.5\textwidth]{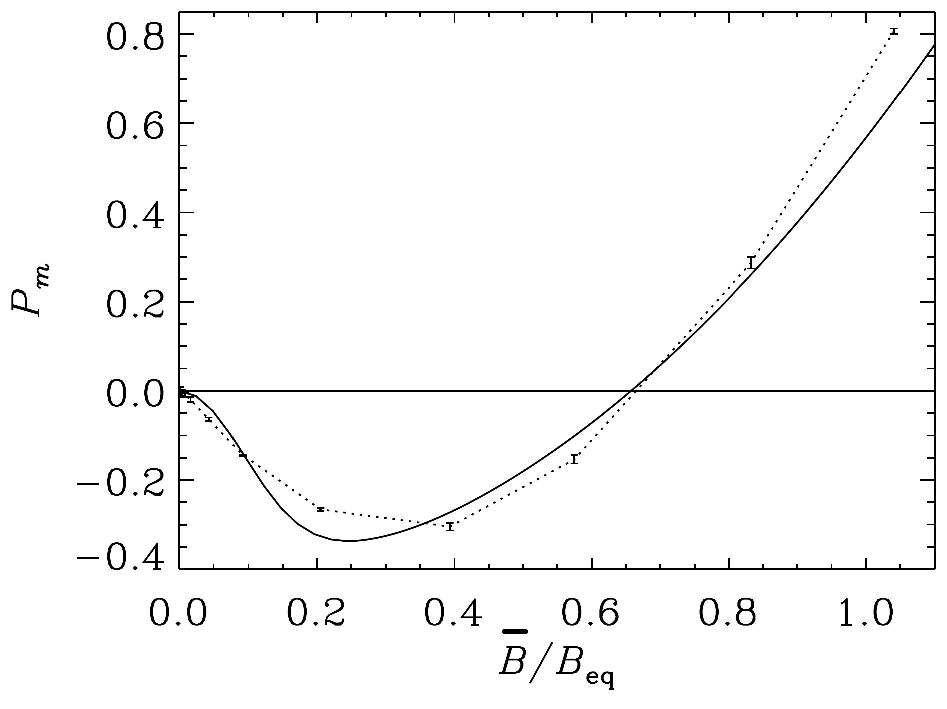}\includegraphics[width=0.5\textwidth]{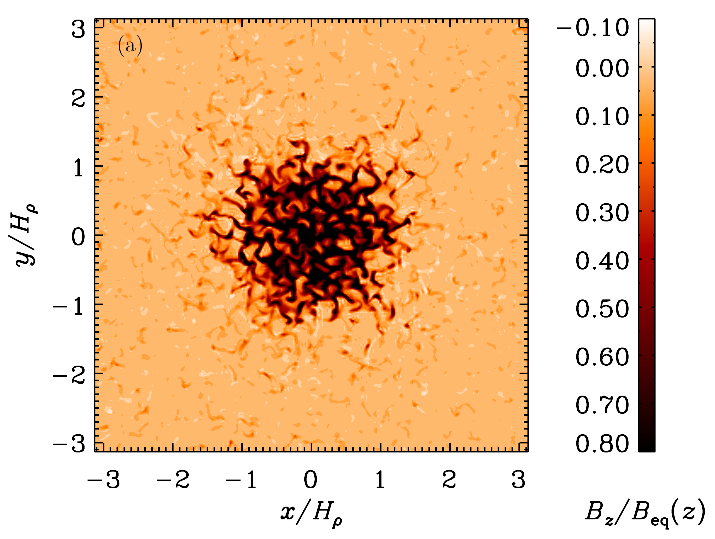}
\end{center}
\caption[]{Left: Effective magnetic pressure as a function of
  $\mBBB/\Beq$ from 3D simulations of forced turbulence (dotted line)
  and an analytic profile (solid line). Adapted from
  \cite{BKR10}. Right: Magnetic field concentration produced by NEMPI
  in density-stratified turbulence. Adapted from \cite{BKR13}.}
\label{fig:NEMPI}
\end{figure}

Sunspot formation in flux-transport solar dynamo models is usually
attributed to flux tubes rising from deep layers of the convection
zone or from the tachocline. On the other hand, if the magnetic field
of the Sun is generated within the convection zone by a distributed
dynamo, and consists of relatively diffuse fields, another mechanism
is needed to form magnetic field concentrations. Numerical studies of
sunspots operate on much smaller scales than the global dynamo and do
not typically consider the origin of the magnetic field but assume an
initial condition that leads to spot formation
\citep[e.g.][]{RSCK09,2014ApJ...785...90R,Fang_Fan_2015_ApJ_806_79}. In
the model of \cite{SN12} a bipolar spot pair forms from flux advected
through the lower boundary due to processing by convection, but again
the spot-forming magnetic field was not a produced by a dynamo. A few
3D dynamo simulations have shown buoyantly rising flux tubes but these
structures are on much larger scales than sunspots
\citep[e.g.][]{GK11,NBBMT13,NBBMT14}.

A possible candidate arises when a large-scale magnetic field is
superimposed on a turbulent background, leading to negative effective
magnetic pressure \citep[e.g.][]{KRR89,KRR90,RK07,BKR10,BKKR12}. This
effect arises from a mean-field analysis of the momentum equation
involving Reynolds and Maxwell stress tensors. This is measured by the
effective magnetic pressure
\begin{eqnarray}
\Peff = \onehalf (1 - q_p)\beta^2,
\end{eqnarray}
where $\beta^2 = \mBBB^2/\Beq^2$, and where $q_p$ is the contribution
of the mean field on the magnetic pressure. $\Peff < 0$ is a necessary
condition for the negative effective magnetic pressure instability
(NEMPI), which causes initially uniform fields to form flux
concentrations. This can be understood such that a large-scale (mean)
magnetic field itself has a positive pressure but it suppresses
turbulence and leads to decreased small-scale (turbulent) pressure. If
the latter effect is larger, corresponding to $q_p>1$, the
\emph{effective} magnetic pressure is negative leading to
instability. Negative contribution to magnetic pressure and the
related instability have been detected from simulations of forced
turbulence given that the scale separation between the scale of the
turbulence and the system size is sufficiently large
\citep[e.g.][]{BKKMR11,KeBKMR12,BKR13,KeBKMR13}; see
\Figa{fig:NEMPI}. Furthermore, density stratification is another
crucial element for NEMPI. Since the negative effective pressure
effect vanishes for $|\mBBB| \gtrsim 0.4 \Beq$, the concentrations it
produces are likely only progenitors of active regions and sunspots
and that, e.g., the buoyancy instability
\citep{Pa55a,Parker_1975_ApJ_198_205} is still needed to produce
superequipartition fields as in sunspots. Furthermore, rotation has
been shown to suppress NEMPI
\citep[e.g.][]{2012A&A...548A..49L,2013A&A...556A..83L,Losada_et_al_2019_AA_621_61},
such that it may only operate near the surface of the Sun, e.g., in
the NSSL.

Convection simulations also show $\Peff < 0$, but no instability has
been detected as of yet \citep{KBKMR12,KBKKR16}. The most likely cause
for this is that the scale separation in convection simulations is
much poorer than in the forced turbulence models where this is an
input to the model. This can also be related to the convective
conundrum which is the issue that deep convection in the Sun appears
to behave drastically differently than what current simulations and
theoretical models predict \citep[e.g.][and references
  therein]{2016AnRFM..48..191H,2023SSRv..219...77H}. Nevertheless, it
remains unclear whether NEMPI is possible in convection.

\section{Outstanding issues}
\label{sec:issues}

\subsection{Self-consistent inclusion of large-scale flows}

In practically all of the comparisons between 3D simulations and
mean-field models the large-scale flows are assumed to be given, e.g.,
by the time-averaged flows from the original DNS. Furthermore, these
are often taken from hydrodynamic runs although the dynamo
coefficients are extracted from the magnetically saturated
regimes. This is unsatisfactory from a rigorous theoretical point of
view because the large-scale flows themselves are self-consistently
generated in the DNS and affected by the dynamo-generated magnetic
fields. Therefore the mean-field modeling should not only relate to
magnetic field generation but also extended to the mean-field
hydrodynamics \citep[e.g.][]{R89,RH04} that govern the large-scale
flows. This complicates the mean-field models considerably because in
addition to the EMF, further turbulent correlations including the
Reynolds stress and turbulent heat flux need to be modeled. This is
exacerbated by the fact that the Navier-Stokes equations are
inherently non-linear. Some mean-field models take all of these effect
into account either in parametric way \citep[e.g.][]{BMT92,Re06} or
via detailed theoretical expressions
\citep[e.g.][]{2017MNRAS.466.3007P,2018ApJ...854...67P}. While some
efforts have been made to compare hydrodynamic 3D simulations with
mean-field hydrodynamics \citep[e.g.][]{RBMD94,2021A&A...655A..79B},
approaches where both, the dynamo and the large-scale flow generation,
would be interpreted in the mean-field framework have yet to appear.
An additional complication is the convective conundrum which hinders
current 3D simulations from reproducing the solar differential
rotation, possibly hinting at fundamental issues in the theory of
stellar convection \citep[e.g.][]{Sp97,Br16,Kapyla2025}. Finally, the
NSSL may play a role in shaping the solar dynamo \citep[e.g.][]{Br05}
as well as in sunspot formation, but current simulations struggle to
incorporate it self-consistently due to high computational demands
deriving from the large gap in spatial and temporal scales near the
surface in comparison to the deep convection zone. Thus, the current
NSSL-resolving simulations cannot be run long enough for the
large-scale dynamo to grow and saturate
\citep[e.g.][]{2025ApJ...985..163H} and therefore the impact of the
NSSL on the dynamo in these models is unclear.

\subsection{Nonlinearity and non-locality}

In a general theory of dynamos, non-linearity due to magnetic fields
needs to be incorporated from the outset. This includes the
back-reaction of the magnetic field on the large-scale flows such as
differential rotation as well as the small-scale flows that ultimately
enter $\mEMF$. This is challenging especially in cases where a
small-scale dynamo is excited and changes the dynamics
\citep[e.g.][]{2017A&A...599A...4K,2022ApJ...933..199H}. In the
comparisons of 3D simulations and corresponding mean-field models, the
large-scale flows are often assumed to be stationary. However, the
large-scale flows are time-dependent especially if the large-scale
dynamo is cyclic
\citep[e.g.][]{KKOBWKP16,2018ApJ...863...35S}. Furthermore, the
turbulent transport coefficients are typically computed from saturated
regimes of dynamos with the often implicit assumption that the
measured coefficients are already affected by the magnetic
fields. However, such quenched coefficients are likely specific to the
magnetic field configuration in the simulation from which they were
extracted and their use in other settings is likely problematic.

Another aspect that is particularly important in dynamos driven by
convection is non-locality. The turbulent transport coefficients from
the test field method are always scale dependent, even in cases where
the scale separation is assumed at least moderate from the outset
\citep[e.g.][]{BRS08,2012A&A...539A..35B,2020A&A...636A..93K}. For
convection the situation is even worse with some coefficients changing
sign as a function of the spatial scale \citep[e.g.][]{KKB09a}; see
\Figa{fig:KKB09_nonloc}. None of the current comparisons between 3D
simulations and mean-field models take non-locality into account, yet
mean-field models often capture the large-scale fields of simulations
remarkably well \citep[e.g.][]{2021ApJ...919L..13W}. Even more
remarkably, even if the turbulent transport coefficients in the
mean-fields models are estimated using highly idealized approximations
such as FOSA, the large-scale fields of the simulations are
nevertheless often recovered
\citep[e.g.][]{2013ApJ...775...69D,2014ApJ...794L...6M}. This seems to
suggest that the dominant dynamo modes excited in the simulations are
largely insensitive to the details of the mean-field models. However,
rigorous studies of this issue are currently missing.

\section{Conclusions and outlook}
\label{sec:conclusions}

Significant progress has been made in both the mean-field models and
3D simulations of astrophysical dynamos in the last two decades or
so. Simulations corresponding to classical $\alpha^2$ and
$\alpha\Omega$ are the most well studied and their behavior, including
aspects of non-linear evolution, are now quite well understood in the
framework of mean-field dynamos including magnetic helicity
conservation. The situation is significantly less straightforward for
spherical shell dynamos aiming to reproduce solar and stellar
dynamos. Comparisons between such simulations and corresponding
mean-field models are still challenging and often marred with issues
in the computation of turbulent transport coefficients and issues
related to non-linearity and non-locality. It is remarkable that the
mean-field models reproduce the simulations results to the degree that
they do despite the various shortcomings in the modeling
approaches. However, in many cases the dynamos in simulations can be
interpreted in terms of relatively simple classical $\alpha\Omega$ or
$\alpha^2\Omega$ dynamos.

The main challenge in the interpretations of 3D simulations with
mean-field theory and models continues to be the computation of
turbulent transport coefficients. The numerical cost of computing all
of the (possibly scale dependent) coefficients with, e.g., test field
methods is high and increases further when non-linearity is taken into
account. Furthermore, taking all of these effects into account
increases the complexity of the resulting mean-field models
immensely. The increasing complexity works against the main premise of
mean-field modeling that the salient large-scale physics can be
represented by a model with far fewer degrees of freedom than in the
original simulation. While test field methods are likely to be
developed further, another possibility is to use machine learning
methods on numerical data that can perhaps extract the
interdependencies in a more compact form in the future.

\begin{acknowledgements}
I gratefully acknowledge the detailed comments from two referees that
led to significant improvement of the review. I wish to thank Igor
Rogachevskii and Nishant Singh for their valuable comments on the
manuscript. This work was partly supported by the Deutsche
Forschungsgemeinschaft Heisenberg programme (grant No.\ KA 4825/4-1)
and by the Munich Institute for Astro-, Particle and BioPhysics
(MIAPbP) which is funded by the DFG under Germany's Excellence
Strategy – EXC-2094 – 390783311. I also thank the Isaac Newton
Institute for Mathematical Sciences, Cambridge, for support and
hospitality during the programme DYT2 where part of the work on this
review was undertaken. This work was supported by EPSRC grant
EP/R014604/1. This work was also partially supported by a grant from
the Simons Foundation. Finally, I acknowledge the stimulating
discussions with participants of the Nordita Scientific Program on
``Stellar Convection: Modeling, Theory and Observations”, in August
and September 2024 in Stockholm.
\end{acknowledgements}

\section*{Conflict of interest}

The author declares that he has no conflict of interest.

\phantomsection
\addcontentsline{toc}{section}{References}
\bibliographystyle{spbasic-FS}
\bibliography{bib_global}

\end{document}